\documentclass[twocolumn]{aastex631}

\usepackage[dvipsnames]{xcolor}

\newcommand{\bt}{\textcolor{blue}}

\begin{document}

\title{Nancy Grace Roman Space Telescope Wide Field Instrument: Bright Point Source Saturation Response and Persistence Properties from Thermal-Vacuum Testing}

\author[0000-0002-2457-272X]{Dana R. Louie}
\affiliation{Catholic University of America, Department of Physics, Washington, DC, 20064, USA}
\affiliation{Exoplanets and Stellar Astrophysics Laboratory (Code 667), NASA Goddard Space Flight Center, Greenbelt, MD 20771, USA}
\affiliation{Center for Research and Exploration in Space Science and Technology II, NASA/GSFC, Greenbelt, MD 20771, USA}
\correspondingauthor{Dana R. Louie}
\email{dana.r.louie@nasa.gov}

\author[0000-0002-4235-6369]{Robert F. Wilson}
\affiliation{Department of Astronomy, University of Maryland, College Park, MD 20742, USA}
\affiliation{Exoplanets and Stellar Astrophysics Laboratory (Code 667), NASA Goddard Space Flight Center, Greenbelt, MD 20771, USA}
\affiliation{Center for Research and Exploration in Space Science and Technology II, NASA/GSFC, Greenbelt, MD 20771, USA}

\author[0000-0001-7139-2724]{Thomas Barclay}
\affiliation{Exoplanets and Stellar Astrophysics Laboratory (Code 667), NASA Goddard Space Flight Center, Greenbelt, MD 20771, USA}

\author[0000-0001-5347-7062]{Joshua E. Schlieder}
\affiliation{Exoplanets and Stellar Astrophysics Laboratory (Code 667), NASA Goddard Space Flight Center, Greenbelt, MD 20771, USA}


\author[0009-0006-1538-6286]{Nicholas Bond}
\affiliation{NASA Goddard Space Flight Center, Greenbelt, MD 20771, USA}
\affiliation{ADNET Systems, Inc., 6720B Rockledge Drive, Suite 504, Bethesda, MD 20817, USA}

\author[0000-0003-0430-3335]{Evan Bray}
\affiliation{KBR, 8800 Greenbelt Road, Greenbelt, MD 20771, USA}

\author[0000-0002-6621-8921]{Mario Cabrera}
\affiliation{Teledyne Imaging Sensors, Camarillo, CA 93012, USA}

\author[0009-0007-7745-0113]{Stephanie Cheung}
\affiliation{Nancy Grace Roman Space Telescope (Code 448), NASA Goddard Space Flight Center, Greenbelt, MD 20771, USA}
\affiliation{General Atomics Electromagnetic Systems, Space Systems Division, Englewood, CO 80112, USA}

\author[0000-0002-5636-233X]{Ami Choi}
\affiliation{Observational Cosmology Laboratory (Code 665), NASA Goddard Space Flight Center, Greenbelt, MD 20771, USA}

\author{Analia Cillis}
\affiliation{University of Maryland Baltimore County, Baltimore, MD 21250, USA}
\affiliation{Detector Characterization Laboratory (Code 521), NASA Goddard Space Flight Center, Greenbelt, MD 20771, USA}
\affiliation{Center for Research and Exploration in Space Science and Technology II, NASA/GSFC, Greenbelt, MD 20771, USA}

\author{Nicholas Collins}
\affiliation{Telophase Corporation, 7501 Forbes Boulevard, Suite 101, Lanham, MD  20706}
\affiliation{Heliospheric Physics Laboratory (Code 672), NASA Goddard Space Flight Center, Greenbelt, MD 20771, USA}

\author[0000-0001-5978-3247]{Tyler D. Groff}
\affiliation{Optics, Lasers, and Photonics Branch (Code 522), NASA Goddard Space Flight Center, Greenbelt, MD 20771, USA}

\author{Robert J. Hill}
\affiliation{221 Hardwood Ct, Hardy, VA 24101, USA}

\author[0000-0002-5861-7236]{Jeffrey Kruk}
\affiliation{Observational Cosmology Lab (Code 665), NASA Goddard Space Flight Center, Greenbelt, MD 20771, USA}

\author[0000-0002-5982-566X]{Gregory Mosby}
\affiliation{Observational Cosmology Lab (Code 665), NASA Goddard Space Flight Center, Greenbelt, MD 20771, USA}

\author[0000-0002-7517-9223]{Jennie Paine}
\affiliation{Center for Space Sciences and Technology, University of Maryland, Baltimore County, Baltimore, MD 21250, USA}
\affiliation{Observational Cosmology Laboratory (Code 665), NASA Goddard Space Flight Center, Greenbelt, MD 20771, USA}
\affiliation{Center for Research and Exploration in Space Science and Technology II, NASA/GSFC, Greenbelt, MD 20771, USA}

\author[0000-0003-2662-6821]{Bernard J. Rauscher}
\affiliation{Rauscher Scientific LLC, Sykesville, MD, 21784, USA}

\author[0009-0000-6700-8144]{Timothy A. Reichard}
\affiliation{Nancy Grace Roman Space Telescope (Code 448), NASA Goddard Space Flight Center, Greenbelt, MD 20771, USA}
\affiliation{ADNET Systems, Inc., 6720B Rockledge Drive, Suite 504, Bethesda, MD 20817, USA}

\author[0000-0002-5073-3806]{Maxime Rizzo}
\affiliation{NASA Goddard Space Flight Center, Greenbelt, MD 20771, USA}

\author{Scott Rohrbach}
\affiliation{Optics, Lasers, and Photonics Branch (Code 522), NASA Goddard Space Flight Center, Greenbelt, MD 20771, USA}

\author[0000-0002-9526-3780]{Nicole Schanche}
\affiliation{Department of Astronomy, University of Maryland, College Park, MD 20742, USA}
\affiliation{Exoplanets and Stellar Astrophysics Laboratory (Code 667), NASA Goddard Space Flight Center, Greenbelt, MD 20771, USA}

\author{Eric R. Switzer}
\affiliation{Observational Cosmology Laboratory (Code 665), NASA Goddard Space Flight Center, Greenbelt, MD 20771, USA}

\begin{abstract}

The Nancy Grace Roman Space Telescope's Wide Field Instrument (WFI) will observe hundreds of thousands of bright stars across its Core Community Surveys, particularly in the dense stellar fields of the Galactic Bulge Time Domain Survey (GBTDS). Sources brighter than $\sim$17th magnitude will saturate WFI detector pixels in typical survey exposures, with the brightest stars deeply saturating large pixel regions and potentially producing persistence signals that may impact subsequent observations. Prior detector characterization did not explore the regime of deep point source saturation. To address this gap, we conducted a bright star saturation test during WFI's second Thermal Vacuum test campaign (TVAC2) at BAE Space \& Mission Systems in Boulder, CO. Using the Stimulus of Ray Cones (SORC) telescope simulator, we projected nine in-focus point sources through the F146 filter onto two Sensor Chip Assemblies (SCAs), with fluxes tuned to approximate stellar magnitudes ranging from $\sim$4 to $\sim$18 in $\sim$170~s exposures. We present analyses of the saturation response and persistence properties of these detectors. We find that the saturated region of a $\sim$4 mag source grows to $\sim$150 pixels in diameter after $\sim$170~s of illumination, compared to $\sim$15 pixels for a $\sim$12 mag source. Pixels adjacent to the expanding saturation front exhibit pronounced non-linear behavior consistent with charge leakage from saturated neighbors. For persistence, we find that the median signal in the first post-illumination dark exposure is broadly consistent across source magnitudes spanning $\sim$4 to $\sim$17, and that persistence decays to detector background levels ($\lesssim$0.05 e$^{-}$ s$^{-1}$) within approximately 20 minutes, consistent with flat field persistence measurements from the same TVAC2 campaign. These pre-flight characterization results inform the astronomical community's understanding of WFI detector response in preparation for Roman science operations. 
We make our analysis products publicly available and encourage the community to pursue further analyses.

\end{abstract}


\section{Introduction} \label{sec:intro}

\begin{figure} [ht!]
\centering
\includegraphics[width=0.5\textwidth]{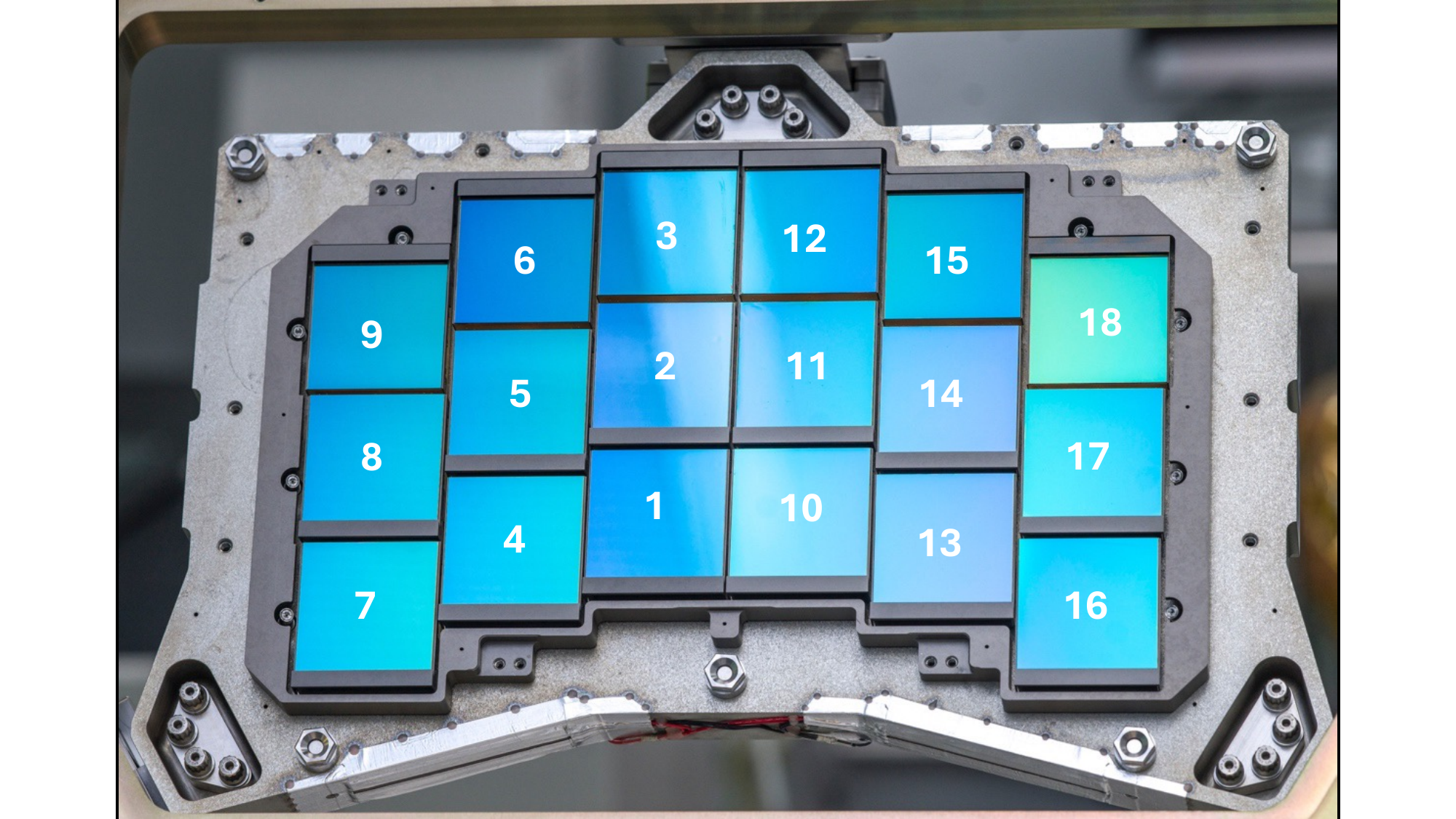}
\caption{ Wide Field Instrument (WFI) Focal Plane Array (FPA). Sensor Chip Assemblies (SCAs), or detectors, are referred to by number. Our test analyzed the saturation response and persistence properties of detectors 4 and 11 to nine different bright point sources designed to approximate stars ranging in magnitude from $\sim$4 to $\sim$18. }
 \label{fig:WFI_FocalPlaneArray}
\end{figure}
\begin{figure}[ht!]
\centering
\includegraphics[width=0.5\textwidth]{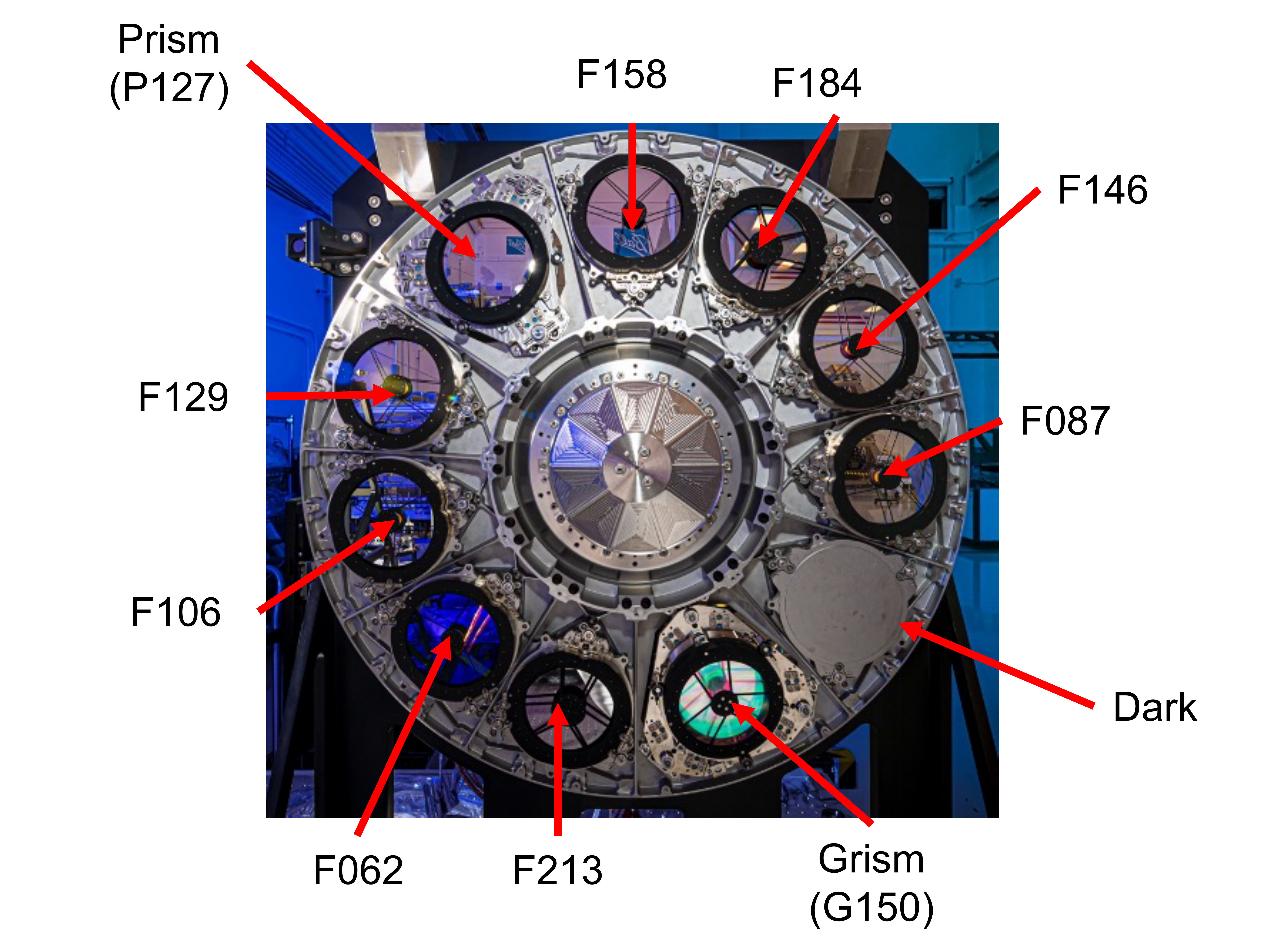}
\caption{ Wide Field Instrument (WFI) Element Wheel Assembly (EWA). The EWA is equipped with 8 filters, the G150 grism (1.0 -- 1.93 $\mu$m), and the P127 prism (0.75 -- 1.80 $\mu$m). 
 \label{fig:WFI_ElementWheelAssembly}}
\end{figure}

The Nancy Grace Roman Space Telescope (Roman) is NASA's next Astrophysics Flagship Mission and is designed to conduct large scale astrophysical surveys. Roman's primary science instrument is the Wide-Field Instrument (WFI), a 300 megapixel optical-to-near-infrared (0.48 to 2.3 $\mu$m) camera equipped with 18 Teledyne H4RG-10 Sensor Chip Assemblies \citep[SCAs,][]{Mosby2020} spread across a 6 x 3 Focal Plane Array (FPA, Figure \ref{fig:WFI_FocalPlaneArray}). The WFI has a 0.28 square-degree field-of-view, which is approximately 200 times greater than that of the Hubble Space Telescope Wide-Field Camera 3 (WFC3) IR channel,\footnote{The Roman FOV is 100 times that of Hubble's Advanced Camera for Surveys (ACS). See, for example, \url{https://svs.gsfc.nasa.gov/13583}.} while retaining an 0.11$^{\prime\prime}$/pix spatial resolution and achieving greater sensitivity.\footnote{For further technical details, refer to \url{https://github.com/RomanSpaceTelescope} and \url{https://science.nasa.gov/mission/roman-space-telescope/wfi-technical/}} Additionally, Roman is approximately 1000 times faster than Hubble when performing large area surveys due to faster slew and settle times, as well as unobstructed views from a Sun-Earth L2 orbit. Each of Roman's HgCdTe SCAs measures 4096 x 4096 pixels, with a 10 $\mu$m pixel pitch, and is tuned to a 2.5 $\mu$m wavelength cut-off \citep{Mosby2020,Mosby2025}.  An Element Wheel Assembly provides 8 imaging filters, a grism and prism for slitless spectroscopy, and a dark element for calibrations \citep[][Figure \ref{fig:WFI_ElementWheelAssembly}]{Bray2024,Cromey2023,Cromey2025,Eegholm2025}. \cite{Schlieder2024} provide more detailed information regarding WFI's design, subsystems, and performance. 

In this work, we examine saturation response of Roman's H4RG-10 detectors after exposure to bright point sources, as well as the ensuing persistence properties. A given detector pixel under constant illumination does not accumulate charge at a constant rate, and also has a finite amount of total charge that can be accumulated, referred to as its full well capacity. \textit{Saturation} occurs when a given pixel reaches its full well capacity.\footnote{See \url{https://roman-docs.stsci.edu/roman-instruments/the-wide-field-instrument/wfi-detectors/sources-of-pixel-to-pixel-variation}.} As a given pixel is subjected to saturating illumination, charge traps within the pixel fill. When the illumination is removed, a faint lingering signal, \textit{persistence}, can be detected as the traps slowly release their charge.\footnote{See \url{https://roman-docs.stsci.edu/roman-instruments/the-wide-field-instrument/wfi-detectors/non-ideal-detector-effects}.} 

Past work has explored persistence characteristics of the H2RG predecessor to Roman's detectors. For example,  \cite{RauscherPASP2014} reported persistence characteristics for Teledyne H2RG detectors designed for use on the James Webb Space Telescope (JWST) Near Infrared Camera (NIRCam) based upon ground performance testing. \cite{Regan_SPIE2018} developed a model to correct persistence in JWST H2RG detectors based upon a decay model of traps. Similarly, \cite{TullochJATIS2019} built a predictive model of persistence based upon a trap characterization method for Teledyne H2RG detectors located at the European Southern Observatory (ESO). The Euclid mission, which launched in 2023, hosts 16 H2RG detectors in the Near Infrared Spectrometer and Photometer (NISP) focal plane array. \cite{Kubik2024} present the NISP empirical persistence model derived from ground characterization data, and compare it to persistence measured during on-orbit calibration. 
Additionally, recent work has examined persistence characteristics of Roman H4RG detectors in laboratory conditions, as well as on ground-based telescopes \citep[e.g.,][]{Mosby2020, Mosby2025}.

\begin{figure*}[ht!]
\centering
\includegraphics[width=1.0\textwidth]{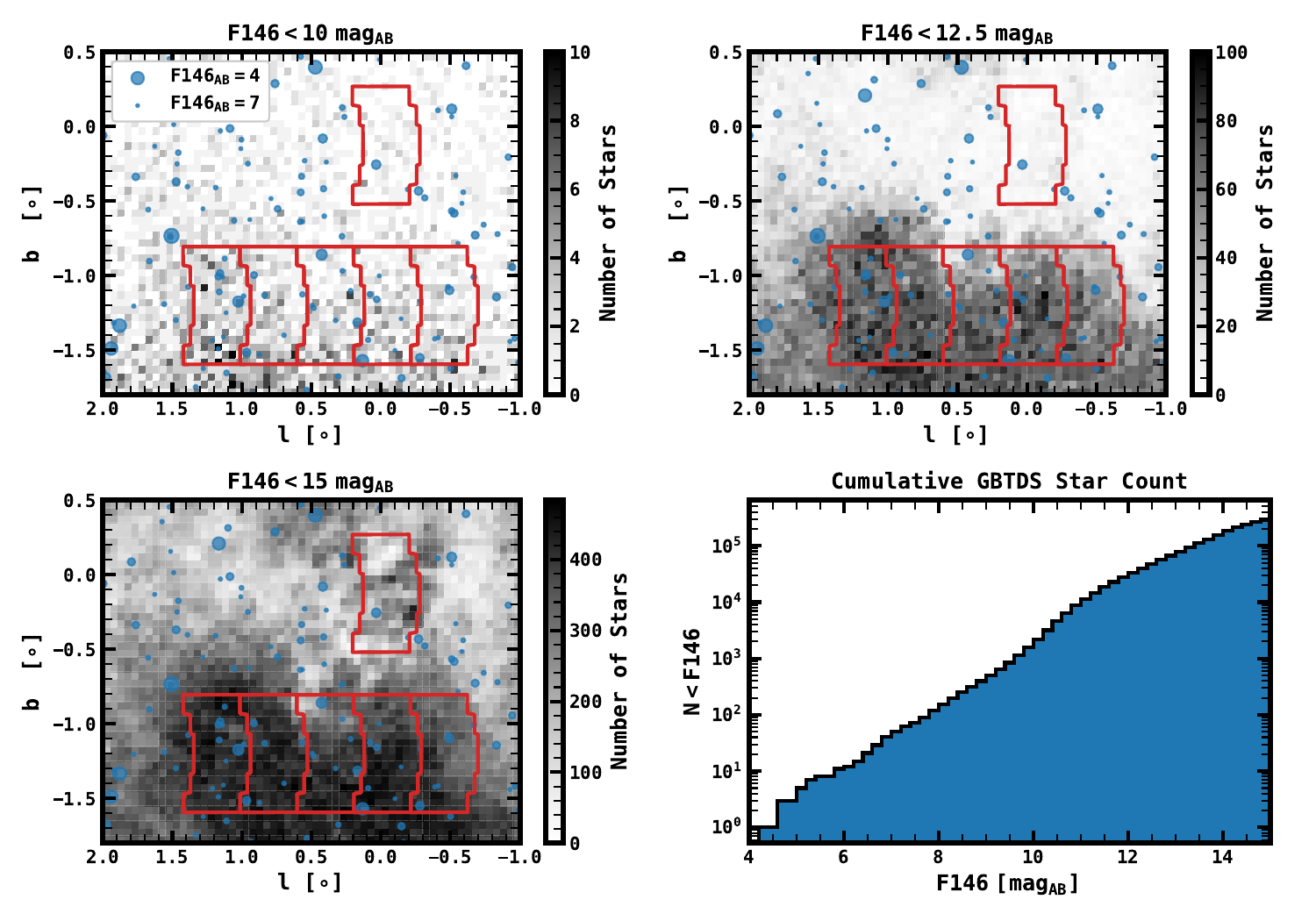}
\caption{ The approximate number and distribution of saturated stars in the GBTDS based on archival 2MASS data. The shading shows the number of expected stars with $F146 < 10~\mathrm{mag_{AB}}$ (top left), $F146 < 12.5~\mathrm{mag_{AB}}$ (top right), and $F146 < 15~\mathrm{mag_{AB}}$ (bottom left) in 0.0025 deg$^2$ bins. The blue points denote the locations of stars with $F146<7.5\,\mathrm{mag_{AB}}$, with the size of the points indicative of their brightness. The red outlines denote the Roman footprint in the six GBTDS pointings \citep{ROTAC2025}. The cumulative total number of stars in the GBTDS footprint brighter than a given magnitude are shown in the bottom right panel. 
 \label{fig:gbtds}}
\end{figure*}

During its 5-year primary mission, Roman's WFI observations will include three Core Community Surveys (CCSs) focused on mission science objectives of dark energy cosmology and exoplanet demographics \citep{ROTAC2025} and additional General Astrophysics Surveys defined by the community\footnote{The first of these, the Galactic Plane Survey, has been defined by a community led process \citep{GalacticPlaneSurvey2025}}. Sky locations and integration times vary widely across each of the three CCSs, leading to a large diversity of source fluxes. When combined with the high sensitivity resulting from Roman's 2.4m aperture and next generation, cryogenic detectors, source saturation and the resulting persistence signal must be considered. Roman SCA testing in the Detector Characterization Lab (DCL) at NASA's Goddard Space Flight Center (GSFC) did not characterize SCA pixel response to deeply saturated point sources. Saturation and persistence were investigated as part of detector flight acceptance testing \citep{Mosby2025}, but used flat field illumination and only reached flux levels equivalent to approximately 15 or 16th magnitude stars. Therefore, the data necessary to characterize deep saturation and resulting persistence in a small region of an SCA were not available. Additionally, a lab set-up that could perform such tests did not exist in the DCL. We consequently considered tests during WFI Thermal Vacuum Testing (TVAC) to characterize these effects.

To bound the problem and inform test planning, we used archival data to investigate the number of bright stars in Roman's CCS with the highest density stellar fields, the Galactic Bulge Time Domain Survey \cite[GBTDS,][]{Penny2019}.\footnote{See interim report at  \url{https://roman.gsfc.nasa.gov/science/ccs/Core_Community_Survey_Reports-rev03-compressed.pdf}} Starting with the proposed GBTDS fields described in \cite{Penny2019}, we used the 2MASS point source catalog \citep{skrutskie2006} and an approximate relation to convert from $JHK$ magnitudes to $F146$ \citep{wilson2023}\footnote{$F146_{\rm AB} = (J_{\rm AB}+H_{\rm AB}+0.5 K_{\rm AB})/2.5$ }, to estimate that Roman will observe several hundred thousand stars brighter than 15th mag in that survey. In a typical $\approx$1 min GBTDS exposure, Roman SCAs will reach saturation for $\approx$17th mag stars\footnote{\url{https://roman.gsfc.nasa.gov/science/WFI_technical.html}}. Thus, these bright sources will deeply saturate SCA pixels in the point spread function (PSF) core and beyond, leading to potentially long lasting persistence in those SCA regions. We then revisited the analysis using the GBTDS fields defined by a community survey definition process and formally adopted by the Roman Project \citep{ROTAC2025}  and found $\sim$300,000 stars brighter than 15th mag, $\sim$1600 stars brighter than 10th mag, and $\sim$40 stars brighter than 7th mag (see Figure \ref{fig:gbtds}).  In Roman's largest area surveys, like the High Latitude Wide Area Survey (HLWAS) CCS, there will be stars significantly brighter than 7th mag, with some sources as bright as 3rd or 4th mag. 

With the expectation of significantly saturated sources in Roman surveys, we planned a Bright Star Saturation test during WFI's second TVAC (TVAC2), which took place in the Titan chamber at BAE Space \& Mission Systems in Boulder, CO over two months between March and May 2024. TVAC2 was designed to verify the previously established TVAC1 performance baseline and to perform instrument characterization and calibration in a flight-like environment. The bright star test was designed as a risk reduction effort to better understand detector performance in flight surveys, and was not used to verify any instrument requirements.

In the remaining sections of this manuscript, we first describe our methods, including technical details of the test, followed by our analysis approach to explore baseline saturation and persistence behavior ($\S$\ref{sec:methods}). We then describe the results of these analyses ($\S$\ref{sec:results}), and finally discuss the general implications of these results for Roman survey science and discuss insights for future work ($\S$\ref{sec:DiscussionConclusion}). We also make our interim data products available to the astronomical community for continued analysis to better understand how SCA pixels respond to point source saturation and provide details on where to find the full TVAC2 test dataset for even deeper analyses.

\section{Methods} \label{sec:methods}

During the test, we used a NASA/GSFC-developed telescope simulator, known as the Stimulus of Ray Cones (SORC, pronounced ``source"), to project simulated point sources onto multiple SCA locations (Wake and Lyons et al., in prep). The test was performed on SCAs 4 and 11 (Figure \ref{fig:WFI_FocalPlaneArray}), which we chose to allow exploration of a diverse number of pixels with varying performance properties over a range of angles of incidence (AOIs), which impacts PSF shape. Additionally, we avoided using SCAs involved in subsequent best focus and wave front error tests during the TVAC2 campaign to preclude any long-lasting persistence impacts. In this section, we first describe the test procedures (\S\ref{subsec:methods_TVAC2_TestProcedure}), followed by data analysis methods for both the saturation and persistence analyses  (\S\ref{subsec:methods_TestAnalysis}).

\subsection{TVAC2 Test Procedures}\label{subsec:methods_TVAC2_TestProcedure}

\begin{figure}[ht!]
\centering
\includegraphics[width=0.5\textwidth]{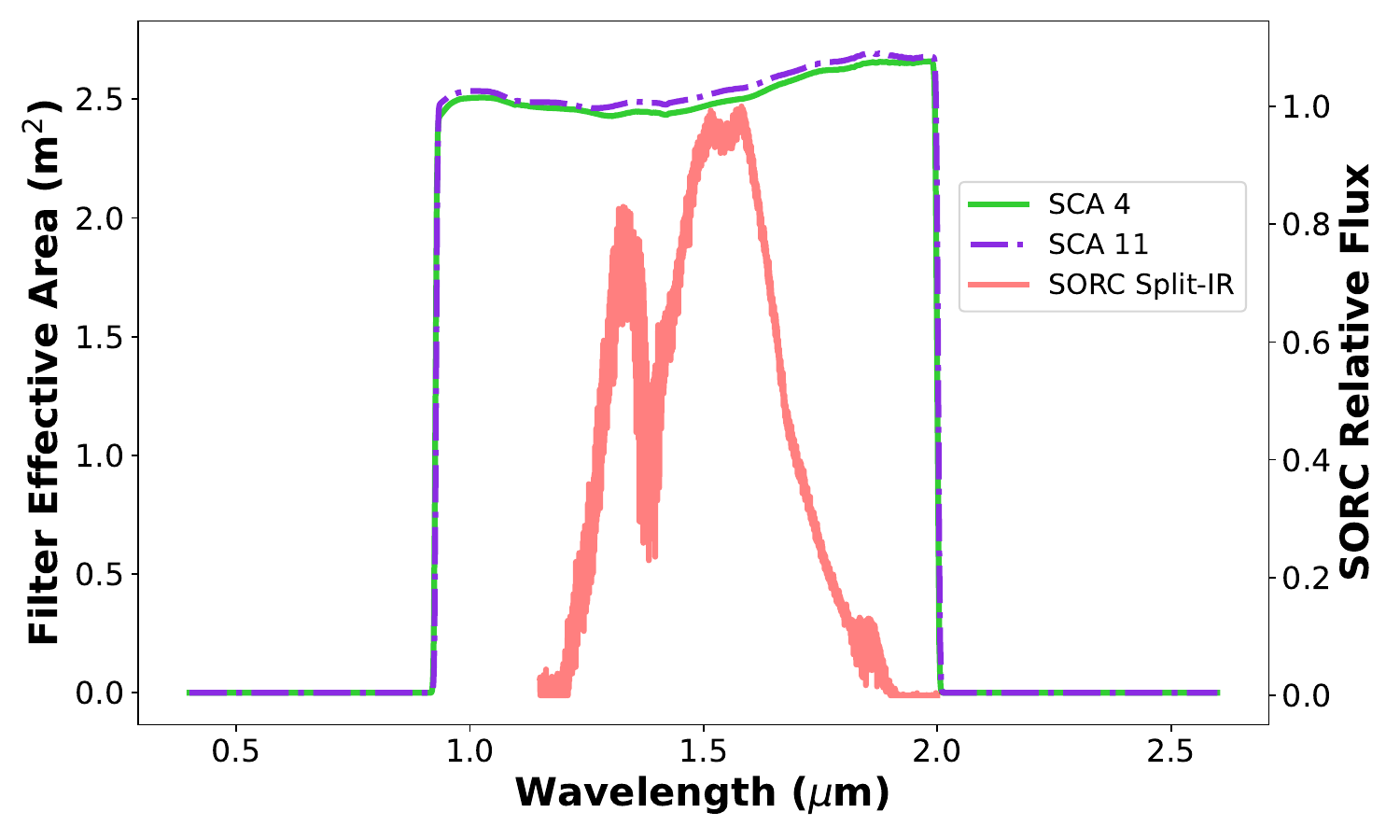}
\caption{ Comparison of the F146 Filter bandpass for SCAs 4 and 11 with the SORC SPLIT-IR input spectrum used in the TVAC2 Bright Star test. The F146 filter spans wavelengths from 0.927 to 2.0 $\mu$m and The SORC SPLIT-IR spectrum fits within the F146 bandpass. The spectrum has a cut-on wavelength of 1.2 $\mu$m and extends to approximately 1.9 $\mu$m on the red end, with significant structure caused by the supercontinuum laser source properties and associated spectral splitter optics.
\label{fig:F146_SORC_Comparison}}
\end{figure}

The SORC was critical for projecting simulated point sources approximating the stellar magnitudes ranging from $\sim$4 to $\sim$18 that were necessary for this test. At the heart of the SORC is a projector matched to the optical prescription of the Roman telescope ($\approx$f/8), mounted into a 5 degree-of-freedom (DOF) gimbal assembly and fed by a light source rack outside the TVAC chamber through optical fibers. The 5DOF projector enabled the SORC to project point sources anywhere on the WFI focal plane. The SORC is also able to flood the WFI with flat field illumination or thermal radiation, but these capabilities were not used during the bright star test. The Source Delivery System (SDS) light source rack provides both pulsed and continuous wave (CW) input across the full WFI bandpass with both monochromatic and spectroscopic output options \citep[][]{Schlieder2024}. To produce the point sources, we specifically used an NKT Photonics SuperK FIANIUM, a supercontinuum white light fiber laser source within the SDS, coupled with an NKT Photonics SuperK SPLIT-VIS/IR spectral splitter. The output of the IR arm of the splitter was a near-IR spectrum, which was projected through the WFI F146 filter to produce a broadband, in focus, PSF. Figure \ref{fig:F146_SORC_Comparison} compares the F146 filter bandpass with the SORC SPLIT-IR spectrum.\footnote{The F146 filter effective area curves for SCAs 4 and 11 were downloaded on 18 Feb 2026 from the following website: \url{https://science.nasa.gov/mission/roman-space-telescope/wfi-technical/}.} The SORC output flux levels were tuned to approximate the desired stellar magnitudes using optical attenuators in the SDS. The illumination was projected in CW mode to best approximate detector response to observations of point sources in flight. The test sequence was performed after a temperature transition from the Nominal Operation (Nom Op) plateau to Hot Qualification (Hot Qual) plateau. At Nom Op the WFI detectors were at 89.5 K, the temperature then rose $\sim$2 K during the transition to Hot Qual. During the Bright Star test, temperatures remained steady at $\sim$91 K.

\begin{figure} [ht!]
\centering
\includegraphics[width=0.5\textwidth]{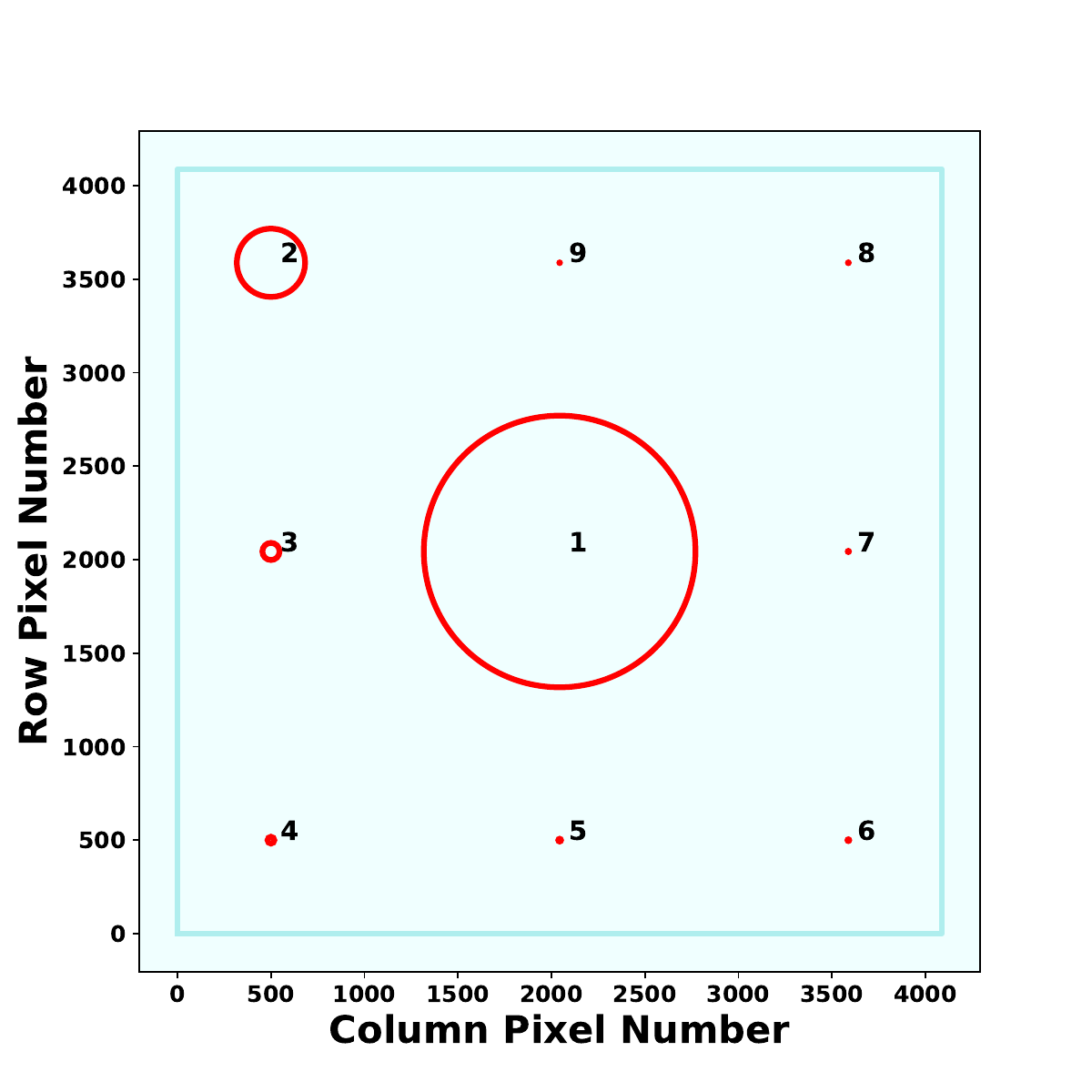}
\caption{ Locations of projected SORC PSFs upon the SCA, with approximate saturated region sizes indicated using red circles. SORC PSFs were projected in sequence from 1 to 9 as tabulated in column 1 of Table \ref{table:test_summary}. Locations of each projection are indicated by the numbers in the figure, with the first projection in the center, the second projection at upper left, and then proceeding counterclockwise around the SCA. 
\label{fig:SCAtest}}
\end{figure}

On both SCAs 4 and 11, we projected a sequence of 9 in-focus PSFs in a 3 x 3 grid, in the order shown and at the positions indicated in Figure \ref{fig:SCAtest}. The flux of the 9 PSFs was tuned to match the anticipated e$^-$/pixel rates of bright stars with magnitudes 4, 7, 10, 12, 14, 15, 16, 17, and 18, as listed in Table \ref{table:test_summary}. We projected each point source separately upon each SCA, following the order listed in Table \ref{table:test_summary}.

The electron counts listed in Table \ref{table:test_summary} refer to the number of electrons falling on the brightest pixel, and are meant to approximate the magnitudes listed in the second column. We estimated these counts during the test planning phase using the following procedures. First, we made use of published data showing that a magnitude 15 star observed using the F146 filter would saturate in 2.5 sec,\footnote{See \url{ https://science.nasa.gov/mission/roman-space-telescope/time-to-saturation-for-point-source-imaging/ }.} and we assumed that WFI pixels would saturate at 80,000 e$^{-}$. With this knowledge, we estimated the total electron count for a magnitude 15 source over $\sim$170 sec, and then scaled that total electron count for other source magnitudes. Finally, we used the STScI \texttt{stpsf} tool\footnote{See \url{https://roman-docs.stsci.edu/simulation-tools/stpsf-space-telescope-point-spread-functions-for-roman/overview-of-stpsf}.} to estimate the brightest pixel flux fraction for a bright point source PSF observed through Roman’s WFI F146 filter. 

All exposures taken in the test consisted of 56 frames to emulate a typical Roman survey exposure. The first frame in each exposure was the reset-read frame, followed by 55 science frames, where signal was integrated in the detector pixels. The reset-read frame marks the beginning of an exposure, when accumulated charge from the previous exposure is cleared from the detector  \citep{casertano2022, Betti2024}. The resulting integration time from the 55 science frames was \(55 {\rm \ frames}\ \times \ 3.16\ \frac{\rm sec}{\rm frame}\ = 173.8 \ {\rm sec} \ \approx \ 170 \ {\rm sec} \).\footnote{3.16 seconds is the approximate exposure time per frame. Please see \url{https://roman-docs.stsci.edu/roman-instruments/the-wide-field-instrument/observing-with-the-wfi/wfi-multiaccum-ma-tables/imaging-multiaccum-tables}} In this paper we assign index 0 to the reset-read frame, with indices 1 through 55 assigned to the science frames. The exposures were acquired in the following sequence:
\begin{enumerate}
\item Three dark exposures with the EWA set to F146 and SORC not projecting were collected at the beginning of the test as a baseline. We intended the EWA to be placed in Dark for the initial dark exposures, but the Roman Test Planning Tool (TPT)\footnote{The TPT is a custom software package designed to draft, organize, and optically model the flight and ground system hardware configurations for thermal vacuum tests like this Bright Star Test.} settings had not been updated.
\item With the EWA in F146, the SORC projected the nine illuminated point sources upon SCA 4, following the sequence in Table \ref{table:test_summary}.
    \begin{itemize}
    \item One exposure was taken for the illuminated point source. 
    \item With SORC projection turned off and the EWA left at F146, one interleaved dark exposure was taken following the illuminated exposure. The interleaved dark allowed us to capture persistence decay on a timescale of minutes. 
    \item The SORC projector was moved to the next pose location for the next point source and the illumination~$\rightarrow$~interleaved dark procedure was repeated until completing the sequence with the magnitude $\sim$18 source. 
    \end{itemize}
\item The same exposure sequence was repeated for SCA 11.
\item After the sequences of illuminated exposures plus interleaved darks were completed for both SCAs 4 and 11, eighteen consecutive dark exposures were taken with the EWA set to Dark to monitor any long-term persistence and capture a post-test baseline.
\end{enumerate}

\begin{table*}[htpb]
\caption{Point Source Saturation Test Summary} \label{table:test_summary}
        \centering  
  	\begin{tabular}{ c c c c }        
  	\hline\hline
    SORC Projection & Point Source &  Electrons at SCA & Location \\
     Order Upon SCA & Simulated (mag) & (e$^-$ in $\sim$170 sec)  & (Figure \ref{fig:SCAtest}) \\
    \hline
    1 & 4  & 4.7 $\times$ 10$^{10}$ & Center \\
    2 & 7  & 2.9 $\times$ 10$^9$ & Upper left \\
    3 & 10  & 1.9 $\times$ 10$^8$ & Center left \\
    4 & 12  & 2.9 $\times$ 10$^7$ & Lower left \\
    5 & 14  & 4.7 $\times$ 10$^6$ & Lower center \\
    6 & 15  & 1.9 $\times$ 10$^6$ & Lower right \\
    7 & 16  & 7.4 $\times$ 10$^5$ & Center right \\
    8 & 17  & 2.9 $\times$ 10$^5$ & Upper right \\
    9 & 18  & 1.17 $\times$ 10$^5$ & Upper center \\
    \\ [-2.5 ex]
    \hline
    \end{tabular}
\end{table*}

\subsection{Data Analysis}\label{subsec:methods_TestAnalysis}

We begin in Section \ref{subsubsec:methods_General} by describing data processing and preparation methods applied to both illuminated and dark data. Then, we describe specific methods used to investigate saturation in Section \ref{subsubsec:methods_Saturation} and persistence in Section \ref{subsubsec:methods_Persistence}, respectively.

\subsubsection{General Procedures}\label{subsubsec:methods_General}

\begin{figure*}[ht!]
\centering
\includegraphics[width=1.0\textwidth]{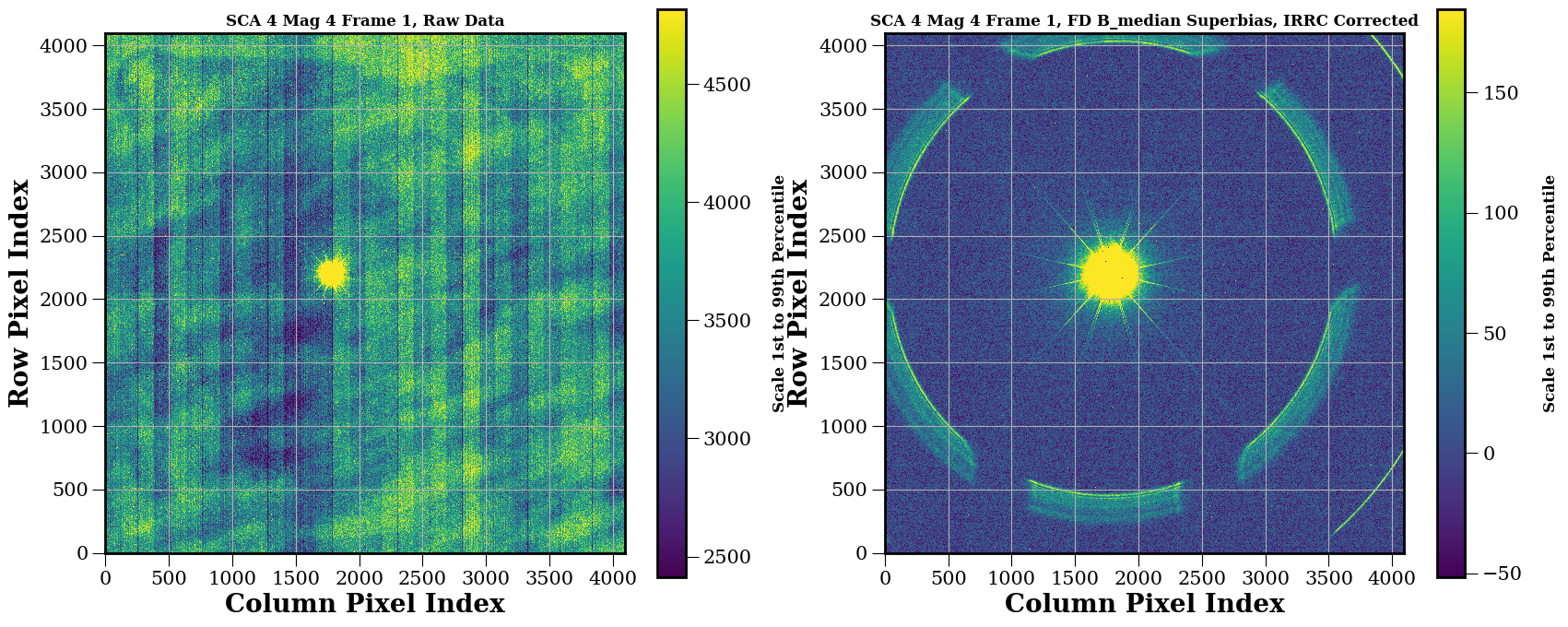}
\caption{ SCA 4 first science frame (after reset-read frame) for the $\sim$4 mag source. We show the raw data at \textbf{\underline{left}}, and the IRRC-corrected and superbias-subtracted frame on the \textbf{\underline{right}}, showing significant improvement. The arcs surrounding the central bright source in the right-hand panel are stray light artifacts caused by the SORC projector optics. The same structured stray light is not expected from the Roman telescope assembly in flight.
 \label{fig:methods_BeforeAfterIRRC}}
\end{figure*}

To mitigate $1/f$ noise, we began our analysis by applying the Improved Roman Reference Correction algorithm \citep[IRRC, ][Rauscher et al., in prep]{Betti2024} to all exposures. We subtracted a superbias frame from all data as part of the IRRC correction. In Figure \ref{fig:methods_BeforeAfterIRRC}, we compare a science frame before and after the IRRC correction and superbias subtraction.

A superbias is a high-quality, low-noise reference bias frame created by combining multiple individual bias frames. We created a superbias frame by taking the median of the bias frames for the final three post-test dark exposures, which were found to be free from any lingering persistence so long after illumination. To create the individual bias frames for each of the three final dark exposures, we applied a linear fit of the form $s = a + b\times t$ to each pixel, where $s$ is the recorded signal, $a$ is bias, $b$ is slope, and $t$ is time. Thus, the bias for each post-test dark exposure is given by the value of $a$ computed using this fit for each exposure. Following IRRC application, we multiplied each individual frame by per-pixel photon transfer gain values. In Appendix Figures \ref{fig:methods_Superbias} and \ref{fig:methods_Gain}, we show the superbias frames and the pixel-level gain maps used in our analysis for both SCAs 4 and 11.

\subsubsection{Saturation Data Analysis}\label{subsubsec:methods_Saturation}

For this initial analysis we focus on the $\sim$12th mag simulated star on SCA 11. To evaluate the per-pixel effects of saturation, we generated cutouts of the super-bias and IRRC corrected up-the-ramp sampled data which contained the entirety of the pixels which eventually saturated by the end of the 56 frame exposure. We then applied classical non-linearity corrections to the data, using the linearity coefficients measured from TVAC2\footnote{\url{https://roman-crds-tvac.stsci.edu}}. To perform these corrections, we use the per-pixel gain map referenced above to convert back to DN. We then use the classical non-linearity corrected data, in units of DN, for the majority of the saturation data analysis.

After applying the classical non-linearity correction to the Mag 12 SCA 11 dataset, we noticed clear non-linearities in the ramps of saturated pixels. To further study this phenomenon, we calculated masks for each pixel in each frame, to define four saturation regimes: not saturated, adjacent \emph{diagonally} to a saturated pixel, adjacent in row/column to a saturated pixel, and saturated. 
To define the regime for each pixel in each frame we utilized the frames before any linearity corrections or other treatments are applied. We first started with a convenient definition for saturation, for which we define both of two scenarios are met: the accumulated signal is constant in the subsequent frame, and the total accumulated signal, in the uncorrected frames, is above 63,500 DN, which serves as a baseline for lower limit of the well depth in the SCA. For the purpose of this analysis, we used 55 frames (we exclude the initial reset frame) to define the saturation regimes, but only the first 54 frames for analysis. As a result, pixels which saturate in frame 55 would not be labeled as saturated, but pixels which saturate in frame 54 would be identified by the constant signal from frames 54 to 55.

After all the saturated pixels are defined, we then identified the pixels that were diagonal and adjacent in row/column to the saturated pixels by convolving the mask of saturated pixels by 2d arrays of [[1,0,1],[0,0,0], [1,0,1]] and [[0,1,0],[1,0,1],[0,1,0]], respectively. While this analysis can be refined, the basic algorithms perform adequately to define the saturation regimes of interest for this work.

Finally, the next step is to characterize the slopes of the accumulated signal in each pixel in each frame and compare the behavior across the four saturation regimes defined. For this purpose, we utilized the classical linearity corrected data. We define two different slopes on per-pixel bases: mean slope, and the instantaneous slope. The instantaneous slope of a pixel is defined per frame, and is taken to be the difference in accumulated signal of the current frame from the subsequent frame. The mean slope, on the other hand, is defined only per pixel and is taken to be the mean value of all the instantaneous slopes of that pixel calculated only from the frames in the ``not saturated" regime. We then use these definitions and framework to compare the instantaneous slope to the mean slope across all of these differing saturation regimes.

\subsubsection{Persistence Data Analysis}\label{subsubsec:methods_Persistence}

We relied upon the following data to complete our persistence analysis on SCAs 4 and 11 for each approximated source magnitude listed in Table \ref{table:test_summary}: 

\begin{itemize}
    \item the final frame of illuminated data for each source magnitude,
    \item the final three post-test dark exposures (exposures 16, 17,  and 18), and
    \item the interleaved dark exposures following illumination for each source magnitude.
\end{itemize} 

\begin{figure*}
\includegraphics[width=0.49\textwidth]{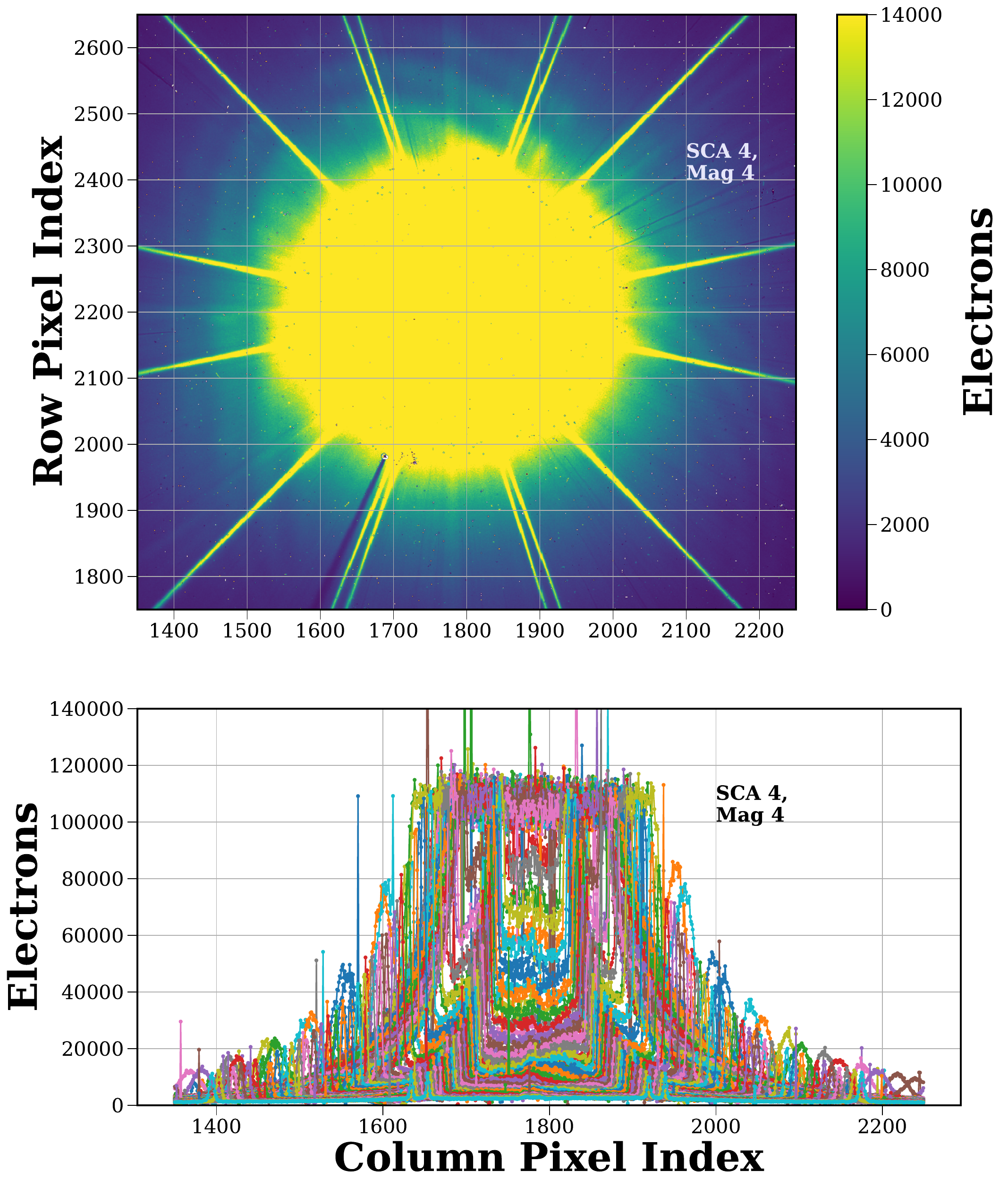}
\hspace{0.5 cm}
\includegraphics[width=0.49\textwidth]{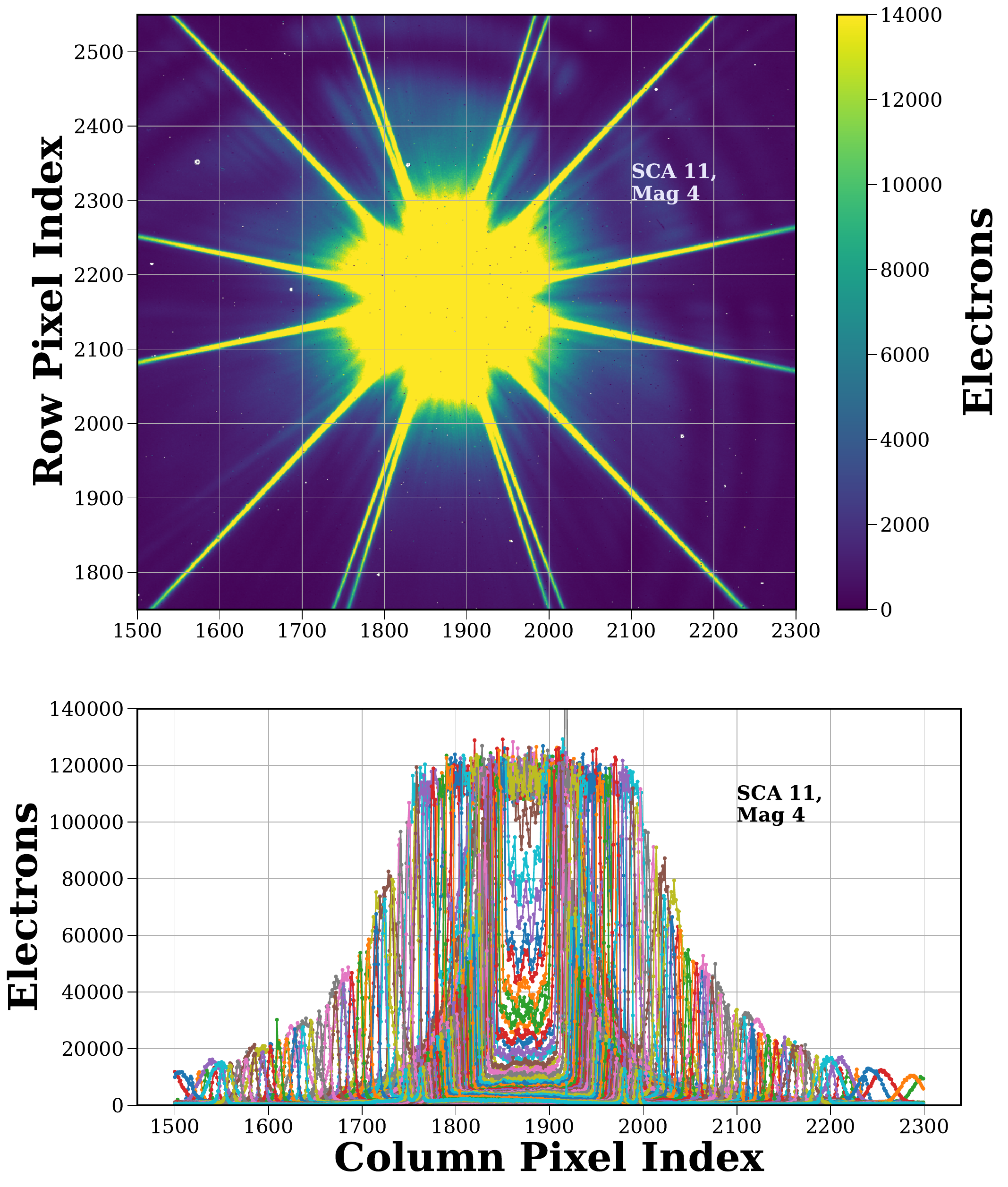}
\caption{ Regions of Interest (ROIs) for SCA 4 (\textbf{\underline{left}}) and 11 (\textbf{\underline{right}}) magnitude $\sim$4 sources. The \textbf{\underline{top}} panels show the final illuminated frames for the bright sources upon each SCA, while the \textbf{\underline{bottom}} panels depict the counts in electrons through each row of the saturation region. Our \textbf{\textit{saturation mask}} flagged those pixels with electron counts between 100,000 and 130,000, as evidenced by the flat top structure in the bottom panels. The PSF shape varies for each source due to the AOI differences between SCAs 4 and 11. SCA 4 is in the top right for the WFI focal plane array, leading to a broader PSF. SCA 11 is near the center, leading to a sharper PSF.}
\label{fig:frame56_4panel_ROIimage_rowCrossSec}
\end{figure*}

\begin{figure}
\includegraphics[width=0.45\textwidth]{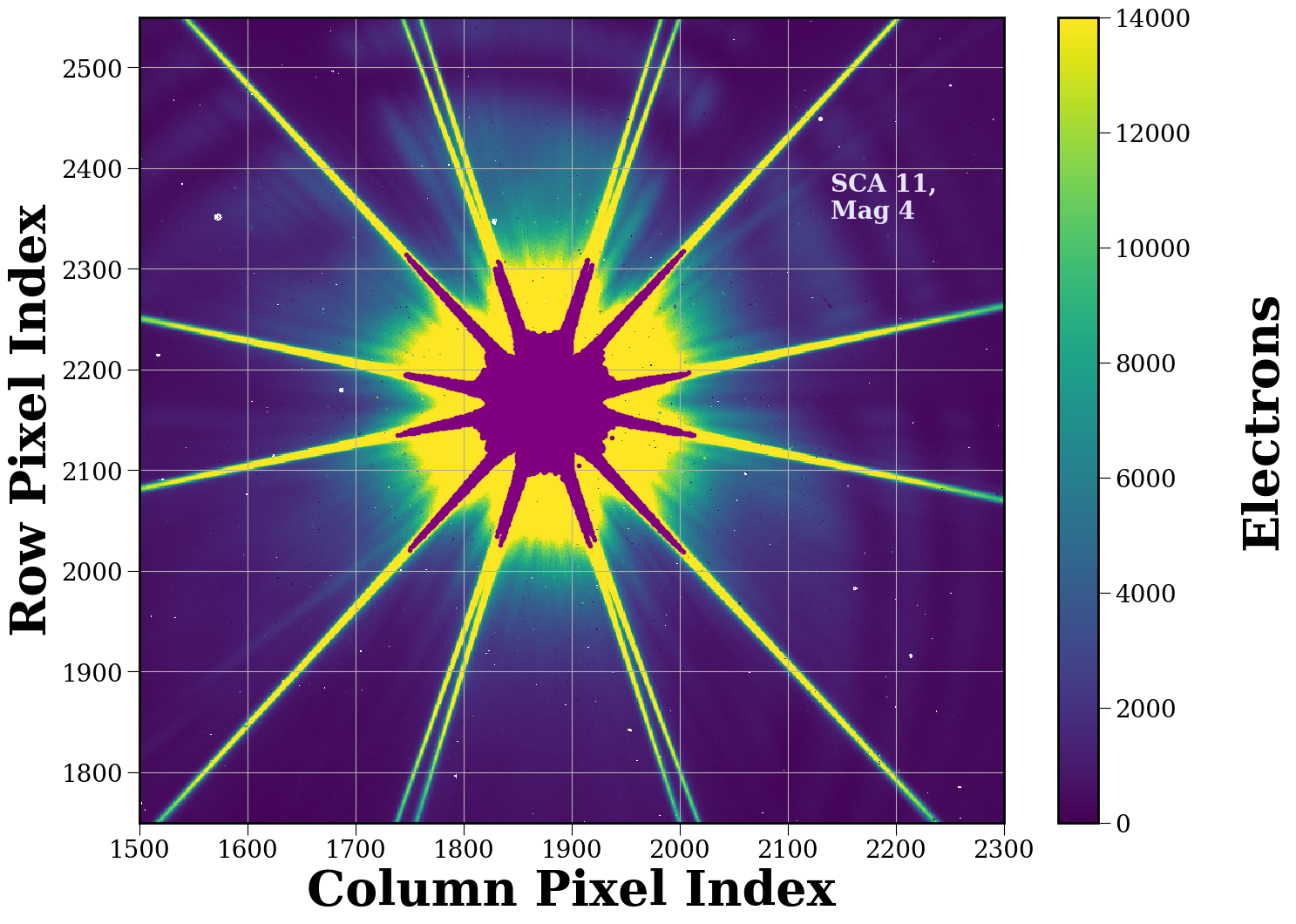}
\caption{ Region of Interest (ROI) for the SCA 11 magnitude 4 source overplotted with the saturation mask used in our persistence analysis. Those pixels that are part of the saturation mask are colored purple. Our \textbf{\textit{saturation mask}} flagged those pixels with electron counts between 100,000 and 130,000 (See Fig.~\ref{fig:frame56_4panel_ROIimage_rowCrossSec}). We eliminated bad pixels from our persistence analysis using a bad pixel mask derived from additional test data obtained during TVAC2. }
\label{fig:frame56_ROIimage_SatMaskoverplotted}
\end{figure}

We used the final frame of illuminated data to create a \textbf{\textit{saturation mask}} flagging those pixels that had reached saturation for each source magnitude in Table \ref{table:test_summary}. In Figure \ref{fig:frame56_4panel_ROIimage_rowCrossSec}, we show the final illuminated frames for SCAs 4 and 11 in a region of interest (ROI) surrounding the $\sim$4 mag source, as well as cross section views through the rows of the images. The saturated PSFs in the top panels have different shapes due to AOI effects across the WFI focal plane. SCA 4 is in the top-right portion of the array (broader PSF), and SCA 11 is toward the center (sharper PSF). In our final analysis, we flagged as saturated those pixels that reached between 100,000 and 130,000 e$^-$ in the final illuminated frame. We chose this range because it approximately encompasses the variation in full well depth across pixels in these SCAs, as demonstrated by the flat top structure seen in the bottom panels of Fig.~\ref{fig:frame56_4panel_ROIimage_rowCrossSec}. In Figure \ref{fig:frame56_ROIimage_SatMaskoverplotted}, we show the saturation mask for the SCA 11 $\sim$4 mag source overplotted onto the ROI for the final illuminated frame. We show similar plots for all sources projected onto SCAs 4 and 11 in Appendix Figures \ref{fig:SCA4frame56_OverplottedSatMasks_AllMags} and \ref{fig:SCA11frame56_OverplottedSatMasks_AllMags}.

A dark current frame was required to correct for dark current and compute persistence in each of the interleaved dark frames. Similar to creation of the superbias frame described in Section \ref{subsubsec:methods_General}, we again used the three final post-test dark exposures 16, 17, and 18 to derive the dark current for each pixel. Here, for each post-test dark exposure, we again applied a linear fit of the form $s = a + b\times t$ to each pixel, but in this case used the resulting \textit{slope} $b$ as the dark current value (e$^-$/sec) for a given pixel in the resultant frame. The dark current frame used in our persistence analysis was computed by taking median values of the final three post-test dark current frames. Appendix Figure \ref{fig:methods_SCA11DarkCurrent} shows the dark current frames for both SCAs 4 and 11. 

Finally, we computed persistence values in e$^-$/sec for all pixels within each interleaved dark exposure. We again derived the slope values for a given exposure by applying a linear fit of form $s = a + b \times t$ to each pixel, taking $b$ as the \textit{slope} for each pixel. We then subtracted the per pixel dark current values from the per pixel slopes for each interleaved dark exposure to find the resultant persistence frame following illumination for a given magnitude. 

Appendix Figures \ref{fig:methods_SCA4PersistenceFrames} and \ref{fig:methods_SCA11PersistenceFrames} show the SCA 4 and 11 persistence frames for all nine interleaved dark exposures. Our choice to leave the EWA in F146 during the interleaved dark exposures presented a few challenges that we treated in our data analysis. First, as is evident upon examination of Figure \ref{fig:methods_SCA11PersistenceFrames}, thermal emission and reflections from SORC subsystem mechanical elements caused faint, structured stray light contamination within regions of the persistence frames. At times, SORC contamination impacted areas within the ROI for a given projected source, including regions within the \textbf{\textit{saturation mask}}. We overcame this issue by computing the median local background level in the associated dark within the ROI for each source, taking as background the region \textit{within} the ROI but outside the \textbf{\textit{saturation mask}}.

\begin{figure}
\includegraphics[width=0.45\textwidth]{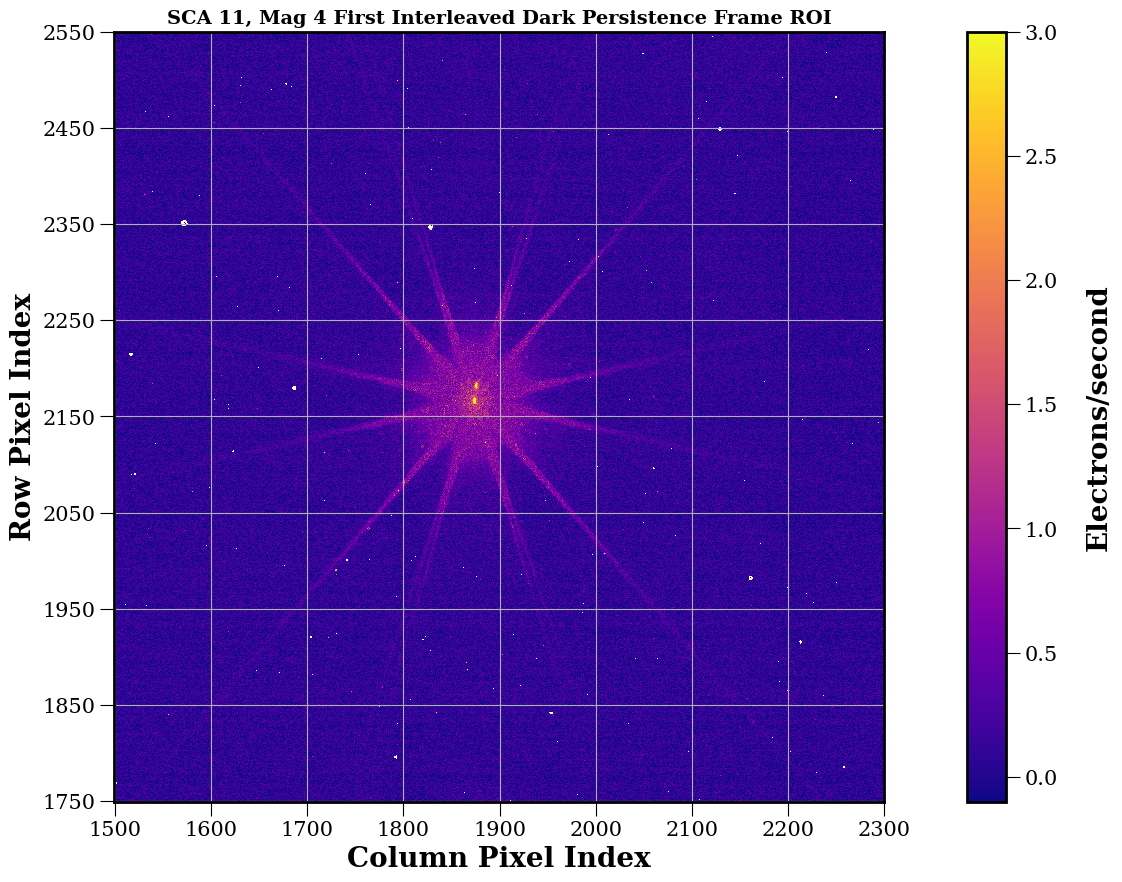}
\caption{SCA 11 First interleaved dark persistence frame. The diffuse, PSF shaped signal region in the center of the ROI is persistence caused by the $\sim$4 mag projected source. The two bright spots near the center of the persistence distribution are caused by SORC fiber light contamination. We developed a method to subtract the light fiber contamination, as described in Section \ref{subsubsec:methods_Persistence}.
    }
\label{fig:SCA11Mag4_FiberContamination}
\end{figure}

An additional challenge was caused by two SORC projector fibers that emitted a constant, low-flux, contaminating signal close to or within the \textbf{\textit{saturation mask}} in the interleaved darks for each magnitude. This fiber leak should have been shuttered in the SORC SDS rack during the darks, but a setting was inadvertently missed in the Roman TPT software and the associated shutter was not closed post-illumination (such is the nature of a risk reduction style, test of opportunity). Figure \ref{fig:SCA11Mag4_FiberContamination} shows the SORC fiber contamination in the SCA 11 first interleaved dark frame for the $\sim$4 mag source. The flux, shape, size, and relative orientation of the fiber contamination remained approximately constant for the first interleaved dark following each illuminated exposure for a given SCA. Only the position upon the SCA differed, as the SORC pose moved to the different relative locations shown in Fig.~\ref{fig:SCAtest}. We used the steps in the numbered paragraphs that follow to develop a method to subtract the fiber contamination from the first interleaved dark exposure after each simulated source. Steps numbered 1 to 3 refer to our procedures for SCA 11. In step 4, we describe slight modifications necessary for SCA 4.

\begin{figure}
\includegraphics[width=0.47\textwidth]{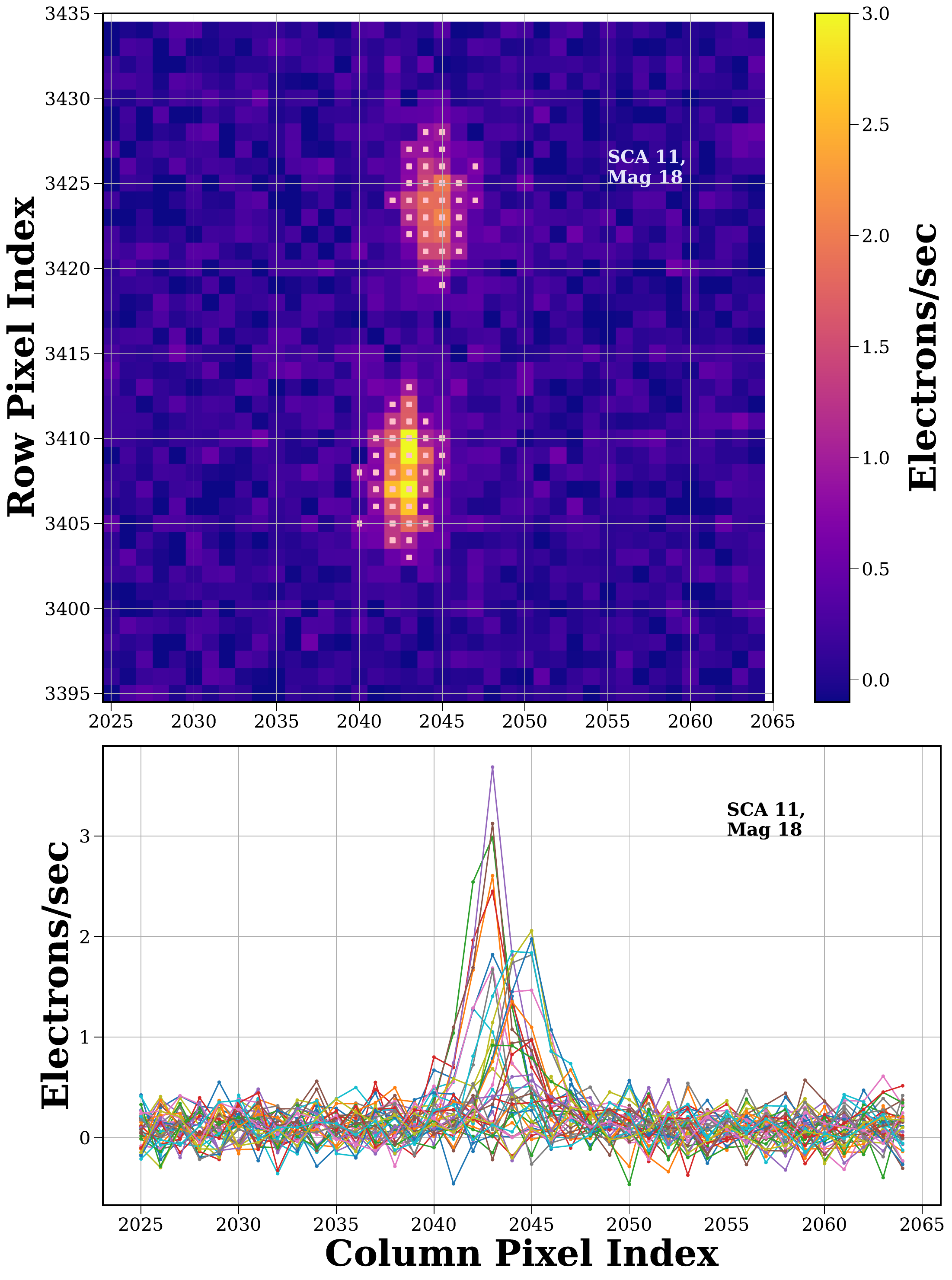}
\caption{ ROI for the SCA 11 $\sim$18 mag first interleaved dark persistence frame. The \textbf{\underline{top}} panel shows the image of the fiber contamination upon SCA 11, with pink squares showing the 70 pixels contained within our \textbf{\textit{light fiber mask}}. Fiber light contamination is the primary cause of the flux values for the `upper' contaminant. The `lower' signal is comprised of both fiber contamination and lingering persistence from the recent $\sim$18 mag illumination. The \textbf{\underline{bottom}} panel depicts the flux values in e$^{-}$ s$^{-1}$ through each row of the persistence frame. Our \textbf{\textit{light fiber mask}} flags those pixels with values greater than 0.62 electrons/sec, the approximate maximum pixel value of the background in the ROI.}
\label{fig:SCA11Mag18_2Panel_PersistenceFrameWFiberMask_RowCrossSection}
\end{figure}

\begin{figure}
\includegraphics[width=0.47\textwidth]{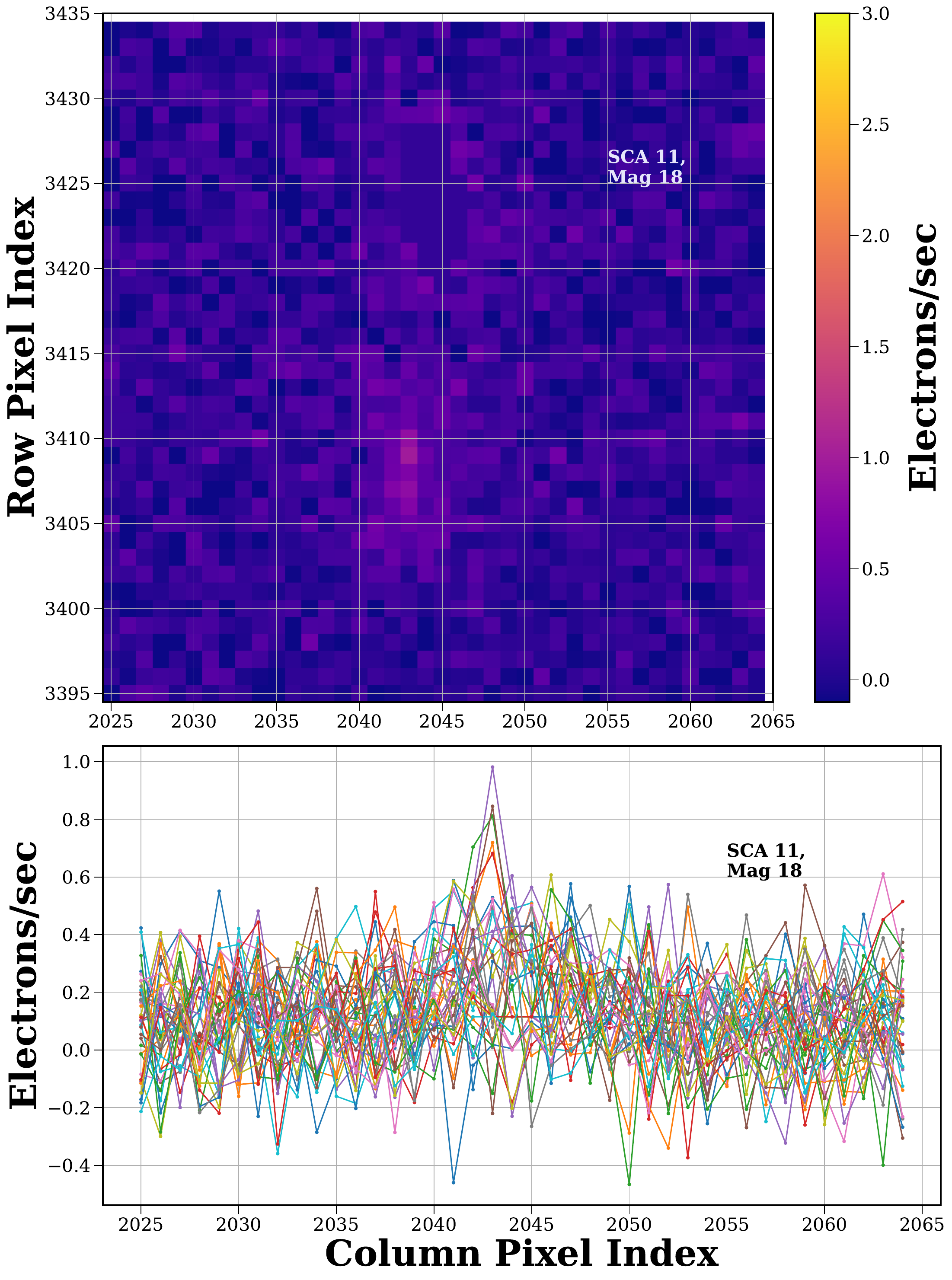}
\caption{ ROI for the SCA 11 $\sim$18 mag first interleaved dark persistence frame---same as Figure \ref{fig:SCA11Mag18_2Panel_PersistenceFrameWFiberMask_RowCrossSection}---but with per pixel \textbf{\textit{fiber subtraction}} applied. The \textbf{\underline{top}} panel shows the corrected persistence image frame, while the \textbf{\underline{bottom}} panel depicts the flux values in e$^{-}$ s$^{-1}$ through each row of the persistence frame. The faint lingering persistence from the recent $\sim$18 mag illumination is visible at the position of the previous lower fiber contamination. 
    }
\label{fig:SCA11Mag18_2Panel_PerPixFiberSubtractionApplied}
\end{figure}

\begin{figure}
\includegraphics[width=0.47\textwidth]{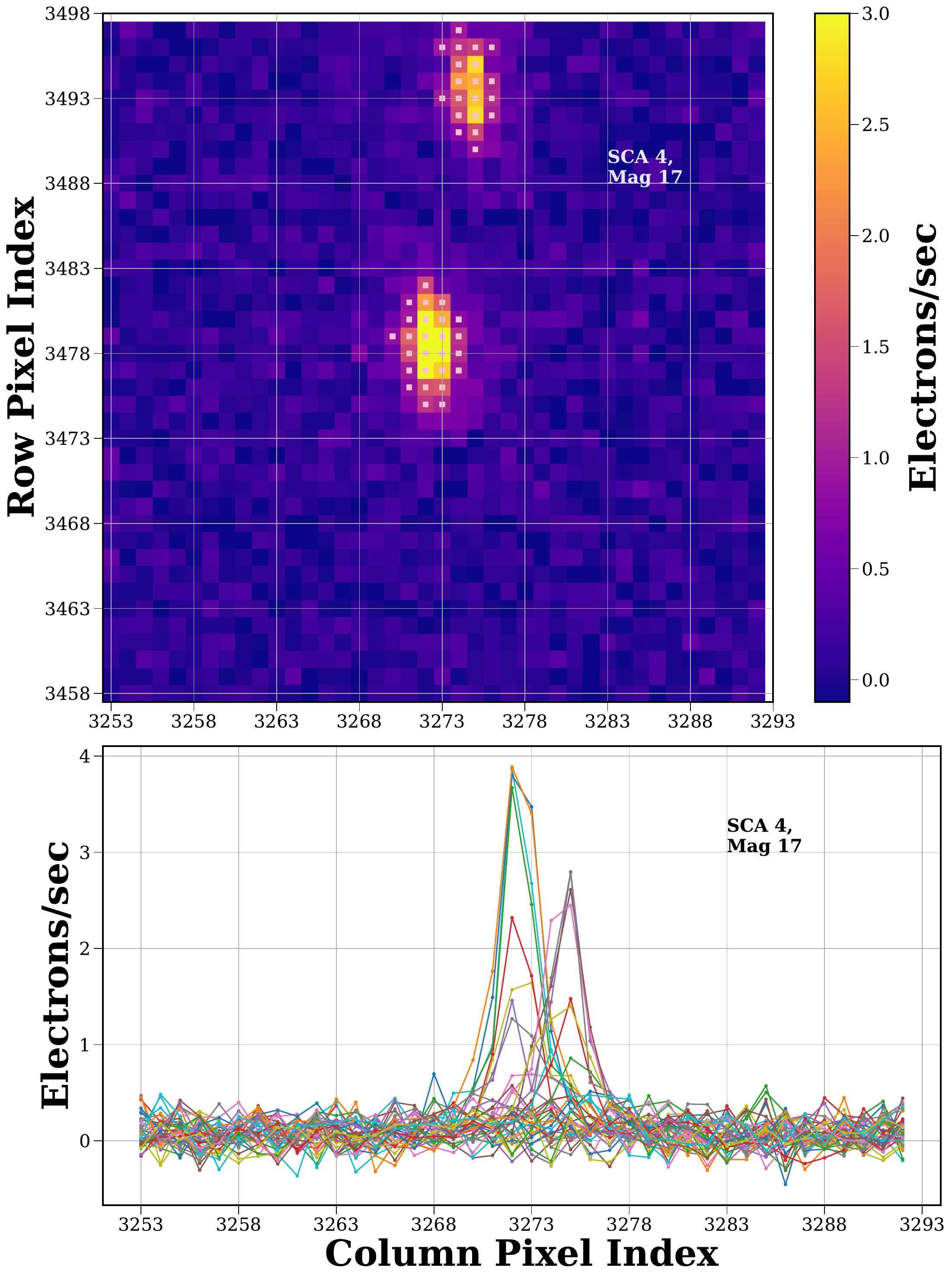}
\caption{ ROI for the SCA 4 $\sim$17 mag first interleaved dark persistence frame. The \textbf{\underline{top}} panel shows the image of the fiber contamination upon SCA 4, with pink squares showing the 46 pixels contained within our \textbf{\textit{light fiber mask}}. Fiber light contamination is the primary cause of the flux values for the `upper' contaminant. The `lower' signal is comprised of both fiber contamination and lingering persistence from the recent $\sim$17 mag illumination. The \textbf{\underline{bottom}} panel depicts the flux values in e$^{-}$ s$^{-1}$ through each row of the persistence frame. Our \textbf{\textit{light fiber mask}} flags those pixels with values greater than 0.75 electrons/sec, the approximate maximum pixel value of the background in the ROI.}
\label{fig:SCA4Mag17_2Panel_PersistenceFrameWFiberMask_RowCrossSection}
\end{figure}

\begin{enumerate}
\item For SCA 11, we created a \textbf{\textit{fiber contamination mask}} (Figure \ref{fig:SCA11Mag18_2Panel_PersistenceFrameWFiberMask_RowCrossSection}) by flagging those pixels within the $\sim$18 mag interleaved dark frame that were greater than 0.62 e$^-$/sec. This threshold for the pixels defining the mask was determined empirically by examining the interleaved dark image in the ROI and selecting pixels that were brighter than maximum background fluctuation (see Fig.~\ref{fig:SCA11Mag18_2Panel_PersistenceFrameWFiberMask_RowCrossSection}). We adopted this approach for both the `upper' and `lower' fiber leak signals. The resulting mask contains 70 pixels. We chose the mag $\sim$18 data to create the mask because none of the pixels in the final illuminated frame had reached our threshold for saturation. Although we expected some lingering persistence due to the mag $\sim$18 source illumination, we anticipate significant contribution from the fiber contamination flux. This expectation is supported by our observation that persistence computations for dimmer magnitudes were only impacted by the lower light fiber contaminant, since only that light fiber was contained within the \textbf{\textit{saturation mask}}. 
\item We subtracted the \textit{light fiber contamination} on a per-pixel basis within the \textbf{\textit{light fiber mask}}. The initial values used for each pixel were the $\sim$18 mag persistence frame flux values recorded for pixels within the \textbf{\textit{light fiber mask}}, minus the background level within the ROI. The background level was computed by finding the median flux value for all pixels within the ROI but outside the \textbf{\textit{light fiber mask}}. 
\item We tested the \textbf{\textit{light fiber subtraction}} values from step 2 across magnitudes $\sim$4 to $\sim$15, with excellent results for the `upper' light fiber contaminant. However, we found evidence for oversubtraction at dimmer magnitudes for the `lower' light fiber leak. Specifically, some corrected persistence values in the Mag $\sim$15 persistence ROI dipped to negative values. We therefore multiplied the \textit{lower light fiber subtraction values} from step 2 by a constant scaling factor, which we computed to ensure that the \textit{lowest} background subtracted corrected persistence value within the fiber mask would be zero. Figure \ref{fig:SCA11Mag18_2Panel_PerPixFiberSubtractionApplied} shows the $\sim$18 mag persistence frame after subtraction of our \textit{refined} per pixel \textbf{\textit{light fiber subtraction}} values. We next applied our \textit{refined} per pixel \textbf{\textit{light fiber subtraction}} values to the first interleaved dark persistence frames for magnitudes $\sim$4 through $\sim$17 for SCA 11.

\item For SCA 4, we applied the same techniques to develop a \textbf{\textit{fiber contamination mask}} optimized for the properties of the detector. However, in this case, the first interleaved dark persistence frame following $\sim$18 mag illumination suffered from severe SORC projector contamination. (See lower right panel of Figure \ref{fig:methods_SCA4PersistenceFrames}.) Contamination occurred when the SORC projector was moved to SCA 11 immediately following the illumination of the final $\sim$18 mag source in the sequence, so there is no fiber contamination apparent in that interleaved dark on SCA 4. We therefore created the \textbf{\textit{fiber contamination mask}} using the interleaved dark following the $\sim$17 mag illuminated exposure. In this case, we flagged pixels within the $\sim$17 mag ROI that were greater than 0.75 e$^-$/sec. The resulting mask contains 46 pixels. In Figure \ref{fig:SCA4Mag17_2Panel_PersistenceFrameWFiberMask_RowCrossSection}, we show the SCA 4 first interleaved dark persistence frame overplotted with the \textbf{\textit{light fiber mask}} (upper panel), as well as a plot showing the persistence values across each row of the frame (lower panel). Figure \ref{fig:SCA4Mag17_2Panel_PersistenceFrameWFiberMask_RowCrossSection} shows the same ROI after subtraction of the derived per pixel \textbf{\textit{light fiber subtraction}} values. 
\end{enumerate} 

We discuss the performance of our applied corrections in Section \ref{subsec:results_Persistence}.

\begin{figure}
\includegraphics[width=0.47\textwidth]{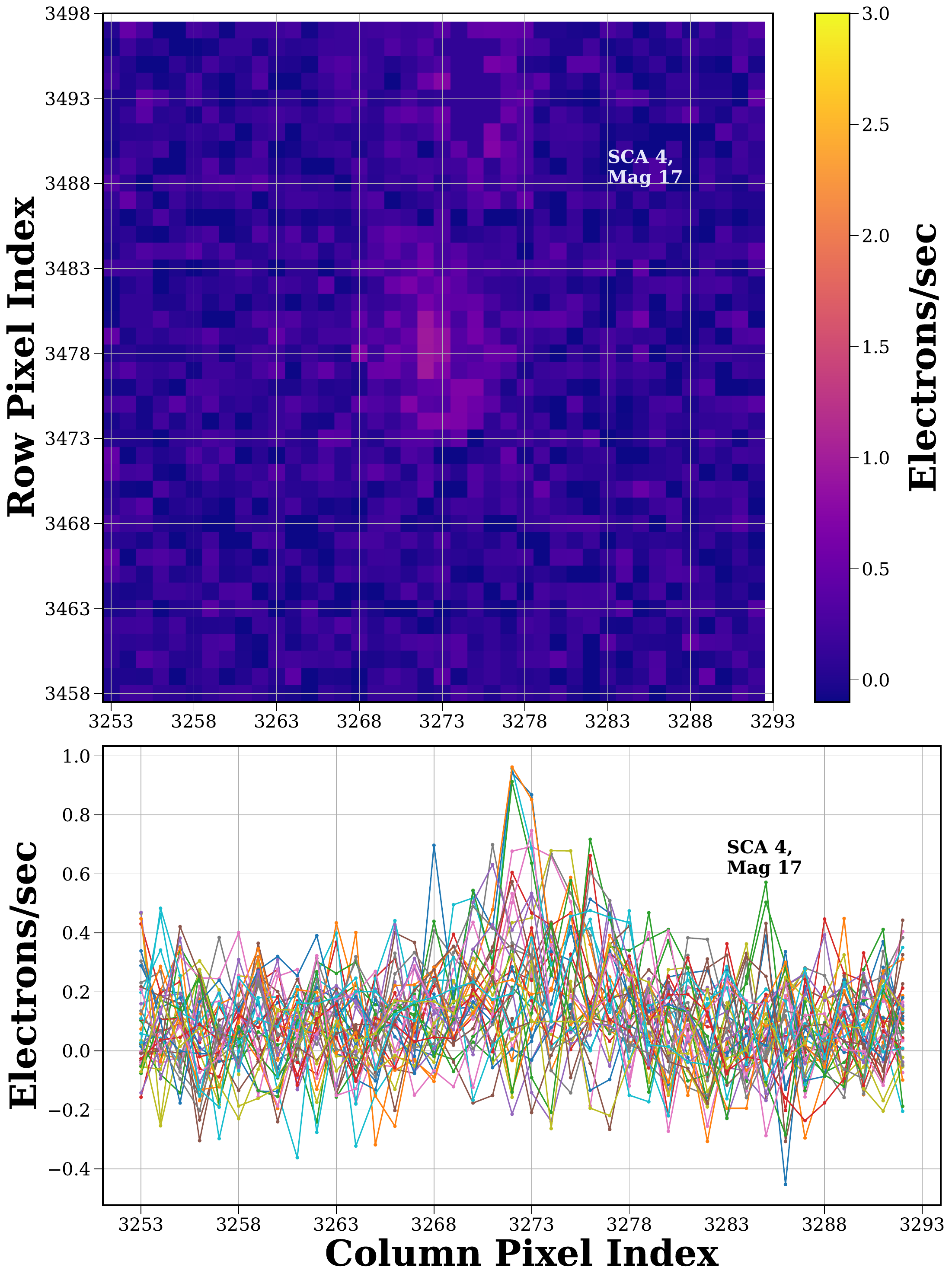}
\caption{ ROI for the SCA 4 $\sim$17 mag first interleaved dark persistence frame---same as Figure \ref{fig:SCA11Mag18_2Panel_PersistenceFrameWFiberMask_RowCrossSection}---but with per pixel \textbf{\textit{fiber subtraction}} applied. The \textbf{\underline{top}} panel shows the corrected persistence image frame, while the \textbf{\underline{bottom}} panel depicts the flux values in e$^{-}$ s$^{-1}$ through each row of the persistence frame. The faint lingering persistence from the recent $\sim$17 mag illumination is visible at the position of the previous lower fiber contamination. 
    }
\label{fig:SCA4Mag17_2Panel_PerPixFiberSubtractionApplied}
\end{figure}

\begin{figure*}
\centering
\includegraphics[width=0.9\textwidth]{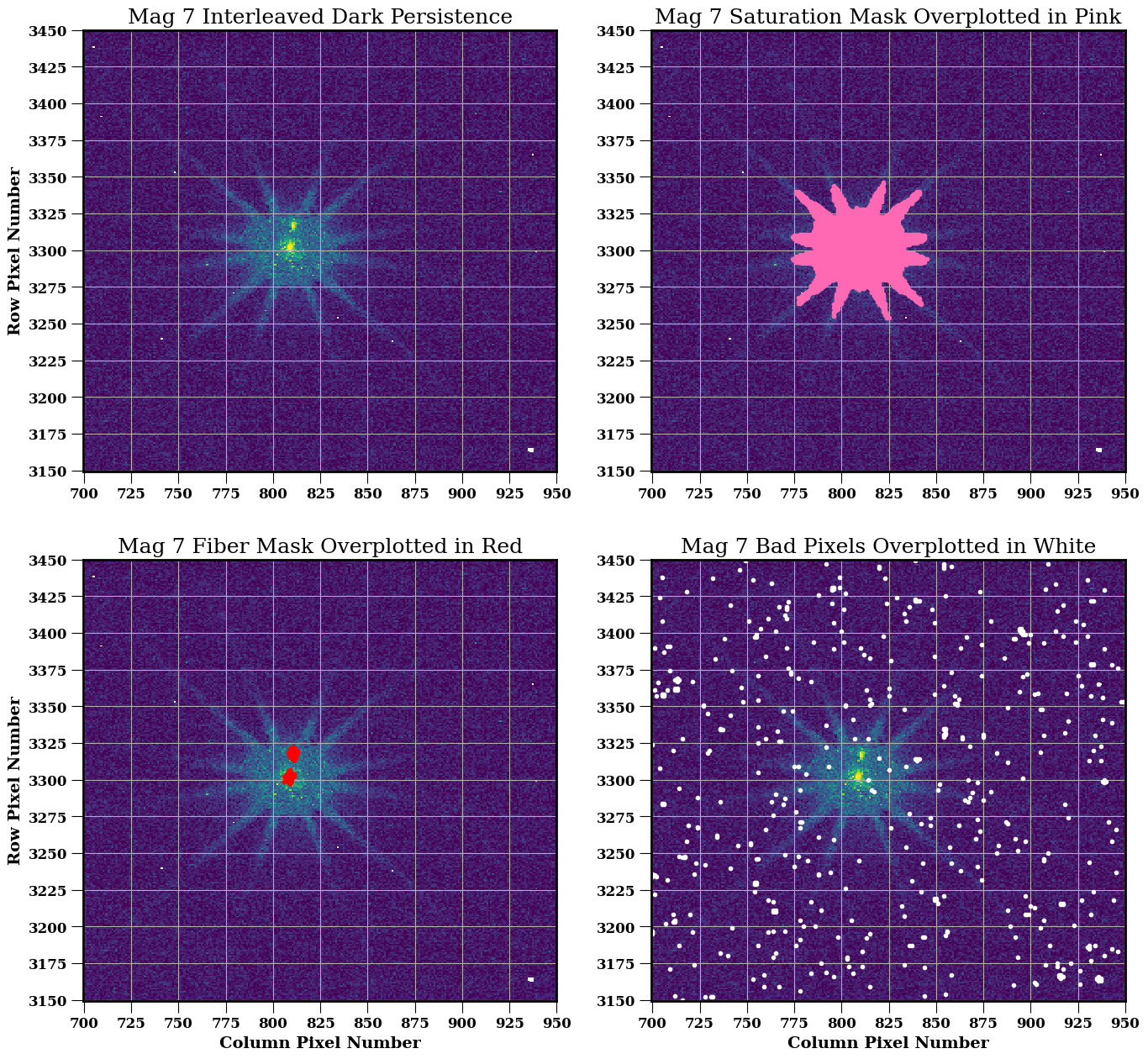}
\caption{ROI for SCA 11 magnitude $\sim$7 interleaved dark persistence frame. The \textbf{\underline{upper left}} panel shows the persistence image, where the 2 bright spots caused by SORC light fiber contamination are clearly visible. In the \textbf{\underline{upper right}} panel we overplot the \textbf{\textit{saturation mask}} in pink. In the \textbf{\underline{lower left}} panel we overplot the \textbf{\textit{light fiber subtraction regions}} in red. Finally, in the \textbf{\underline{lower right}} panel we overplot the \textbf{\textit{bad pixels}} in white. 
    }
\label{fig:SCA11Mag7_4PanelPersistence}
\end{figure*}

\begin{figure*}
\includegraphics[width=0.9\textwidth]{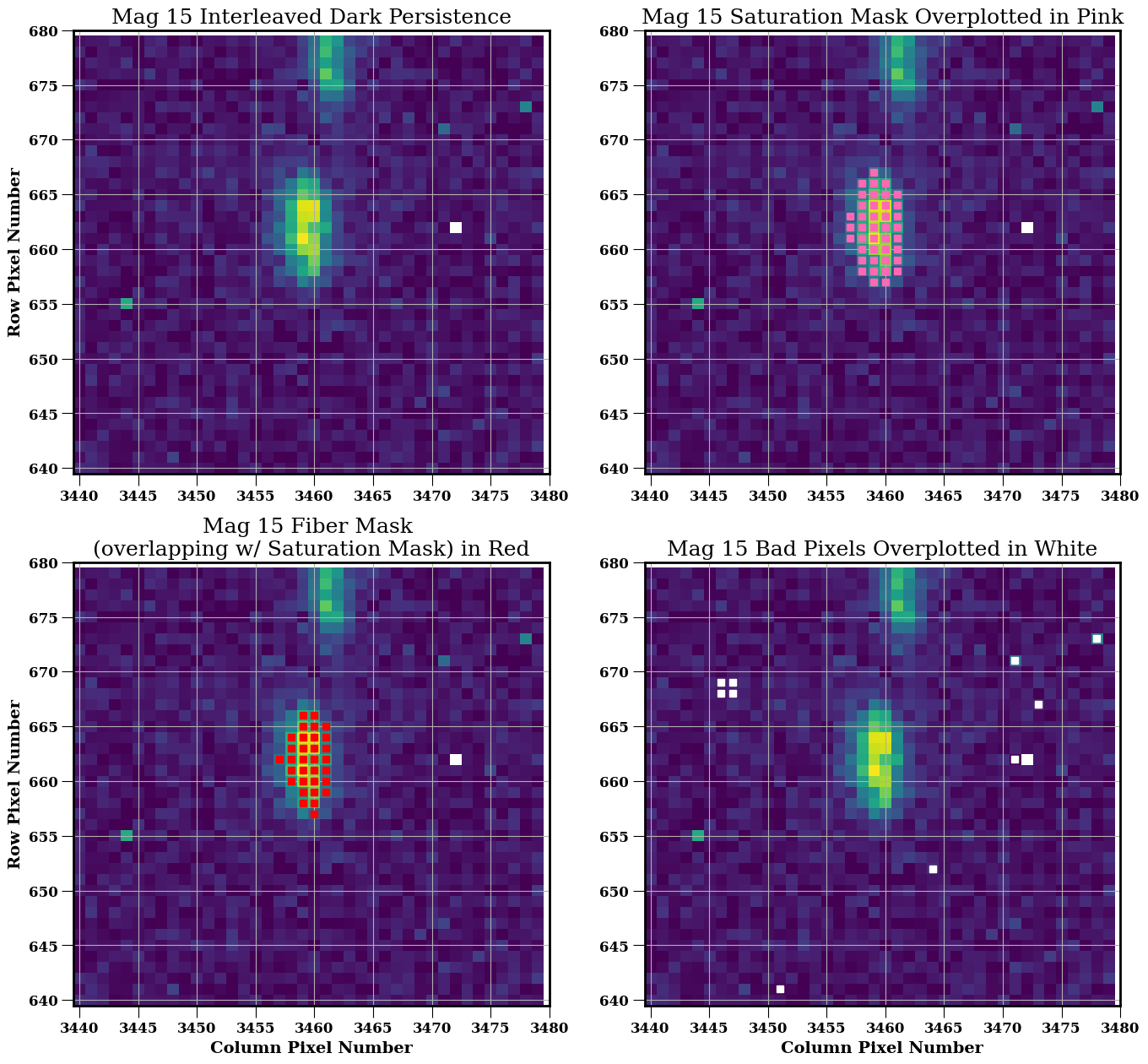}
\caption{ ROI for SCA 11 magnitude $\sim$15 interleaved dark persistence frame. The \textbf{\underline{upper left}} panel shows the persistence image, which contains a large component of SORC light fiber contamination. In the \textbf{\underline{upper right}} panel we overplot the \textbf{\textit{saturation mask}} in pink. Note that the bright spot above the saturation mask is due to the upper component of SORC light fiber contamination. In the \textbf{\underline{lower left}} panel we overplot the \textbf{\textit{light fiber subtraction region}} which overlaps with the saturation mask in red. Only the lower component of SORC light fiber contamination need be taken into account to compute median persistence within the \textbf{\textit{saturation mask}}. Finally, in the \textbf{\underline{lower right}} panel we overplot the \textbf{\textit{bad pixels}} in white. 
    }
\label{fig:SCA11Mag15_4PanelPersistence}
\end{figure*}

In addition to correcting our persistence data for background and fiber contamination, we note that our persistence computations excluded pixels flagged within the \textbf{\textit{bad pixel mask}}. A bad pixel mask was derived for each SCA using data collected in other TVAC2 tests. The bad pixels were flagged using a variety of pixel performance criteria thresholds from flight SCA acceptance testing \citep{Mosby2025}. In Figures \ref{fig:SCA11Mag7_4PanelPersistence} and \ref{fig:SCA11Mag15_4PanelPersistence}, we show 4-panel plots for the SCA 11 first interleaved dark persistence frames for magnitudes $\sim$7 and $\sim$15, where we overplot the \textbf{\textit{saturation masks}}, \textbf{\textit{light fiber subtraction}} regions, and \textbf{\textit{bad pixels}} used in our analysis.

We produced eight persistence decay curves for projected magnitudes $\sim$4 through $\sim$17 by computing the median persistence values within each \textbf{\textit{saturation mask}}, and plotting those values versus time, using the time stamp on the final illuminated frame for each magnitude as $t = 0$. We took the time for each persistence computation as the average time stamp for a given interleaved dark exposure. The error bars are computed as the average of the persistence interquartile range for all pixels within a given \textbf{\textit{saturation mask}}. In producing the persistence decay curves, the number of interleaved darks available ranged from 9 for the 4$^{\rm th}$ mag  source to 2 for the 17$^{\rm th}$ magnitude source.

\section{Results} \label{sec:results}

\subsection{Bright Star Saturation Test Results}\label{subsec:results_Saturation}

To investigate the degree to which saturation affects the derived slope more quantitatively, we compare the instantaneous slopes on a per-pixel basis within the four differing saturation regimes. For the purpose of this analysis, we assume that the slope determined before any neighboring or diagonal pixels saturate is indicative of the true count rate observed by the detector.  In the absence of any calibration procedures, one would need to ignore any signal accumulated in frames after adjacent pixels have saturated, otherwise the inferred count-rate would necessarily be overestimated. However, this choice means that a significant fraction of the accumulated signal cannot be used which severely limits the photometric precision that could, in theory, be extracted with a properly calibrated model for the observed charge leakage. Although developing such a model is out of scope for the current work, we conduct some preliminary analysis in an attempt to identify some promising avenues for such a model to consider.

\begin{figure*}[ht!]
\centering
\includegraphics[width=1.0\textwidth]{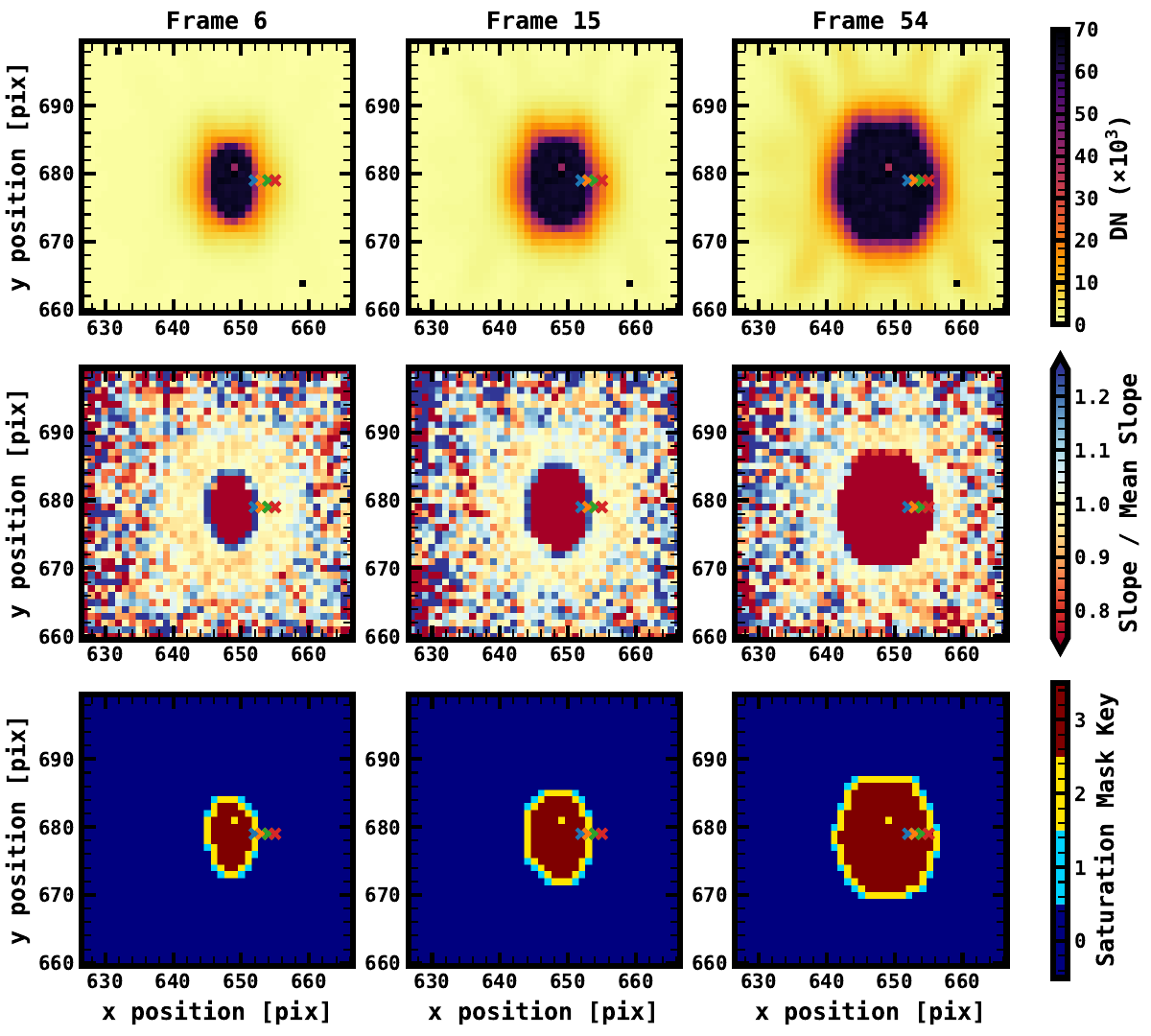}
\caption{Pixels adjacent to saturated regions show pronounced non-linear behavior. The saturation response for an approximately 12 mag source on SCA 11 is shown here. Top row: total accumulated signal in the indicated frame. Middle row: instantaneous slope, computed as the difference between the indicated frame and the preceding frame, divided by the best-fit ramp slope derived from frames acquired before neighboring pixels saturate. Bottom Row: the saturation mask key used to trace the state of each pixel in each frame, with numbers indicating as follows, 0: not saturated, 1: diagonal to a saturated pixel, 2: adjacent in row or column to a saturated pixel, and 3: saturated. 
The local slopes increase when the leading edge of the saturated region -- the ``saturation front'' -- approaches a pixel, producing deviations from linearity. For example, as the pixel marked with the blue X (652, 679) approaches saturation, the pixel marked with the orange X (653, 679) shows a clear change in slope. This effect appears in the left-most middle panels as a ring of elevated instantaneous slopes surrounding the saturated core, which correspond directly to the pixels neighboring saturated pixels in the bottom panels. As the saturation front expands, the instantaneous slope increases compared to the mean slope measured prior to the saturation of adjacent or diagonal pixels. 
 \label{fig:saturationgrowth1}}
\end{figure*}

\begin{figure}[ht!]
\centering
\includegraphics[width=1.0\linewidth]{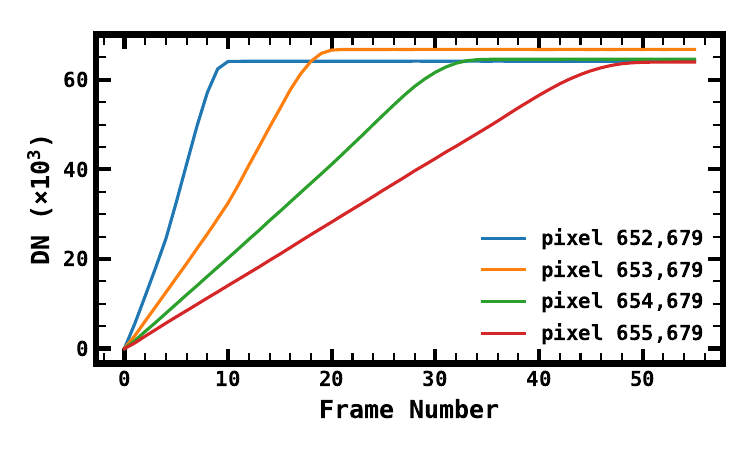}
\caption{
The accumulated signal for specific pixels of an approximately 12 mag source on SCA 11 is shown here across all frames. Each line corresponds to a pixel from Figure \ref{fig:saturationgrowth1} marked by an ``X''. The local slopes increase when the leading edge of the saturated region -- the ``saturation front'' -- approaches a pixel, producing deviations from linearity. For example, as the pixel marked as the blue line (652, 679) approaches saturation, the pixel marked with the orange line (653, 679) shows a clear increase in slope. 
 \label{fig:saturationgrowth2}}
\end{figure}

Figure \ref{fig:saturationgrowth1} shows the growth of saturation over time when the detector is illuminated by the equivalent of a $\sim$12 mag point source on SCA 11. The top panels show three different image frames throughout the exposure, the middle panels show the instantaneous slope compared to the mean slope at each of the three different frames,  and the bottom panel tracks individual pixel fluxes as the saturation region spreads. For pixels marked with ``X", the corresponding slopes are plotted in Figure \ref{fig:saturationgrowth2} with matching colors in each panel of figure \ref{fig:saturationgrowth1}. Hard saturation is indicated for a given pixel when the slope becomes zero. Note that pixel slopes increase when the leading edge of the saturation region is adjacent to them (the ``saturation front''), impacting linearity. This slope change in the vicinity of a saturated pixel is likely due to charge leakage and has been previously observed by \cite{Brandt_CHARIS_JATIS2017} who reported a similar effect in CHARIS H2RG detectors. When a pixel saturated they found the four nearest neighbor pixels received an immediate/compensating influx of electrons, while pixels located diagonally to the saturated pixel also received a significant increase. Such an effect may need to be taken into account in an analysis of saturated source photometry. Figure~\ref{fig:saturationgrowth1} also shows that after 54 frames of illumination ($\approx$170 s), the saturation region of the $\sim$12 mag source grows to about 15 pixels in diameter. For comparison, the saturation region of the $\sim$4 mag source grows to about 150 pixels in diameter after the same number of frames. Around the same time as the analysis presented here, the same behavior was also observed in other Roman WFI TVAC data\footnote{\url{https://roman-docs.stsci.edu/roman-instruments/the-wide-field-instrument/wfi-detectors/sources-of-pixel-to-pixel-variation}}

\begin{figure}
    \centering
    \includegraphics[width=1.0\linewidth]{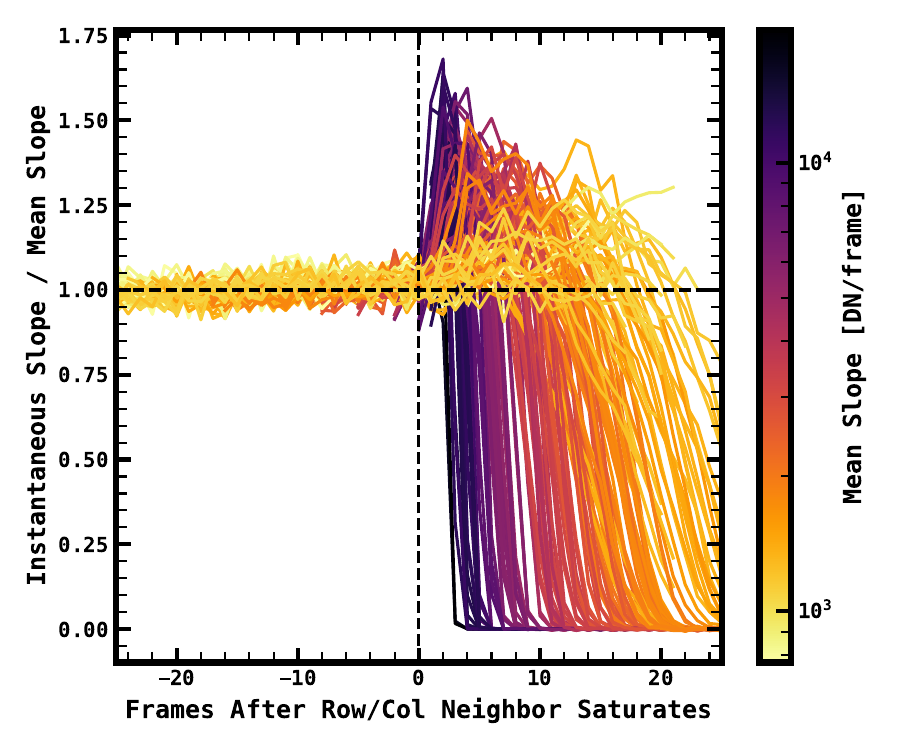}
    \caption{The ratio of the instantaneous slope to the mean slope measured for all pixels for which a neighboring pixel saturates during an exposure. Each line is the time-series of one pixel, with the frame number shifted so that the frame immediately before the neighboring pixel saturates is set to 0. The color of each line is the mean slope measured for that pixel. }
    \label{fig:saturated_slope_ratio}
\end{figure}

Another method to visualize the charge leakage we report in these detectors is shown in Figure \ref{fig:saturated_slope_ratio}. Here, we plot each pixel that has a neighbor which eventually saturates as a line, showing the time-series in the instantaneous slope divided by the mean slope, as defined in \S\ref{subsubsec:methods_Saturation}. We have also shifted the frame number so that Frame 0 coincides with the last frame before a neighboring pixel saturates. The color of each line represents the mean slope of that pixel. In this figure there are several patterns that become clear. First, the slope of each pixel is relatively well behaved, at nearly all brightnesses, up until a neighboring pixel saturates. However, at all ranges of accumulated signal displayed, the corruption in slope occurs immediately after the saturation of a neighboring pixel. The subsequent decrease to a slope of 0 is the result of the classical non-linearity affecting the pixels as you approach saturation and is therefore an expected behavior. 
Another pattern of note is that the corrupted slope that occurs as a result of the saturation of an adjacent pixel is not constant in time. Even before there is a decrease in slope driven by the classical non-linearity, the slope ratio plotted in Figure \ref{fig:saturated_slope_ratio} appears to be increasing after multiple reads, implying that the charge leakage displayed is not a linear function, and instead appears to be dynamic.

In addition to the evolving, dynamic count-rate observed after the saturation of a neighboring pixel, we also observe that the magnitude of the observed change in slope ratio correlates with the mean slope of the pixel being observed, which is evident in Figure \ref{fig:saturated_slope_ratio} as pixels with larger mean slopes have a larger relative change in instantaneous slope after the frame in which an adjacent pixel saturates.

\subsection{Persistence Test Results}\label{subsec:results_Persistence}

\begin{figure*}
\centering
\includegraphics[width=0.9\textwidth]{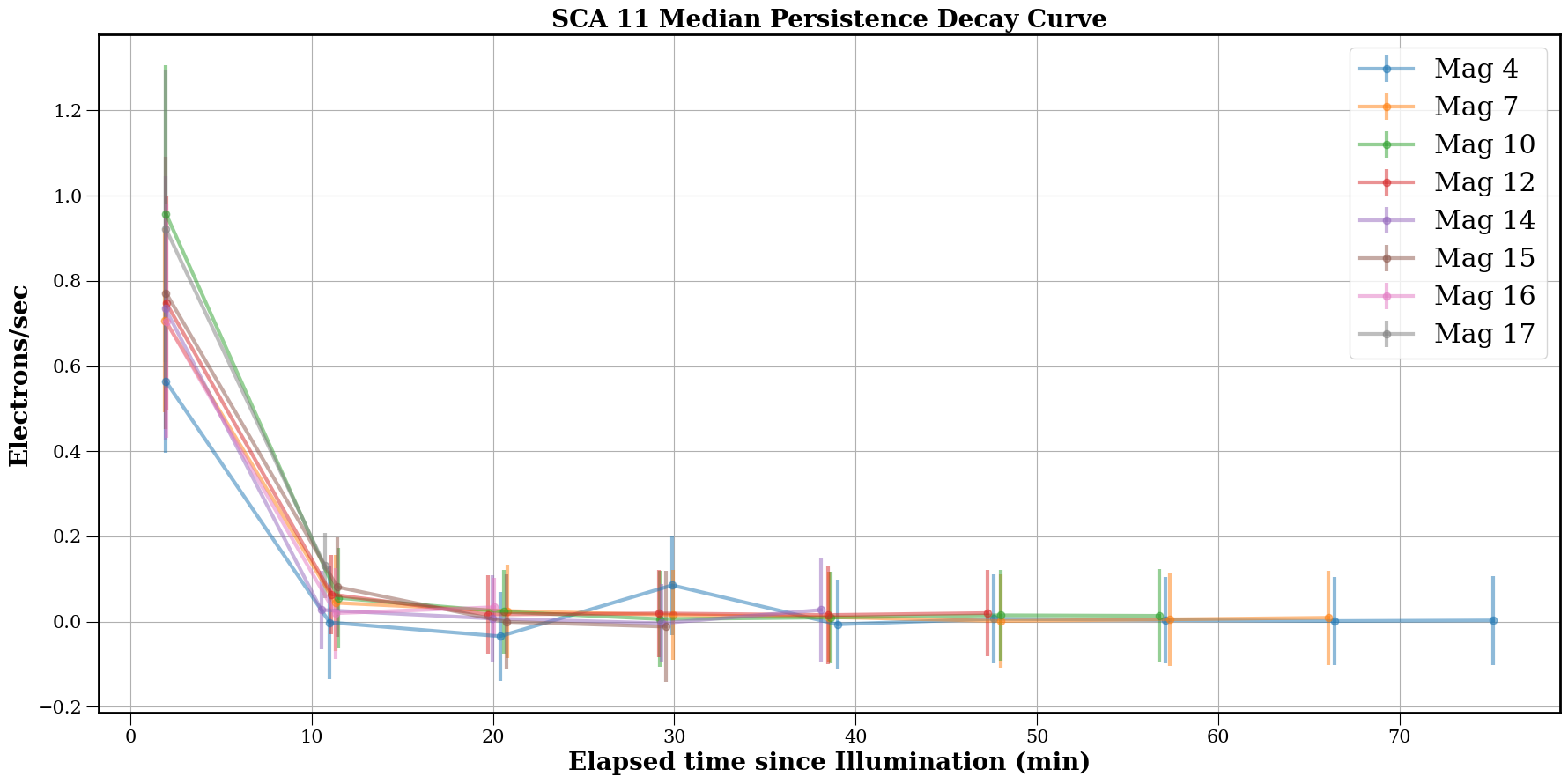}
\caption{SCA 11 Persistence Decay Curves for magnitudes $\sim$4 through $\sim$17. The first interleaved dark persistence values largely agree within the errors bars. However, we note that the mag $\sim$4 persistence value is slightly lower than the value for other magnitudes, while the values for magnitudes $\sim$10 and $\sim$17 are slightly higher. Part of the explanation for this lies in the spatial variation of persistence across the detector. See Section \ref{subsec:results_Persistence} and Appendix Figures \ref{fig:SCA11_FlatFieldIlluminationWSatMasksROIsOverplotted} and \ref{fig:SCA11_ComparePersistence_BrightSourceVsFlatField} for details. Additionally, the decay curve for the mag $\sim$4 source appears to suffer from oversubtraction in the first 3 data points, while stray light biases the 4th data point on the curve. For all sources regardless of magnitude, the persistence decays to detector background levels after about 20 minutes post-illumination.
    }
\label{fig:SCA11_AllMag_PersistenceDecayCurves}
\end{figure*}

\begin{figure*}
\centering
\includegraphics[width=0.9\textwidth]{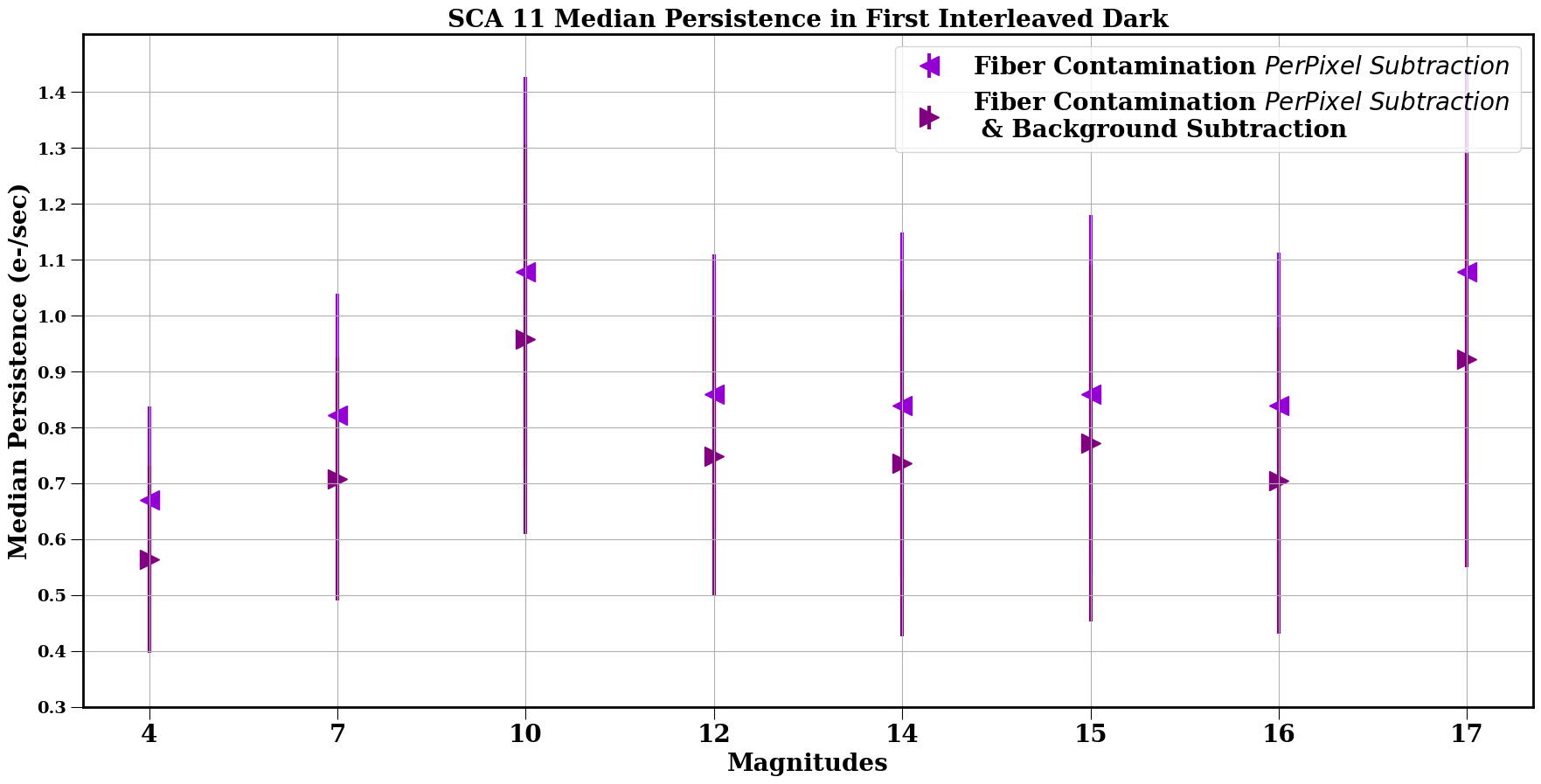}
\caption{SCA 11 comparison of persistence in first interleaved dark exposures, which suffered from SORC light fiber contamination. We show the first interleaved dark frames after subtracting the fiber contamination (see $\S$\ref{subsubsec:methods_Persistence}). Although the Mag $\sim$4 source persistence appears slightly lower, and the Mags $\sim$10 and $\sim$17 source persistence values appear slightly higher, all persistence values agree within the error bars. We also note that only 6 pixels are in the saturation mask for the Mag $\sim$17 source, and all are affected by SORC fiber contamination.  }
\label{fig:SCA11_FirstInterleavedDarkPersistenceComparison}
\end{figure*}

We present our SCA 11 persistence decay curves for magnitudes $\sim$4 through $\sim$17 in Figure \ref{fig:SCA11_AllMag_PersistenceDecayCurves}, and compare the median persistence values for the first interleaved dark exposures in Figure \ref{fig:SCA11_FirstInterleavedDarkPersistenceComparison}. The first interleaved dark persistence values largely agree within the errors bars. However, we note that the mag $\sim$4 persistence value is slightly lower than the value for other magnitudes, while the values for magnitudes $\sim$10 and $\sim$17 are slightly higher. Additionally, the decay curve for the mag $\sim$4 source appears to suffer from oversubtraction in the first 3 data points, while stray light biases the 4th data point on the curve. In Appendix Figures \ref{fig:SCA11_Mags4to12_PersistenceDecayCurves} and \ref{fig:SCA11_Mags7_12_14_15_PersistenceDecayCurves}, we present decay curves showing only the 4 brightest sources, and the best corrected sources (magnitudes 7, 12, 14, and 15), respectively.

\begin{figure*}
\centering
\includegraphics[width=0.9\textwidth]{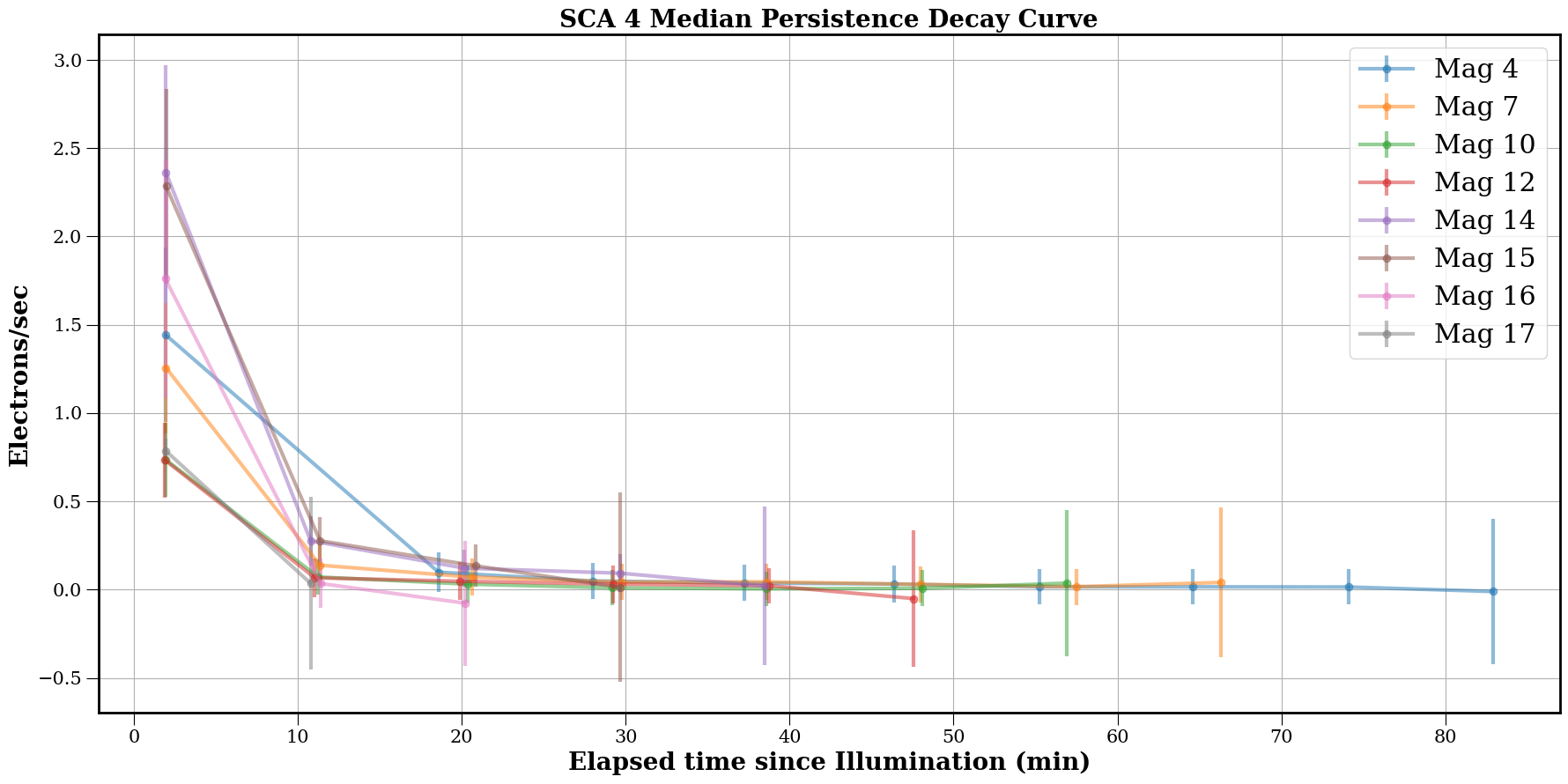}
\caption{SCA 4 Persistence Decay Curves for magnitudes $\sim$4 through $\sim$17. The first interleaved dark values do not agree within the error bars. Underlying persistence performance varies spatially across SCA 4, and thus persistence values in the first interleaved dark exposure were impacted by the region of SCA 4 where the illuminated source was projected. The slightly discrepant persistence values and larger error bars for the last data point in each curve were caused by higher background levels in the Mag $\sim$18 interleaved dark exposure. The larger background levels were caused by stray light from the SORC as it was moved to begin the SCA 11 projection sequence. These larger background levels for the $\sim$18 mag interleaved dark are evident in the bottom right frame of Appendix Figure \ref{fig:methods_SCA4PersistenceFrames}. Despite these issues, for all sources regardless of magnitude, the persistence decays to detector background levels after about 20 minutes post-illumination.
    }
\label{fig:SCA4_AllMag_PersistenceDecayCurves}
\end{figure*}

\begin{figure*}
\centering
\includegraphics[width=0.9\textwidth]{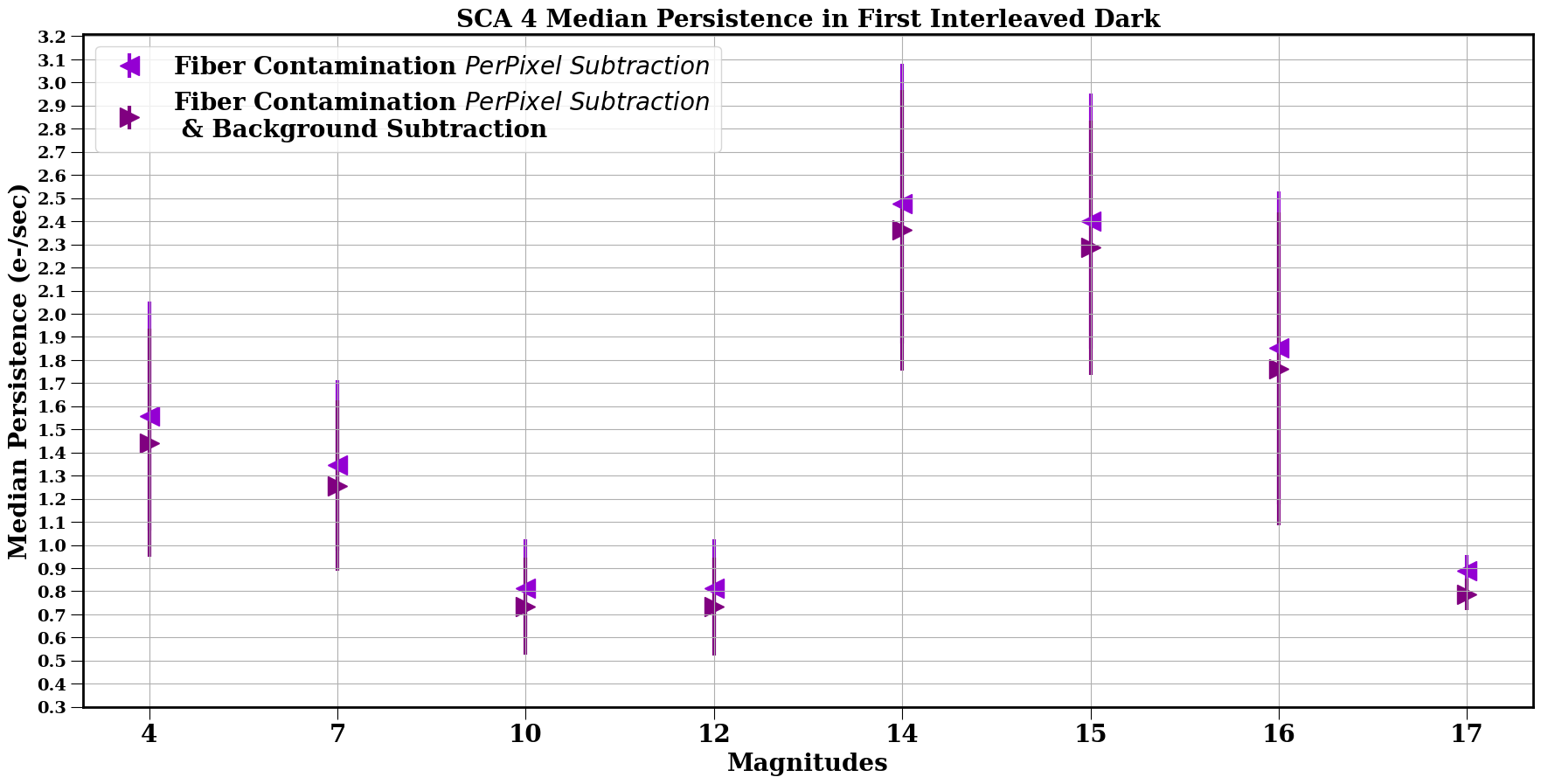}
\caption{SCA 4 comparison of persistence in first interleaved dark, showing that the values do not agree within the error bars. Underlying persistence performance varies spatially across SCA 4, and thus persistence values in the first interleaved dark exposure were impacted by the region of SCA 4 where the illuminated source was projected.
    }
\label{fig:SCA4_FirstInterleavedDarkPersistenceComparison}
\end{figure*}

For SCA 4, we present persistence decay curves for magnitudes $\sim$4 through $\sim$17 in Figure \ref{fig:SCA4_AllMag_PersistenceDecayCurves}, and compare the median persistence values for the first interleaved dark exposures in Figure \ref{fig:SCA4_FirstInterleavedDarkPersistenceComparison}. Both figures indicate that the first interleaved dark persistence values do not agree within the error bars. As explained in the next paragraph, persistence values in the first interleaved dark exposure were impacted by the region of SCA 4 where the illuminated source was projected. Additionally, the last data points in the persistence decay curves show slightly discrepant persistence values and larger error bars, which is caused by higher background levels in the final mag $\sim$18 interleaved dark exposure (see bottom right frame of Appendix Figure \ref{fig:methods_SCA4PersistenceFrames}). These larger background levels were caused by stray light from the SORC as it was moved to begin the SCA 11 projection sequence. In Appendix Figure \ref{fig:SCA4_Mags4to12_PersistenceDecayCurves}, we present decay curves for only the 4 brightest point sources. 

\begin{figure*}
\centering
\includegraphics[width=0.9\textwidth]{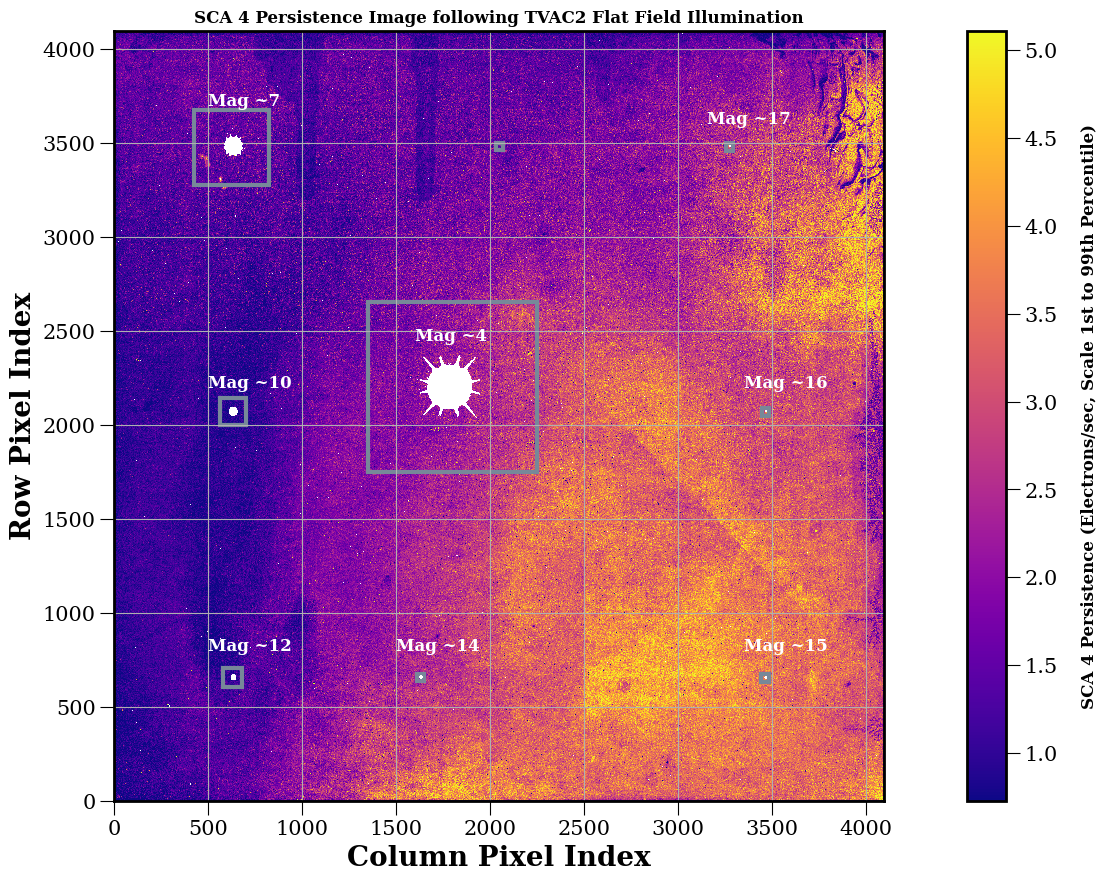}
\caption{SCA 4 persistence measured from a dark exposure following flat field illumination during TVAC2. The bright point source saturation masks for each magnitude are overplotted in white, with the accompanying ROIs outlined using gray boxes. The spatial variation of persistence performance across SCA 4 is evident, and directly affects the persistence measurements during the bright point source test. 
    }
\label{fig:SCA4_FlatFieldIlluminationWSatMasksROIsOverplotted}
\end{figure*}

\begin{figure*}
\centering
\includegraphics[width=0.9\textwidth]{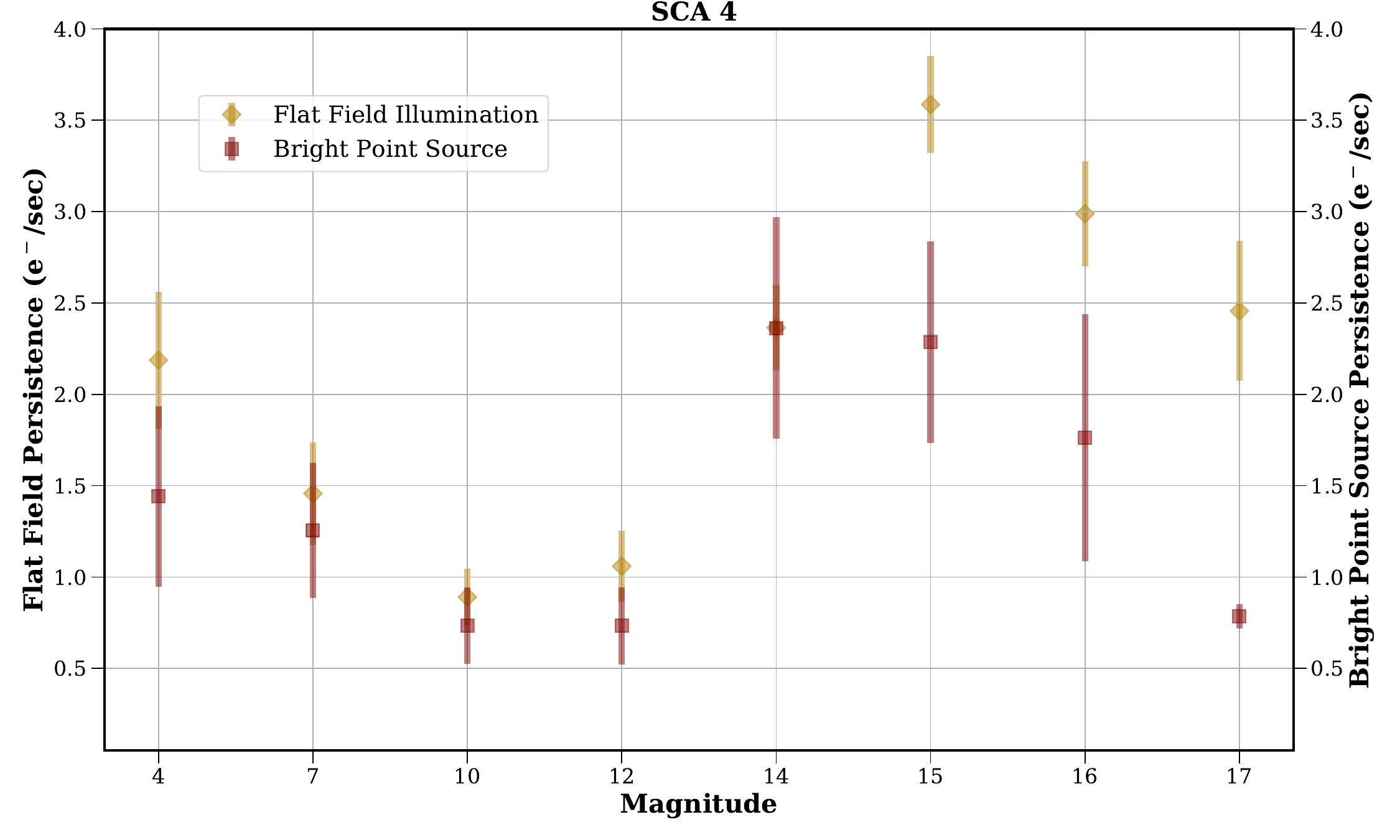}
\caption{Comparison of SCA 4 persistence measurements in subsequent dark frames following two separate TVAC2 tests: 1.) flat field illumination and 2.) bright point source projection. Flat field persistence shown was computed within the ROIs designated in Figure \ref{fig:SCA4_FlatFieldIlluminationWSatMasksROIsOverplotted} for the magnitudes indicated on the x-axis. The bright point source persistence is the same as that plotted in Figure \ref{fig:SCA4_FirstInterleavedDarkPersistenceComparison} for the first interleaved dark frames. The comparison shows that the underlying spatial variation in persistence across SCA 4 affects our persistence measurements during the bright point source test. Projected point sources in regions of the SCA with poorer persistence performance have similarly degraded performance after bright point source illumination. The effect is particularly evident in the first interleaved dark frames shown here, but is also apparent in the second and third points of some persistence decay curves in Figure \ref{fig:SCA4_AllMag_PersistenceDecayCurves}.  
    }
\label{fig:SCA4_ComparePersistence_BrightSourceVsFlatField}
\end{figure*}

During TVAC2, a separate test measured persistence decay for each SCA after exposing the entire Focal Plane Array to 56 frames of $\sim$900 e$^{-}$ s$^{-1}$ flat field illumination.\footnote{See resulting decay curves at \url{https://roman.gsfc.nasa.gov/science/WFI_technical.html}.} The persistence measured for SCA 4 during this test varied spatially across the SCA. This was also true for SCA 11, but to a lesser extent. In Figure \ref{fig:SCA4_FlatFieldIlluminationWSatMasksROIsOverplotted}, we overplot the SCA 4 bright star test ROIs for each magnitude onto the SCA 4 persistence measured during flat field illumination. We computed the median persistence values of flat field illumination within each ROI for SCA 4, and compare the bright star persistence values to the flat field persistence values in Figure \ref{fig:SCA4_ComparePersistence_BrightSourceVsFlatField}.\footnote{For comparison, we provide similar plots for SCA 11 in Appendix Figures \ref{fig:SCA11_FlatFieldIlluminationWSatMasksROIsOverplotted} and \ref{fig:SCA11_ComparePersistence_BrightSourceVsFlatField}. } The comparison shows that the underlying spatial variation in persistence across SCA 4 affects our persistence measurements during the bright point source test. The effect is particularly evident in the first interleaved dark frame, but is also apparent in the second and third points of the persistence decay curve in Figure \ref{fig:SCA4_Mags4to12_PersistenceDecayCurves}. However, despite these intrinsic systematics, and those lingering from imperfect background and fiber contamination correction, we conclude that for all sources tested, regardless of magnitude, on both SCAs, the persistence decays to detector background levels after about 20 minutes post-illumination.

\section{Discussion and Conclusion} \label{sec:DiscussionConclusion}

The TVAC2 bright point source saturation test and data analyses presented here provide a baseline assessment of saturation and persistence properties of Roman WFI's SCAs when exposed to bright point sources.

For saturated sources, we perform some preliminary analysis to characterize the behavior of pixel slope changes as the accumulated signal approaches saturation. We find that the accumulated signal is well-behaved when classical non-linearity corrections are applied to the pixel slopes, except when neighboring pixels saturate. In this limit, we observe that the accumulating signal deviates strongly from linear behavior, and increases immediately after a neighboring pixel saturates. The increased slope observed from a pixel after its nearest neighbor saturates appears consistent with previous observations of charge leakage in H2RG arrays. We identify this change in slope as a dynamic attribute, as the rate of accumulated signal changes with each read, rather than remaining constant until the non-linearity becomes dominated by the classical saturation effects. Therefore, we recommend any potential model that attempts to calibrate this non-linearity will likely need to account for a time-variable derivative in accumulation rate.

The persistence analysis reveals that for all sources regardless of magnitude, on both SCA 4 and SCA 11, the persistence decays to detector background levels ($\lesssim$0.05 e$^{-}$ s$^{-1}$) after about 20 minutes post-illumination. Considering that typical sky backgrounds in Roman surveys are dominated by zodiacal light at the level of a few tenths of an e$^{-}$ s$^{-1}$, the persistence decays to the sky background in a few to 10 minutes, depending on the underlying pixel level persistence performance. 

We compare our results to the persistence performance of H2RG focal planes flown in Euclid's NISP instrument \citep{Kubik2024_EuclidNISP} and NIRCam on JWST \citep{LeisenringSPIE2016_NIRCam,Rieke2023PASP_NIRCam}. For this comparison, we parameterize persistence performance as the median persistence signal 10 mins after saturation. For Roman WFI SCA 11, which has performance typical of the full flight SCA complement, this is $\lesssim$0.1 e$^{-}$ s$^{-1}$. Compared to Euclid NISP's arrays, operating at a similar temperature of 85 K, this is a $\sim$10$\times$ performance improvement. Compared to JWST NIRCam's arrays operating at 40 K, Roman's persistence performance is comparable. This is remarkable given that the WFI's detectors operate 50 K warmer than NIRCam's and in a given detector persistence is a steep function of temperature \citep{Tulloch2018,TullochJATIS2019}. Furthermore, the point source persistence results presented here, where signal levels can greatly exceed full well, are broadly consistent with those measured using flat field illumination to approximately twice full well during the same WFI TVAC2 campaign.\footnote{\url{https://roman.gsfc.nasa.gov/science/WFI_technical.html}~under the \textit{Focal Plane System} tab} Additionally, for SCA 11, which has more uniform pixel level persistence performance, the persistence in the first interleaved dark frame after each source largely agree within the error bars (Figs.~\ref{fig:SCA11_AllMag_PersistenceDecayCurves},~\ref{fig:SCA11_FirstInterleavedDarkPersistenceComparison}), indicating that the initial persistence signal is largely independent of source magnitude. These significant improvements in persistence performance compared to the previous generation of HgCdTe detectors are linked to persistence being a driving requirement for Roman and changes to the detector design to meet it. We note that our analysis for SCA 4 shows that underlying variation in the persistence response across the SCA affects the results (see Fig.~\ref{fig:SCA4_FlatFieldIlluminationWSatMasksROIsOverplotted}), which is particularly evident in the first interleaved dark frame. This spatially dependent persistence response is expected for HgCdTe detectors and will need to be considered and calibrated in Roman surveys. Furthermore, this underlying feature is not unique to SCA 4. To a lesser extent, all 18 SCAs in the flight WFI focal plane array have spatially varying persistence\footnote{See~\url{https://roman.gsfc.nasa.gov/science/WFI_technical.html}~under the \textit{Focal Plane System} tab}. Roman's commissioning and calibration plans include routine observations of bright stars in touchstone fields. These sources will provide a similar diversity of fluxes across all WFI SCAs, enabling deeper analyses of these effects as seen through the flight telescope assembly, and providing further insight into science performance.

\subsection{Observation of a Flux Halo with Defect Shadows}

In data associated with the brightest point source tested, on both SCAs 4 and 11, we observed a radially extended halo of flux surrounding the deeply saturated PSF. This effect is most readily observed in the upper left panel of Figure~\ref{fig:frame56_4panel_ROIimage_rowCrossSec}. Here we present the final frame of illumination for the $\sim$4 mag point source projected onto SCA 4. The halo is observed as a prominent, radially extending flux distribution that exhibits a smoothly decreasing gradient with distance from the saturated PSF core. In an intriguing twist, this halo of flux appears to be blocked or obstructed by some defects in the detector material. A strong shadow is apparent on the anti-PSF side of a detector defect in the upper left inset image of Fig.~\ref{fig:frame56_4panel_ROIimage_rowCrossSec} at pixel location approximately (1680, 1980). Additional less prominent shadows are also observed elsewhere in the image, particularly in the upper right quadrant, in the vicinity of the ``SCA 4, Mag 4'' label. Internal discussions with detector experts indicated that this flux halo and the defect shadows are a known effect and can be simultaneously explained by the process of spontaneous radiative emission of photons near the material bandgap edge. 

In this scenario, the flux halo surrounding the bright point source is caused by light generated inside the detector layer that travels laterally through the material. When the intense input light hits the central pixels, it leads to a flood of excess electrons and holes, effectively saturating the material at a physical level. In this state, the HgCdTe material temporarily loses its ability to absorb its own light, and the electron-hole pairs rapidly recombine to emit new photons at the material's 2.5 $\mu$m bandgap edge. Because pixels in and near the PSF core have become temporarily transparent, the freshly generated photons escape sideways, traveling outward through the detector layer. As these photons propagate outward into darker, unaffected regions of the semiconductor, they are gradually reabsorbed and re-emitted in all directions, creating the smoothly fading radial gradient seen around the core. The appearance of distinct shadows on the far side of material defects is a byproduct of the lateral photon propagation. The material flaws act as obstacles that absorb or scatter the laterally-traveling 2.5 $\mu$m photons. By cutting off this internal light path, the defects prevent downstream pixels from receiving and re-emitting the photons, casting a clear shadow.

These effects have been previously observed in similar semiconductor materials and are not a-priori unexpected under the extreme flux regimes tested here. The physics are relatively well understood but significant work is required to fully characterize the observed flux distribution, how it varies as a function of input source brightness, and how it contributes to flux/charge conservation in a given integration. Such effects may become particularly important for high accuracy flux measurements of highly saturated sources. We note that the same effect has been observed in flight data from the JWST NIRCam instrument and detailed analyses of the flux halo properties and an associated descriptive model are in progress (T. Brandt, private communication). 

\subsection{Future Work and Data Availability}

The TVAC2 Bright Star saturation test is a rich dataset, presenting significant opportunities for additional analysis. Potential avenues for future work include:
\begin{itemize}
    \item Further analysis of saturation behavior to measure photometric properties and systematics for stars at varying levels of saturation and investigate charge conservation. These test data, combined with flat field and other saturation results, may ultimately allow the development of a saturation model.
    \item Deeper persistence analyses to better characterize post illumination persistence baselines and decay timescales as a function of source brightness. This includes frame-by-frame decay analysis and investigating evidence of radial persistence signal gradients in saturated PSF cores (see Fig.~\ref{fig:SCA11Mag4_FiberContamination}). Such analyses, when combined with persistence measurements post flat field illumination, set the stage for the development of a Roman H4RG persistence model.
    \item Analysis of the flux halo mentioned in the previous sub-section to characterize halo properties as a function of point source flux, understand contributions to bright source photometry, and identify additional related properties.
\end{itemize}

To facilitate additional analyses, we provide the IRRC and gain corrected data cubes, as well as animations and additional plots produced during our analysis, under the following Zenodo DOI: \href{https://zenodo.org/records/21358405}{10.5281/zenodo.21358405} \bt{\textbf{Note the Zenodo repository will be published after paper acceptance.}}

Additionally, Space Telescope Science Institute repositories provide access to the TVAC2 raw data files used in this analysis.\footnote{See Table of product availability at: \protect\url{https://roman-docs.stsci.edu/roman-instruments/the-wide-field-instrument/wfi-characterization-activities/wfi-ground-testing-campaigns}} TVAC2 data are available on the Roman Integration \& Test Archive (RITA).\footnote{See \protect\url{https://mast.stsci.edu/cassi/\#/roman} } Additionally, the Roman Research Nexus\footnote{See \protect\url{https://roman-docs.stsci.edu/data-handbook/roman-research-nexus}} hosts a curated dataset of TVAC2 bright star data, along with an accompanying Jupyter notebook tutorial. We encourage interested members of the community to access these data and continue analyses of saturation, persistence, and other effects.

\begin{acknowledgments}

The authors acknowledge use of the Advanced Data Analytics Platform Technologies (ADAPT) Science Cloud and its GPU PRISM cluster provided by the NASA Center for Climate Simulation (NCCS) at Goddard Space Flight Center. 

The material is based upon work supported by NASA under award number 80GSFC24M0006.

\end{acknowledgments}

%

\vspace{5mm}
\facilities{Roman Space Telescope Wide Field Instrument, BAE Titan Thermal Vacuum Chamber}


\software{numpy \citep{numpy2020}, scipy \citep{scipy2020}, astropy \citep{2013Astropy}, matplotlib \citep{Matplotlib}, stpsf \citep{stpsf1, stpsf2}.    
          }

\clearpage

\appendix

\clearpage

\section{Appendix information}

This appendix contains additional Figures to support the analysis presented in the main text.

\clearpage

\begin{figure*}
\centering
\includegraphics[width=0.45\textwidth]{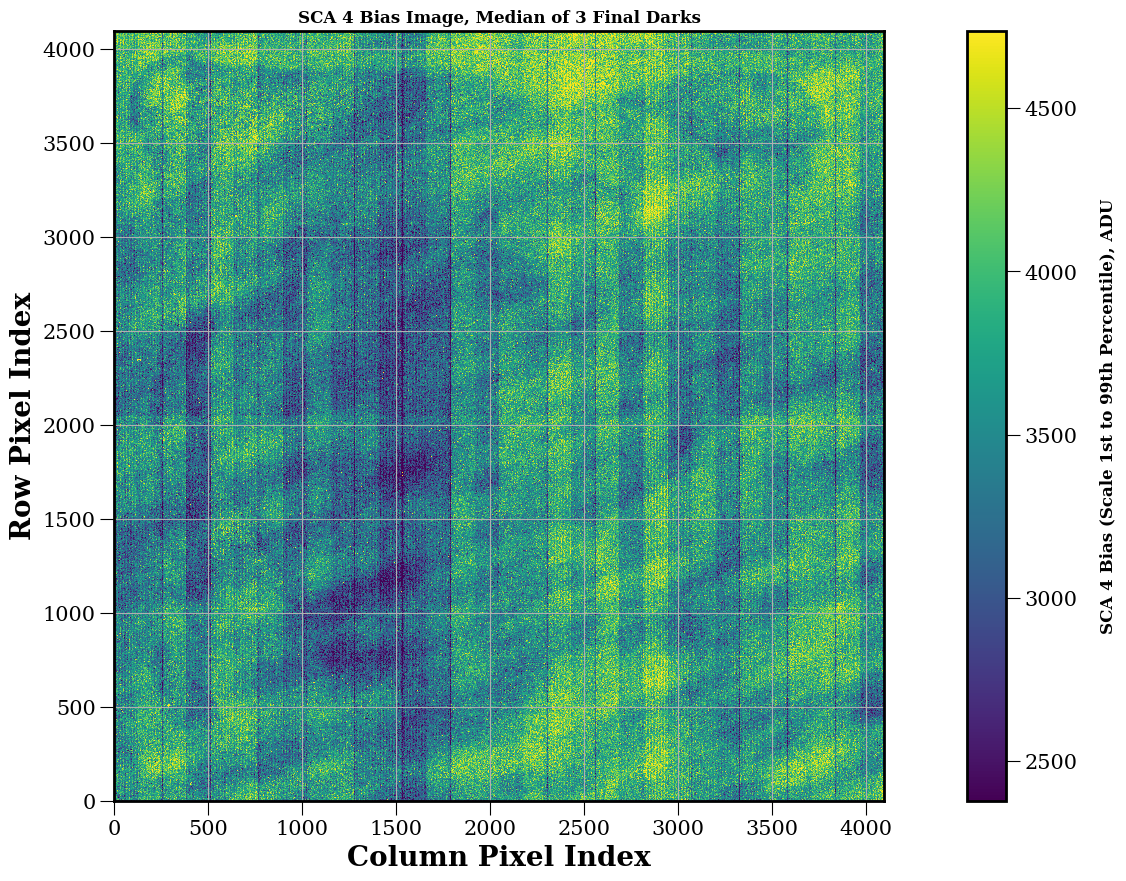}
\hspace{1.0 cm}
\includegraphics[width=0.45\textwidth]{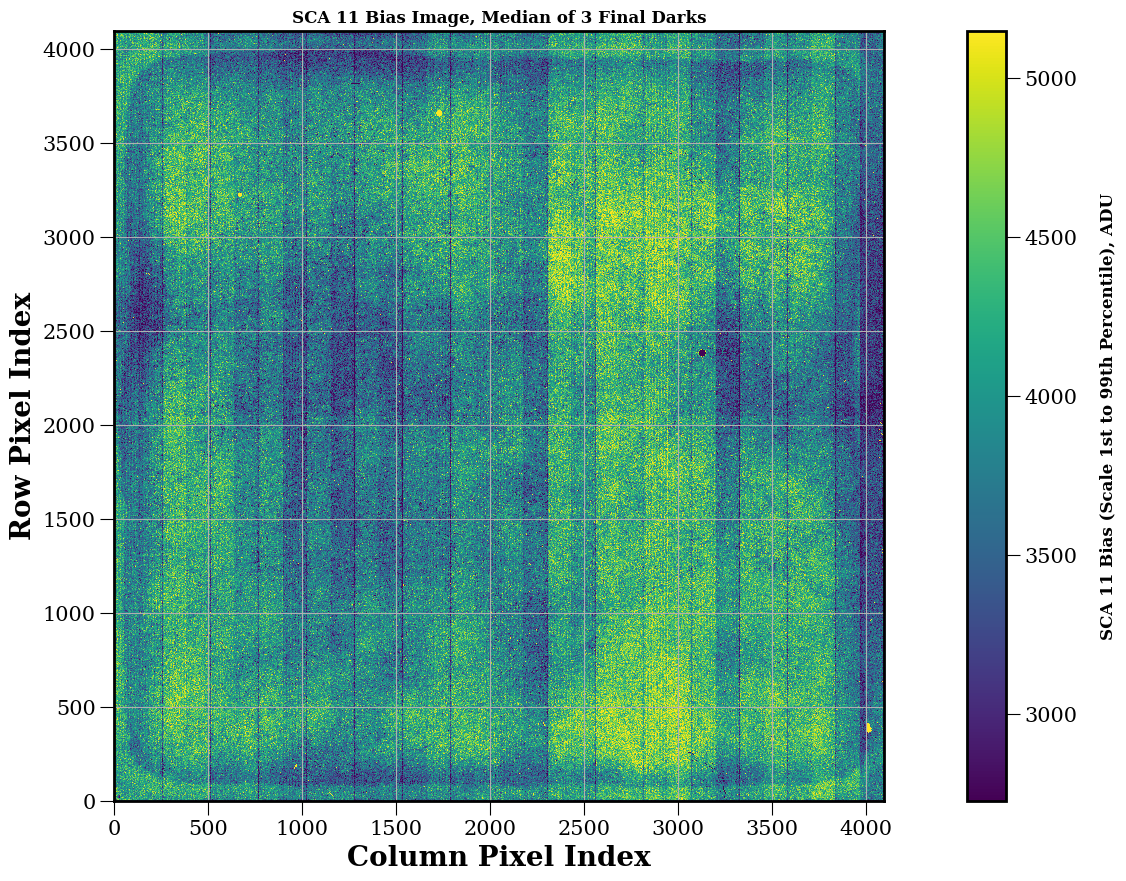}
\caption{ SCA 4 (\textbf{\underline{left}}) and 11 (\textbf{\underline{right}}) superbias frames subtracted during our IRRC correction. 
 \label{fig:methods_Superbias}}
\end{figure*}

\begin{figure*}
\centering
\includegraphics[width=0.45\textwidth]{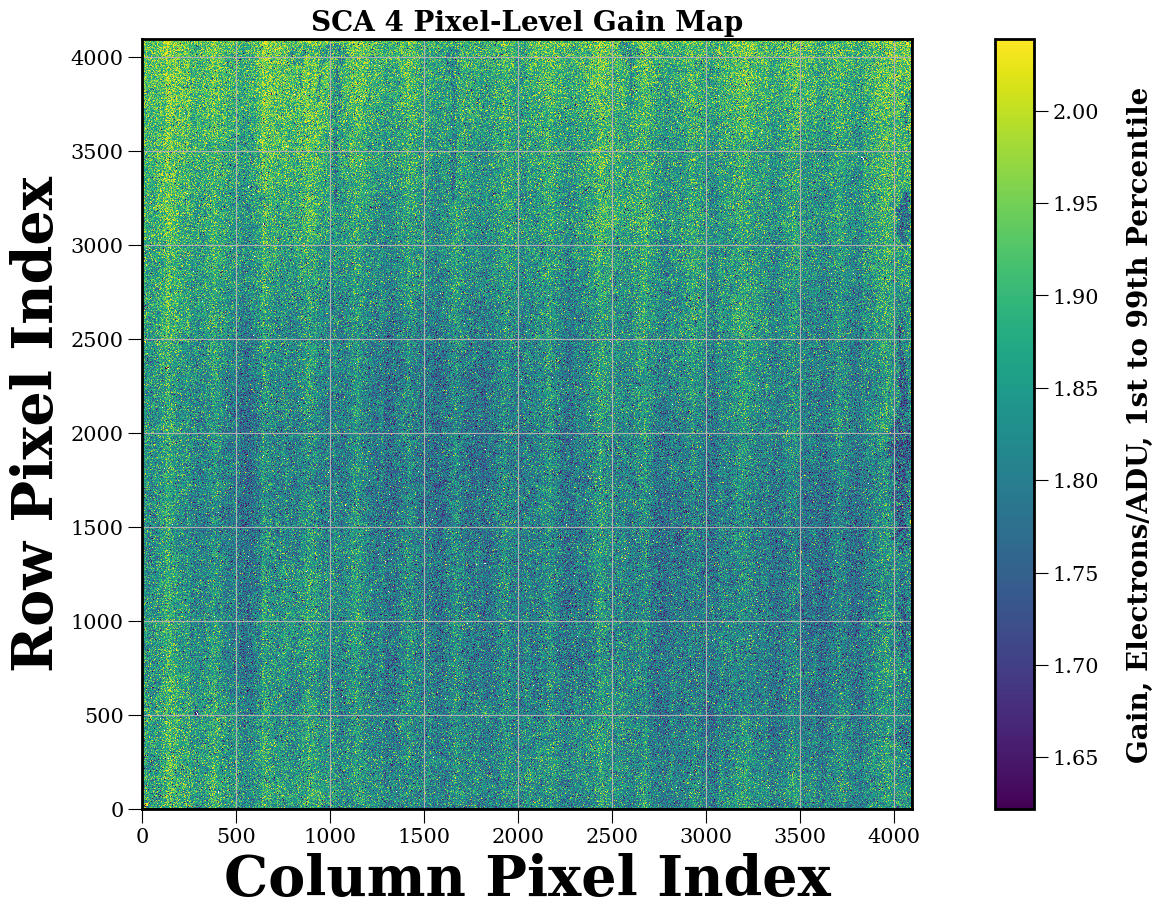}
\hspace{1.0 cm}
\includegraphics[width=0.45\textwidth]{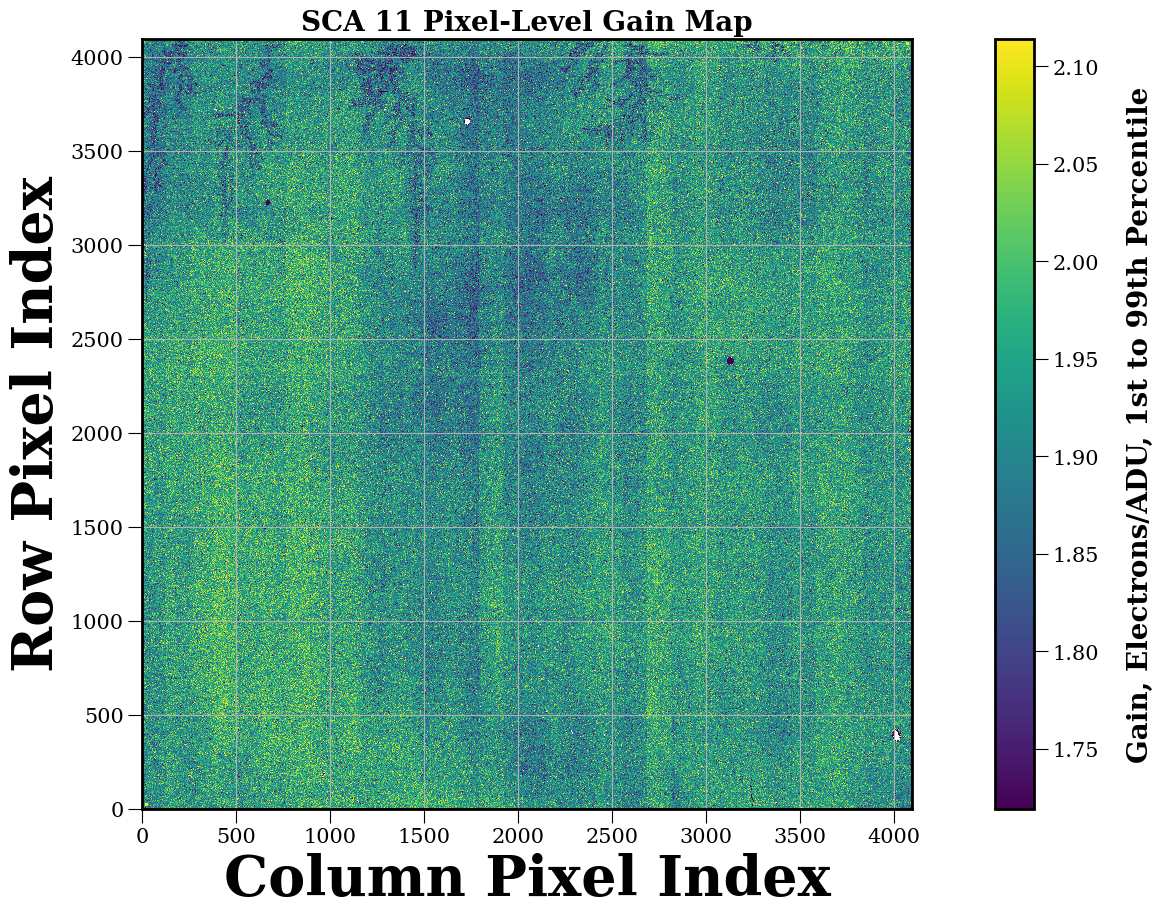}
\caption{ Frames for SCA 4 (\textbf{\underline{left}}) and 11 (\textbf{\underline{right}}) pixel-level photon transfer gains applied following IRRC correction and superbias subtraction. 
 \label{fig:methods_Gain}}
\end{figure*}

\begin{figure}
\includegraphics[width=1.0\textwidth]{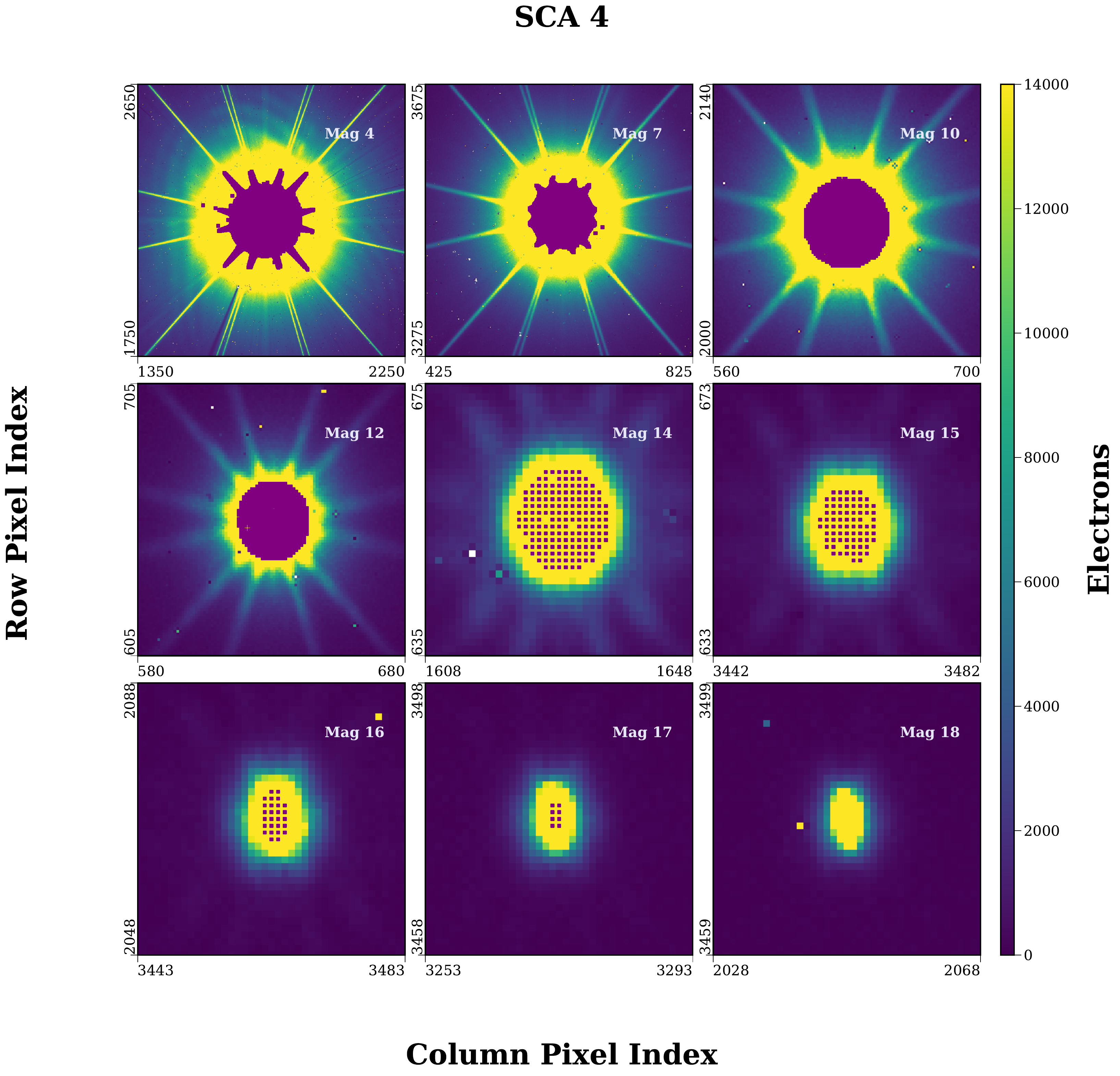}
\caption{ Final frame of SCA 4 illuminated data overplotted with saturation masks for each source magnitude. Those pixels that are part of the saturation mask are colored purple. Sources appear from brightest to dimmest, going from left to right and top to bottom. Our \textbf{\textit{saturation mask}} flagged those pixels with electron counts between 100,000 and 130,000. Note that \textit{none} of the pixels within the $\sim$18 mag source reached saturation. The maximum electron count reached $\sim$99,600 e$^-$ in that ROI. We eliminated bad pixels from our persistence analysis using a bad pixel mask. 
    }
\label{fig:SCA4frame56_OverplottedSatMasks_AllMags}
\end{figure}

\begin{figure}
\includegraphics[width=1.0\textwidth]{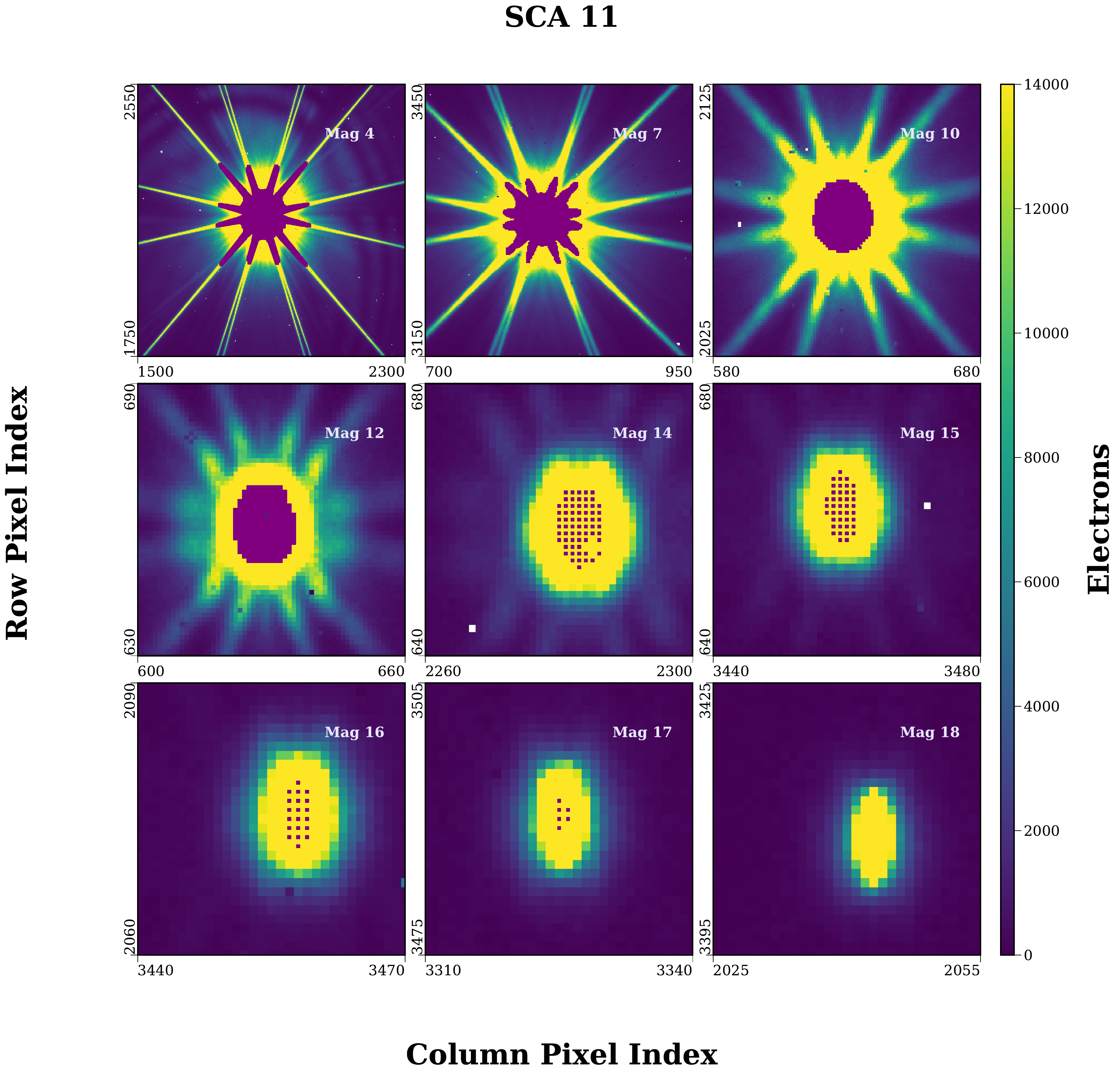}
\caption{Final frame of SCA 11 illuminated data overplotted with saturation masks for each source magnitude. Those pixels that are part of the saturation mask are colored purple. Sources appear from brightest to dimmest, going from left to right and top to bottom. Our \textbf{\textit{saturation mask}} flagged those pixels with electron counts between 100,000 and 130,000. Note that \textit{none} of the pixels within the $\sim$18 mag source reached saturation. The maximum electron count reached in that ROI was $\sim$91,000 e$^-$. We eliminated bad pixels from our persistence analysis using a bad pixel mask. 
    }
\label{fig:SCA11frame56_OverplottedSatMasks_AllMags}
\end{figure}

\begin{figure*}
\includegraphics[width=0.45\textwidth]{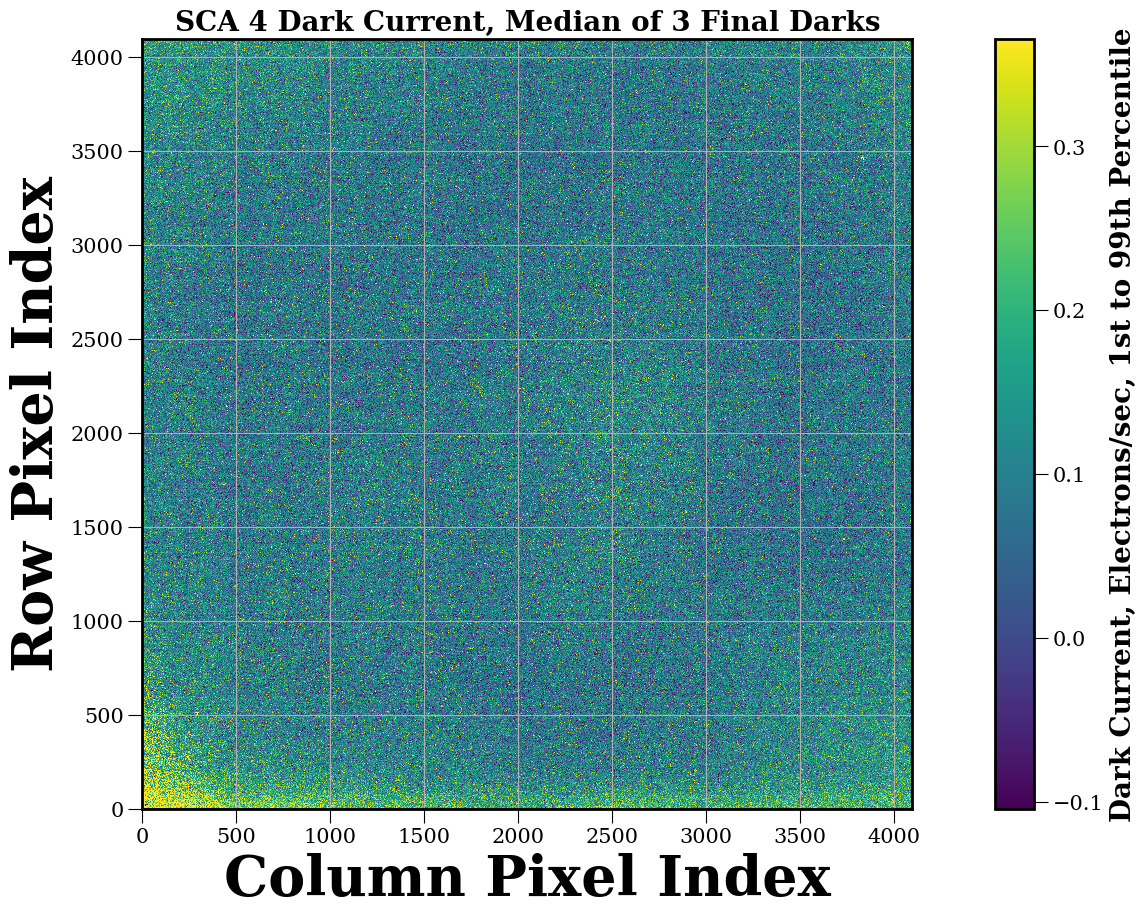}
\hspace{1.0 cm}
\includegraphics[width=0.45\textwidth]{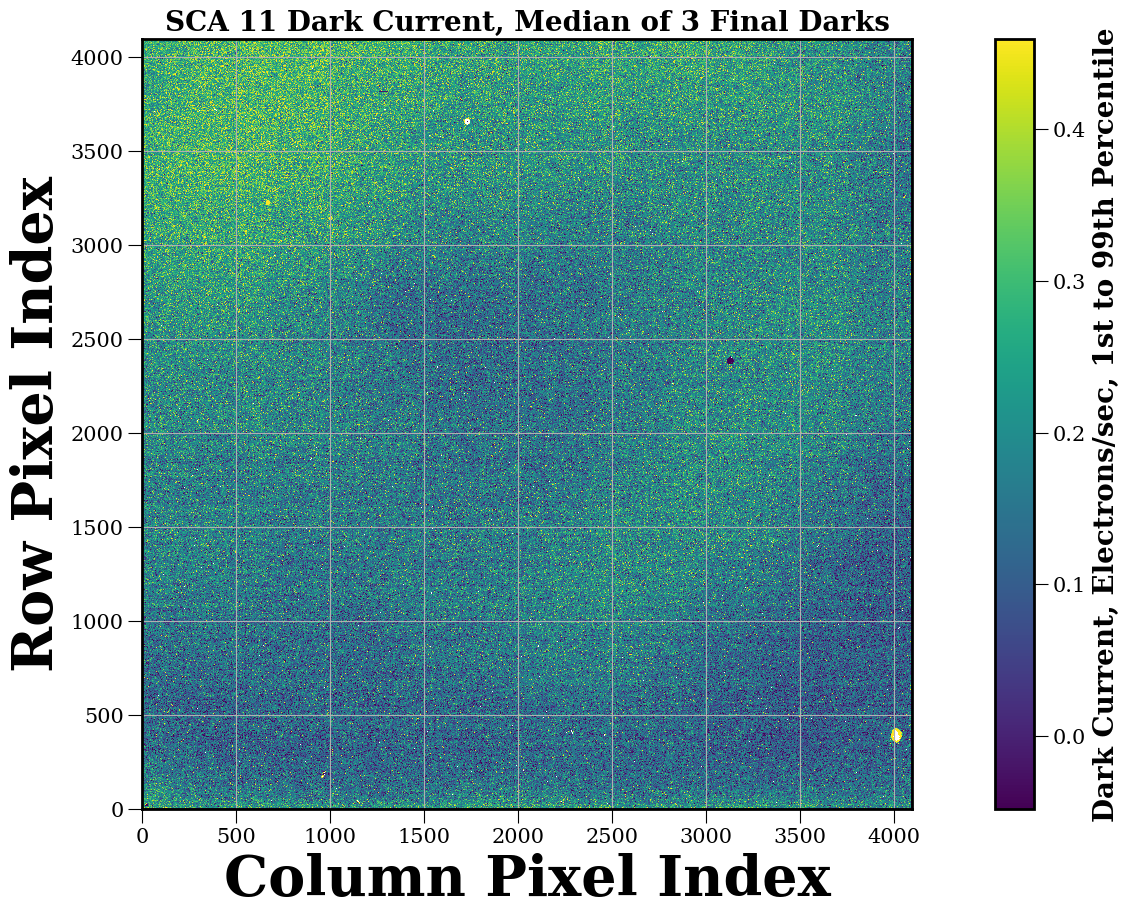}
\caption{ SCA 4 (\textbf{\underline{left}}) and 11 (\textbf{\underline{right}}) dark current frames used in persistence analysis.
    }
\label{fig:methods_SCA11DarkCurrent}
\end{figure*}

\begin{figure}
\includegraphics[width=1.0\textwidth]{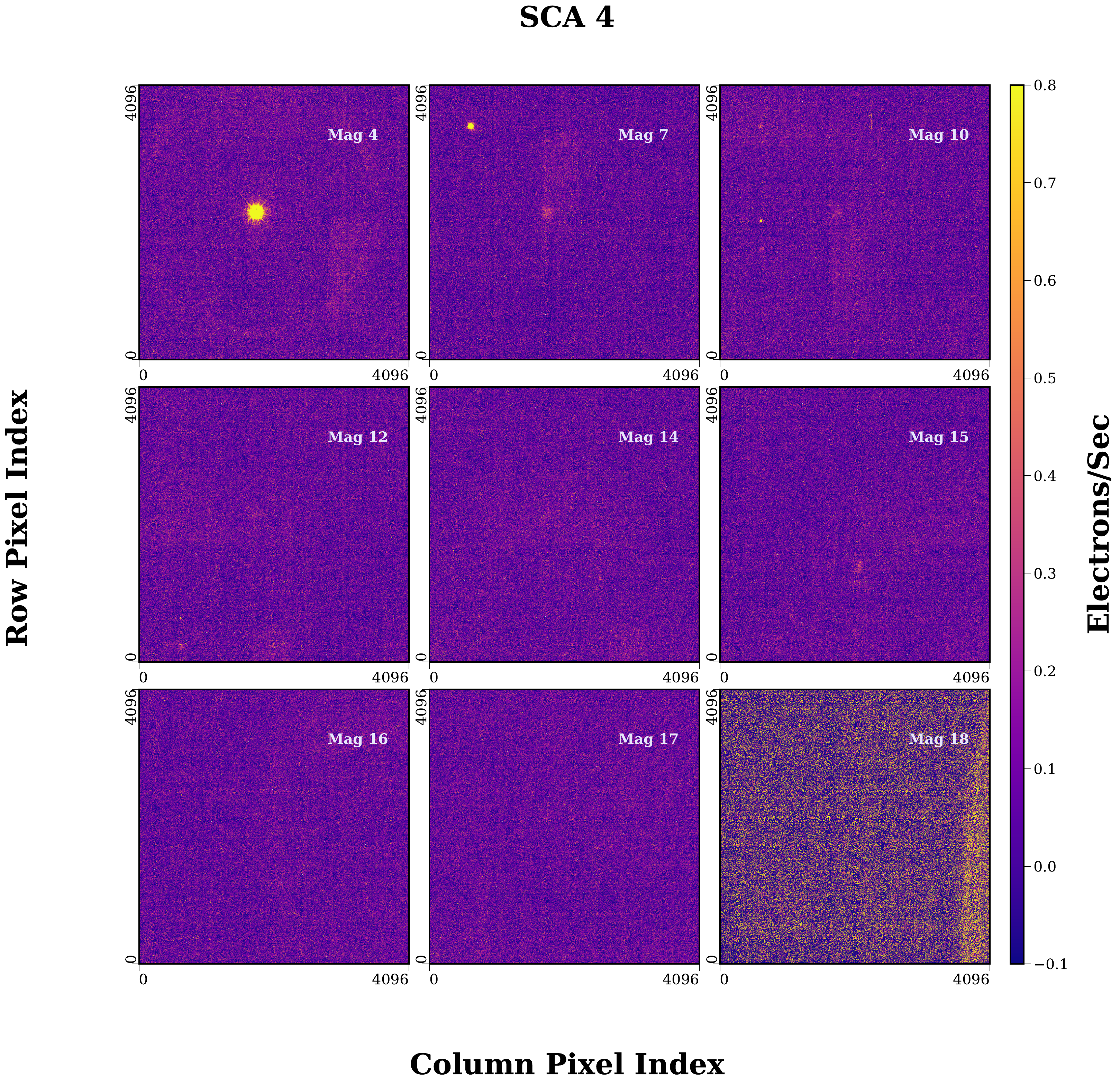}
\caption{ SCA 4 Persistence Frames produced from interleaved dark exposures and the three post-test final darks, as explained in Section \ref{subsubsec:methods_Persistence}. As compared to SCA 11, stray light artifacts from the SORC telescope simulator appear more diffuse and less structured on SCA 4. However, the first interleaved dark persistence frame following $\sim$18 mag illumination (lower right panel) suffered from severe SORC projector contamination, which occurred when the SORC projector was moved to SCA 11 immediately following the illumination of the final $\sim$18 mag source in the sequence.
    }
\label{fig:methods_SCA4PersistenceFrames}
\end{figure}

\begin{figure}
\includegraphics[width=1.0\textwidth]{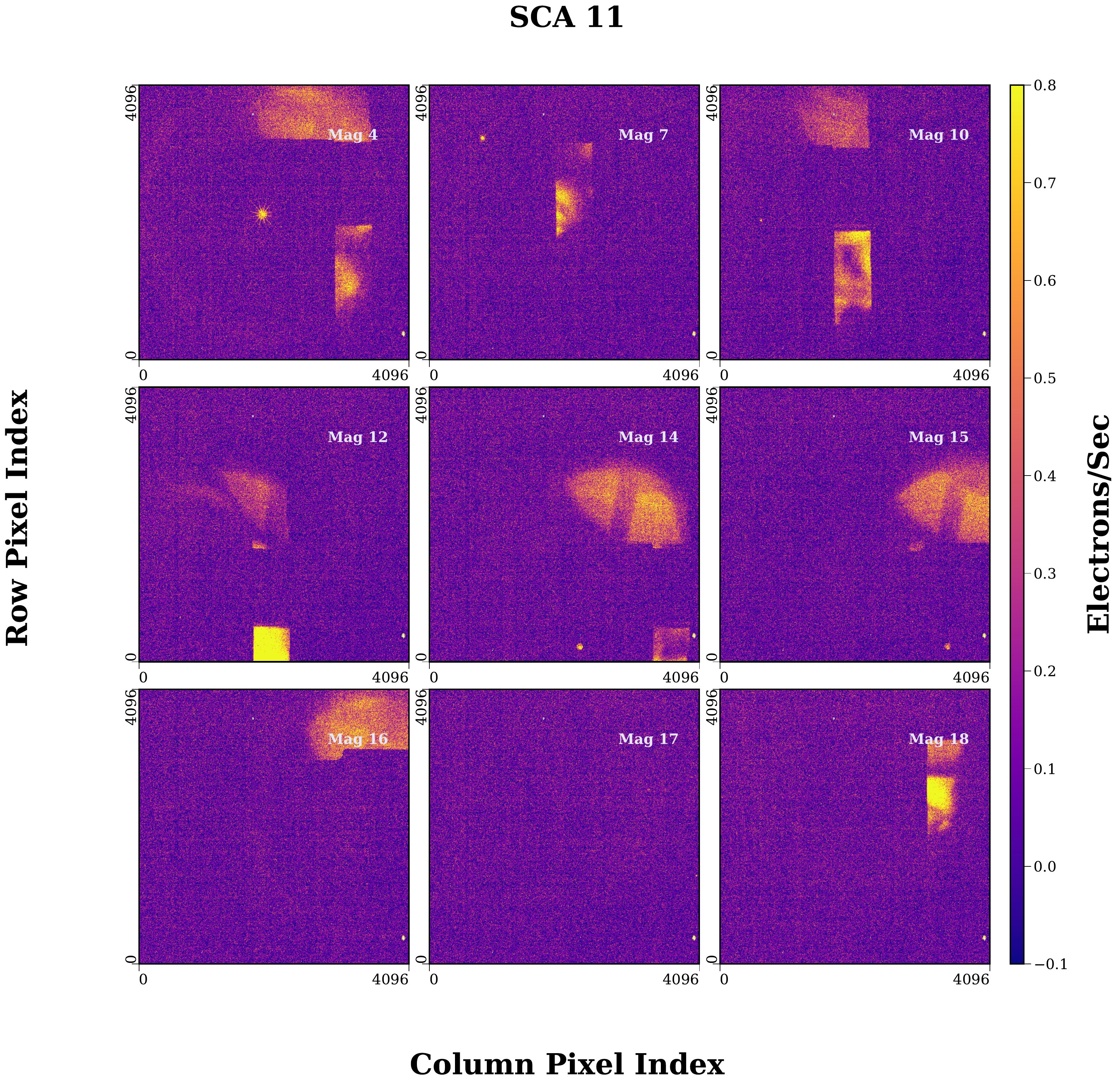}
\caption{ SCA 11 Persistence Frames produced from interleaved dark exposures and the three post-test final darks, as explained in Section \ref{subsubsec:methods_Persistence}. The SORC telescope simulator produced stray light artifacts that our team dealt with to produce our final persistence decay curves. 
    }
\label{fig:methods_SCA11PersistenceFrames}
\end{figure}

\begin{figure*}
\centering
\includegraphics[width=0.9\textwidth]{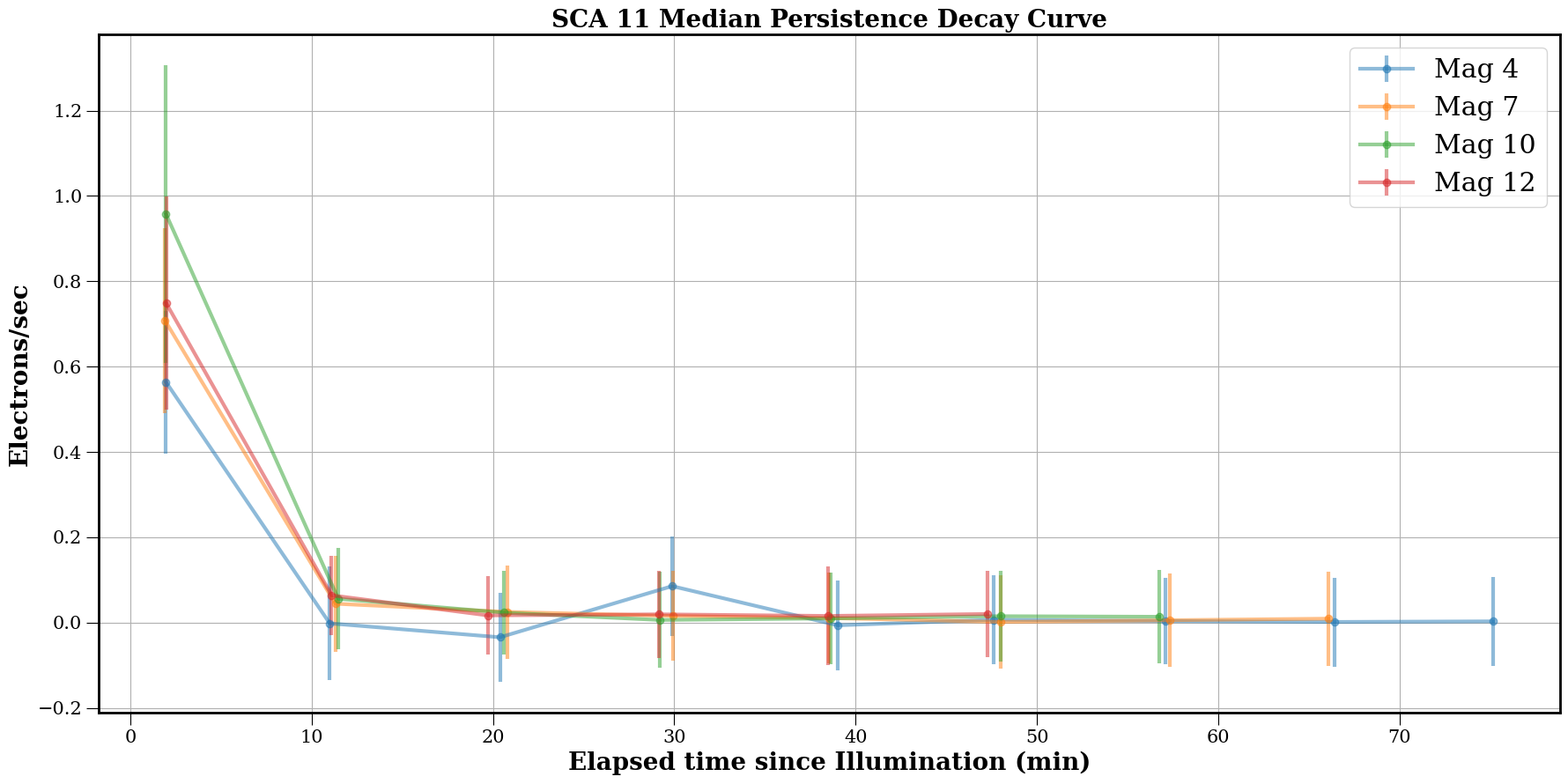}
\caption{SCA 11 Persistence Decay Curves for the 4 brightest magnitude sources in our study. The first interleaved dark persistence values largely agree within the errors bars. However, we note that the mag $\sim$4 persistence value is slightly lower than the value for other magnitudes, while the values for magnitude $\sim$10 is slightly higher. Additionally, the decay curve for the mag $\sim$4 source appears to suffer from oversubtraction in the first 3 data points, while stray light biases the 4th data point on the curve.
    }
\label{fig:SCA11_Mags4to12_PersistenceDecayCurves}
\end{figure*}
\begin{figure*}
\centering
\includegraphics[width=0.9\textwidth]{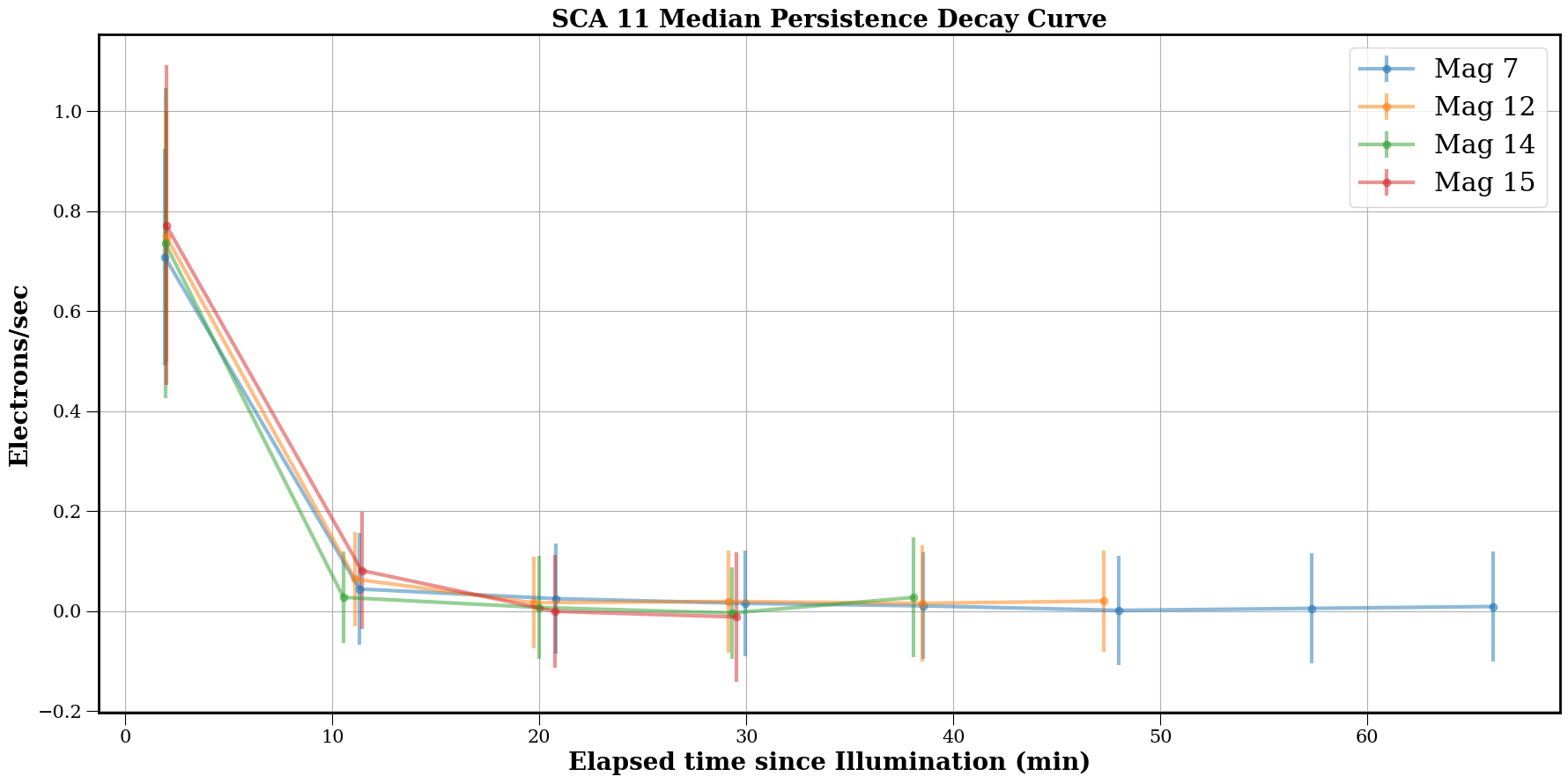}
\caption{SCA 11 Persistence Decay Curves for magnitudes 7, 12, 14, and 15. 
    }
\label{fig:SCA11_Mags7_12_14_15_PersistenceDecayCurves}
\end{figure*}

\begin{figure*}
\centering
\includegraphics[width=0.9\textwidth]{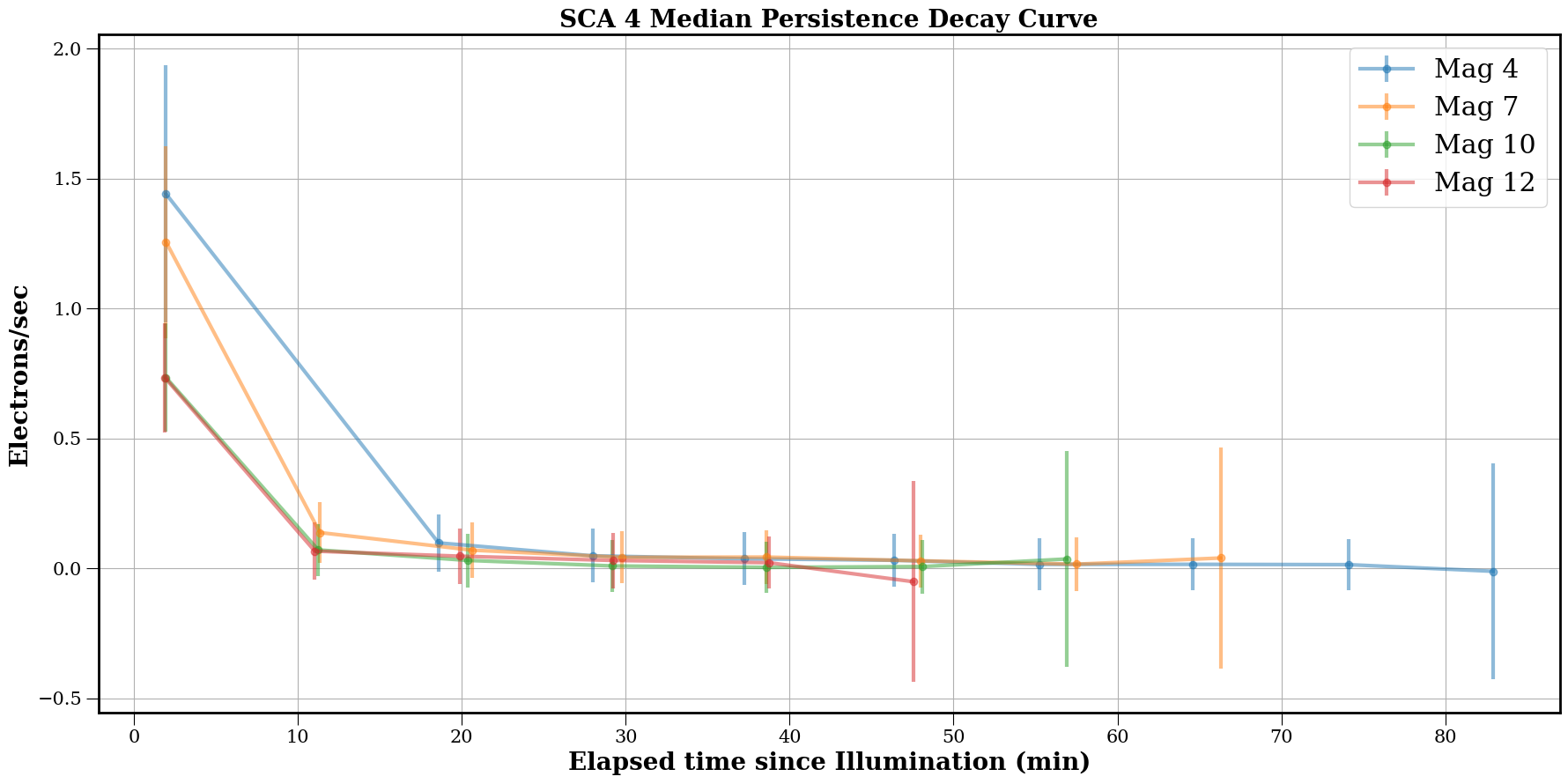}
\caption{SCA 4 Persistence Decay Curves for the 4 brightest magnitude sources in our study. The first interleaved dark values do not agree within the error bars. Underlying persistence performance varies spatially across SCA 4, and thus persistence values in the first interleaved dark exposure were impacted by the region of SCA 4 where the illuminated source was projected. The slightly discrepant persistence values and larger error bars for the last data point in each curve were caused by higher background levels in the final interleaved dark exposure, which took place following the magnitude $\sim$18 illumination. The larger background levels were caused by stray light from the SORC as it was moved to begin the SCA 11 projection sequence. The larger background levels for the $\sim$18 mag source are evident in the bottom right frame of Appendix Figure \ref{fig:methods_SCA4PersistenceFrames}. 
    }
\label{fig:SCA4_Mags4to12_PersistenceDecayCurves}
\end{figure*}

\begin{figure*}
\centering
\includegraphics[width=0.9\textwidth]{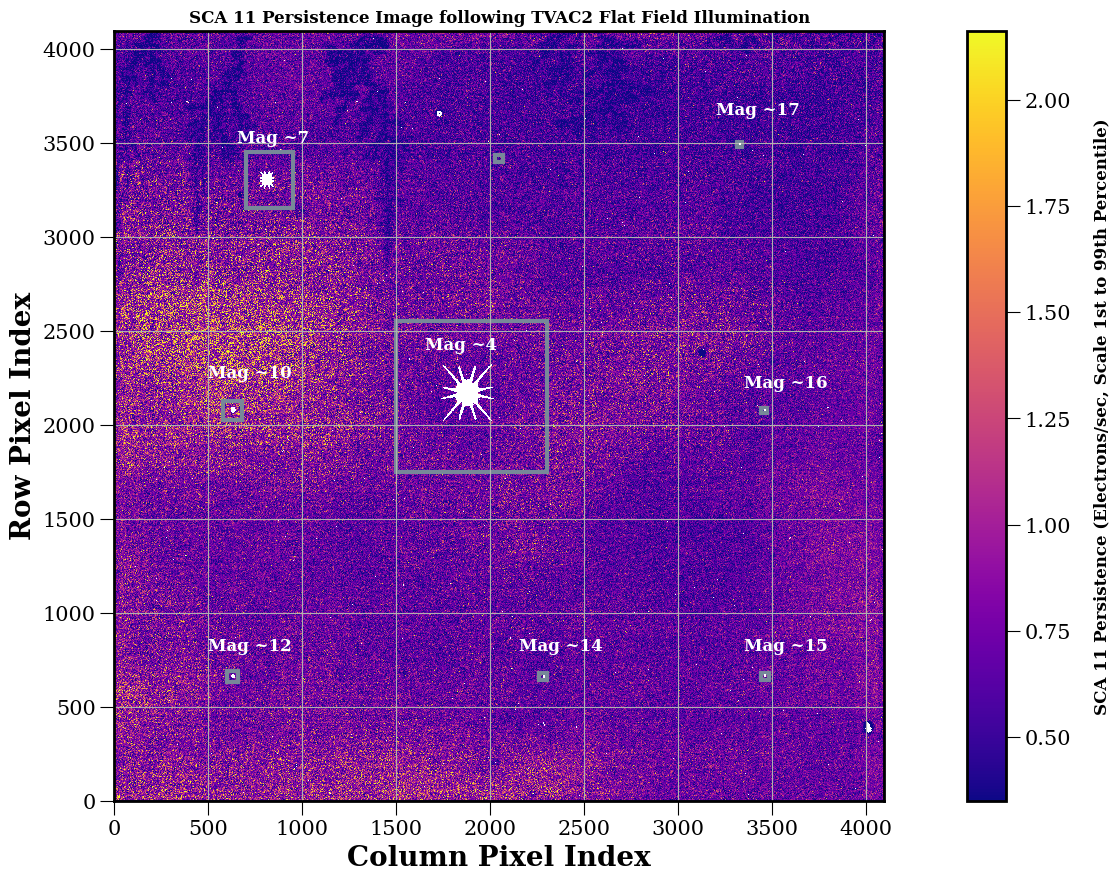}
\caption{SCA 11 persistence measured from a dark exposure following flat field illumination during TVAC2. The bright point source saturation masks for each magnitude are overplotted in white, with the accompanying ROIs outlined using gray boxes. The spatial variation of persistence performance across SCA 11 is evident, but is less severe than that shown for SCA 4 in Figure \ref{fig:SCA4_ComparePersistence_BrightSourceVsFlatField}. 
    }
\label{fig:SCA11_FlatFieldIlluminationWSatMasksROIsOverplotted}
\end{figure*}

\begin{figure*}
\centering
\includegraphics[width=0.9\textwidth]{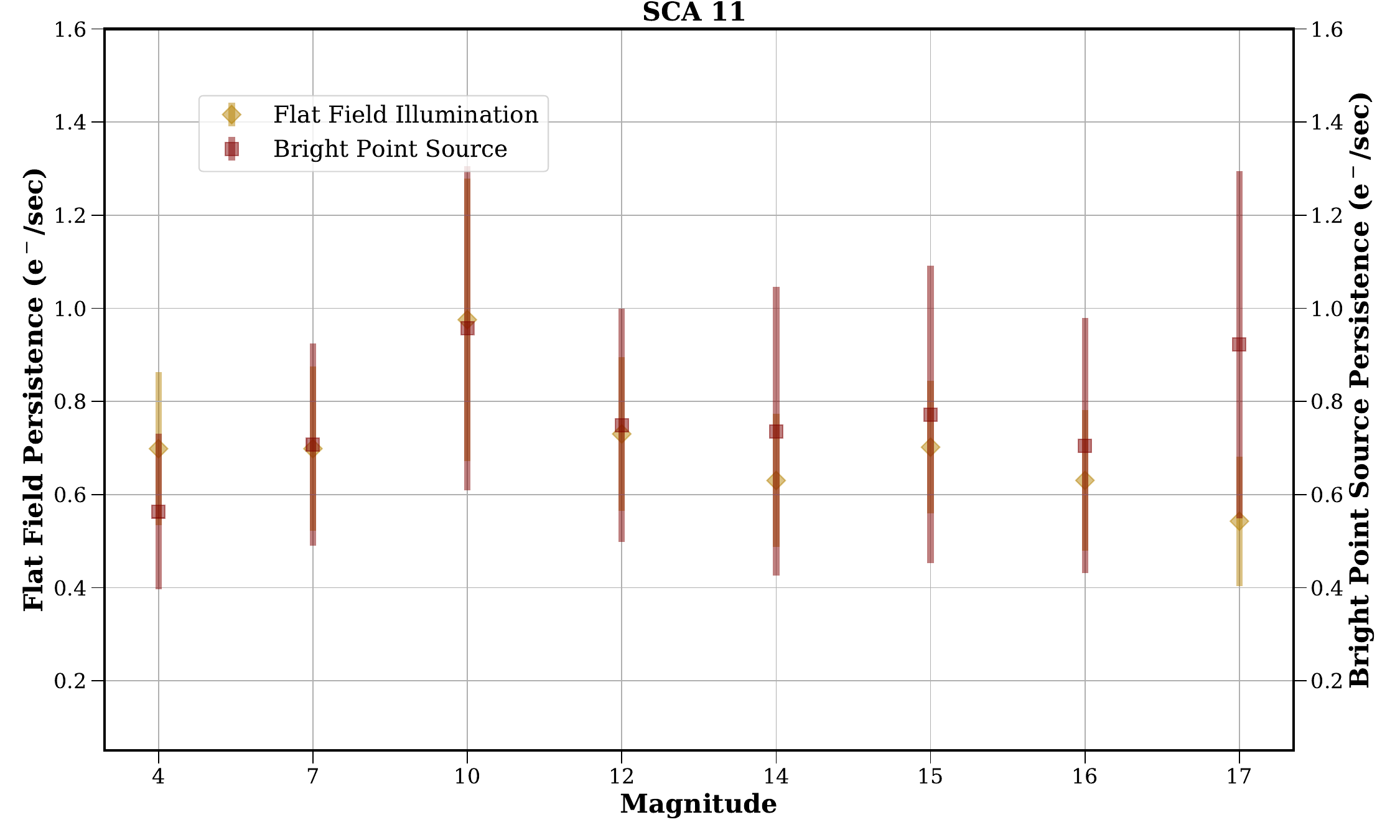}
\caption{Comparison of SCA 11 persistence measurements in subsequent dark frames following two separate TVAC2 tests: 1.) flat field illumination and 2.) bright point source projection. Flat field persistence shown was computed within the ROIs designated in Figure \ref{fig:SCA11_FlatFieldIlluminationWSatMasksROIsOverplotted} for the magnitudes indicated on the x-axis. The bright point source persistence is the same as that plotted in Figure \ref{fig:SCA11_FirstInterleavedDarkPersistenceComparison} for the first interleaved dark frames. The Flat Field and Bright Point Source persistence values agree within the error bars for all magnitudes of Bright Point Source data.  
    }
\label{fig:SCA11_ComparePersistence_BrightSourceVsFlatField}
\end{figure*}

\clearpage

\bibliography{tvac2_brightstar}{}
\bibliographystyle{aasjournal}

\end{document}